\documentclass[modern]{aastex62}

\usepackage{amsmath, amsthm, amssymb, amsfonts}

\newcommand{\Rp}{R_p}

\newcommand{\Rs}{R_{\star}}
\newcommand{\RpRs}{\Rp/\Rs}

\newcommand{\Teq}{T_{\textnormal{eq}}}
\newcommand{\RE}{R_{\Earth}}
\newcommand{\ME}{M_{\Earth}}
\newcommand{\RS}{R_{\odot}}

\newcommand{\muatm}{\mu_{\textnormal{atm}}}
\newcommand{\aRs}{a/\Rs}

\newcommand{\Tatm}{T_{\textnormal{atm}}}
\newcommand{\Pc}{P_{\textnormal{c}}}

\newcommand{\um}{\mu\textnormal{m}}

\newcommand{\GJRS}{0.201_{-0.003}^{+0.004}\,\RS}
\newcommand{\HDRS}{0.872 \pm 0.057\,\RS}
\newcommand{\HDMP}{9.80_{-1.24}^{+1.30}\,\ME}
\newcommand{\KTEQ}{543 \pm 11}
\newcommand{\KMP}{2.1_{-0.8}^{+1.5}\,\ME}
\newcommand{\KRP}{7.1 \pm 0.3\,\RE}

\submitjournal{AJ (now accepted)}

\shorttitle{Atmospheric characterization of HD~3167c}
\shortauthors{Mikal-Evans et al.}

\begin{document}

\title{Transmission spectroscopy for the warm sub-Neptune HD~3167c: evidence for molecular absorption and a possible high metallicity atmosphere}

\correspondingauthor{Thomas Mikal-Evans}
\email{tmevans@mit.edu}

\author[0000-0001-5442-1300]{Thomas Mikal-Evans}
\affil{Kavli Institute for Astrophysics and Space Research, Massachusetts Institute of Technology, 77 Massachusetts Avenue, 37-241, Cambridge, MA 02139, USA}

\author{Ian J.\ M.\ Crossfield}
\affiliation{Department of Physics and Astronomy, University of
  Kansas, Lawrence, KS, USA}

\author{Bj\"orn Benneke}
\affiliation{Departement de Physique, and Institute for Research on Exoplanets, Universite de Montreal, Montreal, Canada}

\author[0000-0003-0514-1147]{Laura Kreidberg}
\affiliation{Max Planck Institute for Astronomy, K{\"o}nigstuhl 17, 69117 Heidelberg, Germany}
\affiliation{Center for Astrophysics $|$ Harvard \& Smithsonian, 60 Garden Street, Cambridge, MA, 02138, USA}

\author[0000-0002-8837-0035]{Julie Moses}
\affiliation{Space Science Institute, 4765 Walnut St Suite B, Boulder, CO 80301, USA}

\author[0000-0002-4404-0456]{Caroline V. Morley}
\affiliation{Department of Astronomy, University of Texas at Austin, Austin, TX 78712, USA}

\author{Daniel Thorngren}
\affil{Institute for Research on Exoplanets (iREx), Universit\'e de Montr\'eal, Montr\'eal, Canada}

\author[0000-0003-4096-7067]{Paul Molli\`ere}
\affiliation{Max-Planck-Institut f\"ur Astronomie, K\"onigstuhl 17, 69117 Heidelberg, Germany}

\author[0000-0003-3702-0382]{Kevin K. Hardegree-Ullman}
\affil{Caltech/IPAC-NExScI, M/S 100-22, 1200 E California Blvd, Pasadena, CA 91125, USA}

\author{John Brewer} 
\affil{SFSU}


\author[0000-0002-8035-4778]{Jessie L. Christiansen}
\affil{Caltech/IPAC-NExScI, M/S 100-22, 1200 E California Blvd, Pasadena, CA 91125, USA}

\author[0000-0002-5741-3047]{David R. Ciardi}
\affil{Caltech/IPAC-NExScI, M/S 100-22, 1200 E California Blvd, Pasadena, CA 91125, USA}

\author[0000-0003-2313-467X]{Diana~Dragomir}
\affiliation{Department of Physics and Astronomy, University of New Mexico, 1919 Lomas Blvd NE, Albuquerque, NM 87131, USA}

\author[0000-0001-8189-0233]{Courtney Dressing}
\affiliation{Department of Astronomy, University of California Berkeley, Berkeley, CA 94720, USA}

\author[0000-0002-9843-4354]{Jonathan J. Fortney}
\affiliation{Department of Astronomy \& Astrophysics, University of California,  Santa Cruz, CA 95064, USA}

\author[0000-0002-8990-2101]{Varoujan Gorjian}
\affiliation{MS 169-506, Jet Propulsion Laboratory, California Institute of Technology, 4800 Oak Grove Drive, Pasadena, CA 91109, USA}

\author[0000-0002-8963-8056]{Thomas P. Greene}
\affiliation{Space Science and Astrobiology Division, NASA Ames Research Center, MS 245-6, Moffett Field, CA 94035, USA}

\author[0000-0001-8058-7443]{Lea A. Hirsch} 
\affil{Kavli Institute for Particle Astrophysics and Cosmology, Stanford University, Stanford, CA 94305, USA} 

\author[0000-0001-8638-0320]{Andrew W. Howard}
\affiliation{Department of Astronomy, California Institute of Technology, Pasadena, CA 91125, USA}

\author[0000-0002-2532-2853]{Steve~B.~Howell}
\affil{NASA Ames Research Center, Moffett Field, CA 94035, USA}

\author{Howard Isaacson}
\affiliation{Astronomy Department, University of California, Berkeley, CA, USA}
\affiliation{University of Southern Queensland, Toowoomba, QLD 4350, Australia}

\author[0000-0002-6115-4359]{Molly R.\ Kosiarek}
\altaffiliation{NSF Graduate Student Research Fellowship}
\affiliation{Department of Astronomy and Astrophysics, University of California, Santa Cruz, CA 95064, USA}

\author[0000-0002-2413-5976]{Jessica Krick}
\affiliation{Caltech/IPAC, 1200 E. California Blvd, Pasadena, CA 91125, USA}

\author[0000-0002-4881-3620]{John~H.~Livingston}
\affiliation{Department of Astronomy, University of Tokyo, 7-3-1 Hongo, Bunkyo-ku, Tokyo 113-0033, Japan}

\author[0000-0003-3667-8633]{Joshua D. Lothringer}
\affiliation{Department of Physics and Astronomy, Johns Hopkins University, Baltimore, MD 21210, USA}

\author[0000-0001-9414-3851]{Farisa Y. Morales}
\affiliation{MS 169-214, Jet Propulsion Laboratory, California Institute of Technology, 4800 Oak Grove Drive, Pasadena, CA 91109, USA}

\author[0000-0003-0967-2893]{Erik A Petigura}
\affiliation{Department of Physics \& Astronomy, University of California Los Angeles, Los Angeles, CA 90095, USA}

\author[0000-0001-5347-7062]{Joshua E. Schlieder}
\affiliation{Exoplanets and Stellar Astrophysics Laboratory, Code 667, NASA Goddard Space Flight Center, Greenbelt, MD 20771, USA}

\author{Michael Werner}
\affiliation{MS 169-506, Jet Propulsion Laboratory, California Institute of Technology, 4800 Oak Grove Drive, Pasadena, CA 91109, USA}

\begin{abstract}
We present a transmission spectrum for the warm (500$-$600\,K) sub-Neptune HD~3167c obtained using the \textit{Hubble Space Telescope} Wide Field Camera 3 infrared spectrograph. We combine these data, which span the $1.125$-$1.643\,\um$ wavelength range, with broadband transit measurements made using \textit{Kepler/K2} ($0.6$-$0.9\,\um$) and \textit{Spitzer}/IRAC ($4$-$5\,\um$). We find evidence for absorption by at least one of H$_2$O, HCN, CO$_2$, and CH$_4$ (Bayes factor 7.4; $2.5\sigma$ significance), although the data precision does not allow us to unambiguously discriminate between these molecules. The transmission spectrum rules out cloud-free hydrogen-dominated atmospheres with metallicities $\leq 100\times$ solar at $>5.8\sigma$ confidence. In contrast, good agreement with the data is obtained for cloud-free models assuming metallicities $>700\times$ solar. However, for retrieval analyses that include the effect of clouds, a much broader range of metallicities (including subsolar) is consistent with the data, due to the degeneracy with cloud-top pressure. Self-consistent chemistry models that account for photochemistry and vertical mixing are presented for the atmosphere of HD~3167c. The predictions of these models are broadly consistent with our abundance constraints, although this is primarily due to the large uncertainties on the latter. Interior structure models suggest the core mass fraction is $>40\%$, independent of a rock or water core composition, and independent of atmospheric envelope metallicity up to $1000\times$ solar. We also report abundance measurements for fifteen elements in the host star, showing that it has a very nearly solar composition. 
\end{abstract}

\section{Introduction}

The Hubble Space Telescope (HST) has proven a productive facility for characterizing the atmospheres of transiting exoplanets. Observational studies released since 2019 alone include: \cite{2019A&A...625A.136A,2019ApJ...887L..14B,2019NatAs...3..813B,2019AJ....158..244C,2019A&A...629A..47D,2019MNRAS.488.2222M,2020MNRAS.496.1638M,2019AJ....158...91S,2019arXiv191108859S,2020AJ....159..239G,2020AJ....159..234W,2020arXiv200511293A,2020arXiv200502568F,2020AJ....159..204W,2020MNRAS.494.5449C,2020MNRAS.491.5361B,2020AJ....159....5S,2020arXiv200605382C,2020arXiv200607444K}; and \cite{2020arXiv200505153C}. Transmission spectroscopy measurements, made during primary transit, allow the composition of the day-night terminator atmosphere to be probed, while at other phases in the orbit when the irradiated dayside hemisphere is visible, the planetary emission can be constrained \citep[for overviews, see][]{2017JGRE..122...53D,2019PASP..131a3001D}. Most HST transmission and emission spectroscopy observations have been performed using either the Space Telescope Imaging Spectrograph (STIS) at near-ultraviolet(UV)/optical wavelengths \citep[e.g.][]{2020arXiv200511293A,2020arXiv200502568F,2019AJ....158...91S} and Wide Field Camera 3 (WFC3) at near-infrared wavelengths \citep[e.g.][]{2019A&A...625A.136A,2019ApJ...887L..14B,2019NatAs...3..813B,2019MNRAS.488.2222M,2020MNRAS.496.1638M,2020MNRAS.494.5449C}. To date, published observations have mainly focused on hot Jupiters, which are especially favorable targets owing to their large radii and high temperatures \citep[e.g.][]{2016Natur.529...59S}. However, the NASA \textit{Kepler} survey revealed planets Neptune-sized and smaller ($\lesssim 4\,\RE$) are far more common than the hot Jupiters, with an abundance distribution that rises with increasing semimajor axis out to orbital periods of at least 100 days \citep{2010Sci...330..653H,2013ApJ...766...81F,2013ApJ...767...95D,2013ApJ...770...69P,2014ApJ...795...64F}. Characterizing the atmospheres of these smaller and cooler planets, although relatively challenging, is therefore key to our overall understanding of the planetary population.

The observed size distribution of planets smaller than Neptune exhibits a distinct minimum, or ``radius valley'', centered around $\sim 1.8\RE$ for orbital periods shorter than 100 days \citep{2015ApJ...801...41R,2017AJ....154..109F,2018MNRAS.479.4786V,2020AJ....159..211C}. The super-Earths fall below this valley, with bulk density measurements indicating predominantly rocky compositions \citep{2014ApJ...783L...6W}. Any H/He atmospheres these close-in super-Earths may have accreted from the protoplanetary nebula during formation must have been lost, likely by thermal escape \citep[e.g.][]{2013ApJ...776....2L,2014ApJ...792....1L,2017ApJ...843..122Z}. For sub-Neptunes with radii above the valley, there is more ambiguity, as their masses and radii can be explained by various proportions of iron, rock, water, and H/He \citep[e.g.][]{2008ApJ...673.1160A,2007ApJ...665.1413V,2013ApJ...775...10V,2010ApJ...712..974R,2010ApJ...716.1208R,2014ApJ...787..173H,2017A&A...597A..37D,2020ApJ...889...42B}. One possibility is that most of these sub-Neptunes possess rock/iron cores and are surrounded by thick H/He envelopes contributing $\sim$1-10\% of the total planet mass \citep{2013ApJ...776....2L,2014ApJ...792....1L,2018ApJ...853..163J}. A number of popular theories posit that many of the rocky planets below the radius valley are in fact the exposed cores of such ``gas dwarfs'', with primordial H/He atmospheres stripped by processes that may include photoevaporation and core-powered mass loss \citep{2013ApJ...775..105O,2017ApJ...847...29O,2019MNRAS.487...24G}. Under this scenario, the sub-Neptunes are planets drawn from the same initial population, only they managed to retain their H/He envelopes. An alternative suggestion is that many of the sub-Neptunes could instead be comprised of rock and water in comparable proportions, with little or no H/He \citep{2019PNAS..116.9723Z,2020ApJ...896L..22M}. Such worlds would need to form beyond the snow line where water ice is abundant and subsequently migrate to their present close-in orbits. Adding further intrigue to the sub-Neptunes, population synthesis simulations are now capable of reproducing many of the properties of the observed exoplanet sample, but continue to significantly underpredict the frequency of sub-Neptunes \citep{2019ApJ...887..157M}. Clearly, there is much remaining to be learned about how the sub-Neptunes form and what they are composed of, which is all the more significant given the prominent place they occupy in the planetary population.

Transmission spectroscopy observations performed with HST provide a means of directly probing the atmospheric compositions for the most favorable sub-Neptunes. To date, there have been HST transmission spectra published for only four such targets with radii $\sim 1.8$-$4\,\RE$: GJ~1214b \citep{2012ApJ...747...35B,2014Natur.505...69K}; HD~97658b \citep{2014ApJ...794..155K,2020AJ....159..239G}; 55~Cnc~e \citep{2016ApJ...820...99T}; and K2-18b \citep{2019ApJ...887L..14B}. There have also been four HST transmission spectra published for planets with radii somewhat larger than Neptune ($\sim 4$-$5\,\RE$): GJ~436b \citep{2014Natur.505...66K,2018AJ....155...66L}; HAT-P-11b \citep{2014Natur.513..526F,2019AJ....158..244C}; GJ~3470b \citep{2014A&A...570A..89E,2019NatAs...3..813B}; and HD~106315c \citep{2020arXiv200607444K}. Of this combined sample, most have produced detections of spectral features at varying levels of confidence \citep{2017AJ....154..261C}. In particular, statistically strong H$_2$O detections have been made for HAT-P-11b \citep{2014Natur.513..526F}, GJ~3470b \citep{2019NatAs...3..813B}, and K2-18b \citep{2019ApJ...887L..14B}. A more tentative H$_2$O detection has recently been reported for HD~106315c \citep{2020arXiv200607444K}, while HD~97658b shows indications of spectral features, although the interpretation remains uncertain \citep{2020AJ....159..239G}. For 55~Cnc~e --- which with a radius of $1.897_{-0.046}^{+0.044}\,\RE$ \citep{2019ApJ...883...79D} falls at the borderline of the super-Earth and sub-Neptune regimes --- \cite{2016ApJ...820...99T} claimed the detection of a thick atmosphere containing HCN. Only GJ~1214b and GJ~436b have failed to produce spectral feature detections. A deck of obscuring cloud or photochemical haze is required to explain the GJ~1214b spectrum \citep{2014Natur.505...69K}, while for GJ~436b the available data favor high metallicity scenarios \citep[$>600\times$ solar;][]{2017AJ....153...86M}, but are not precise enough to resolve small-amplitude spectral features and rule out a cloud deck. 

In this paper, we present a transmission spectrum for the sub-Neptune HD~3167c measured with HST WFC3, combined with transit photometry obtained with \textit{Kepler K2} and the \textit{Spitzer Space Telescope} Infrared Array Camera (IRAC). The HD~3167 system comprises a bright ($J=7.5$\,mag, $K=7.1$\,mag) early-K-type dwarf located approximately 47 parsec away \citep{2018A&A...616A...1G}, orbited by at least three planets. Two of the known planets transit the host star and were first discovered in \textit{K2} photometry by \cite{2016ApJ...829L...9V}, while the third was detected by radial velocity follow-up measurements and does not transit \citep{2017AJ....154..122C,2017AJ....154..123G}. Our target, HD~3167c, is the outermost of these planets, with a semimajor axis of $\sim 0.18$ astronomical units (AU) and an equilibrium temperature of $\Teq \sim 500$-$600$\,K. It falls squarely in the sub-Neptune regime, with measured mass $9.80_{-1.24}^{+1.30}\,\ME$ and radius $3.01_{-0.28}^{+0.42}\,\RE$ \citep{2017AJ....154..122C}. The innermost planet (HD~3167b) is a super-Earth  with mass $3.58_{-0.26}^{+0.25}\,\ME$ and radius $1.70_{-0.15}^{+0.18}\,\RE$ orbiting at a distance of $\sim 0.02$\,AU from the host star, where it has $\Teq \sim 1700-1900$\,K and photoevaporation would almost certainly have stripped any primordial H/He atmosphere \citep{2019ApJ...879...26K}. The third planet (HD~3167d) orbits at an intermediate distance of $\sim 0.08$\,AU with a minimum mass of $6.90 \pm 0.71\,\ME$ \citep{2017AJ....154..122C}. Notably, radial-velocity measurements made during primary transit indicate via the Rossiter-McLaughlin effect that HD~3167c is on an orbit crossing close to the stellar poles, suggestive of an active dynamical past \citep{2019A&A...631A..28D} or a primordial misalignment of the stellar rotation axis \citep{2010MNRAS.401.1505B,2012Natur.491..418B,2017AJ....154..122C}. Measurements of the stellar H$\alpha$ emission and Ca II H \& K absorption also indicate the host star is relatively quiescent \citep{2017AJ....154..123G}, enhancing the prospects for stable transmission spectroscopy measurements.

The paper is organized as follows. Section \ref{sec:obsdatred} describes the transit observations and data reduction, with light curve fitting presented in Section \ref{sec:lcfits}. Host star elemental abundances derived from high resolution spectra are reported in Section \ref{sec:stellarabund}. The transmission spectrum of the planetary atmosphere and its implications are considered in Section \ref{sec:atmchar}. Models for the planetary interior structure are presented in Section \ref{sec:interior}. We discuss the results in a broader context in Section \ref{sec:discussion} and conclude in Section \ref{sec:conclusion}. We also note that a separate analysis of the HST WFC3 data is presented in a study by Guilluy et al., which we became aware of during the preparation of this manuscript.

\section{Observations and data reduction} \label{sec:obsdatred}

\subsection{HST spectroscopy} \label{sec:wfc3observations}

We observed five primary transits of HD~3167c with HST/WFC3 using the G141 grism, which covers a wavelength range of approximately $1.12$-$1.65\,\um$ with a spectral resolving power of $R \sim 130$ at $\lambda = 1.4\,\um$. The visits were made as part of GO-15333 \citep{2017hst..prop15333C} on 2018 May 22, 2018 July 20, 2019 June 14, 2019 August 12, and 2020 July 5. We refer to these datasets as the G141v1, G141v2, G141v3, G141v4, and G141v5 datasets, respectively. Observations for all visits consisted of seven HST orbits and were made using round-trip spatial scanning with a scan rate of $0.429$\,arcsec\,s$^{-1}$. We adopted the SPARS25 sampling sequence with 4 non-destructive reads per exposure ($\textnormal{NSAMP}=4$) resulting in total integration times of $70\,$s and scans across approximately 240 pixel-rows of the cross-dispersion axis. For each science exposure, only the $512 \times 512$ pixel subarray of the detector containing the target spectrum was read out. With this setup, we obtained 18 science exposures in the first HST orbit following acquisition and 20 exposures in each subsequent HST orbit. Typical peak frame counts were $\sim 47,000$ electrons ($\textnormal{e}^{-}$) per pixel for all visits, which is within the recommended range derived from an ensemble analysis of WFC3 spatial-scan data spanning eight years \citep{2019wfc..rept...12S}. Spectra were extracted from the raw data frames using custom-written Python code,\footnote{https://github.com/thommevans/wfc3} which has been described previously \citep{2016ApJ...822L...4E,2017Natur.548...58E,2019MNRAS.488.2222M}. Further details are provided in Appendix \ref{app:hst:datared}.

\subsection{\textit{K2} and IRAC photometry}

Additional broadband transit measurements were made for HD~3167c at optical wavelengths with \textit{Kepler K2} \citep{2014PASP..126..398H} and at longer infrared wavelengths with \textit{Spitzer} IRAC \citep{2004ApJS..154...10F}. Details of the \textit{K2} observations have previously been reported in \cite{2016ApJ...829L...9V}, \cite{2017AJ....154..122C}, and \cite{2017AJ....154..123G}. For the present study, we used the \dataset[\texttt{K2SFF} photometry]{https://archive.stsci.edu/prepds/k2sff/html/c08/ep220383386.html} \citep{2014PASP..126..948V} available on the Mikulski Archive for Space Telescopes.\footnote{\url{https://archive.stsci.edu/hlsp/k2sff}} The single IRAC observation was made in the $4.5\,\um$ passband on 2016 Oct 31 as part of Program GO-13052 \citep{2016sptz.prop13052W}. Observations were performed in stare mode with exposure times of 0.4\,s and lasted 11.5 hours, including a baseline of 3.8 hours before transit ingress and 2.8 hours after transit egress. Photometry was extracted using the custom-written Python code described in \cite{2015MNRAS.451..680E} with a circular 3-pixel-radius aperture. Below, we present a preliminary analysis of the resulting IRAC light curve to help constrain the HD~3167c transmission spectrum. However, a full analysis of this dataset, along with additional IRAC transit observations for HD~3167b, will be presented in an upcoming study (Hardegree-Ullman et al., in prep).

\begin{figure}
\centering  
\includegraphics[width=\columnwidth]{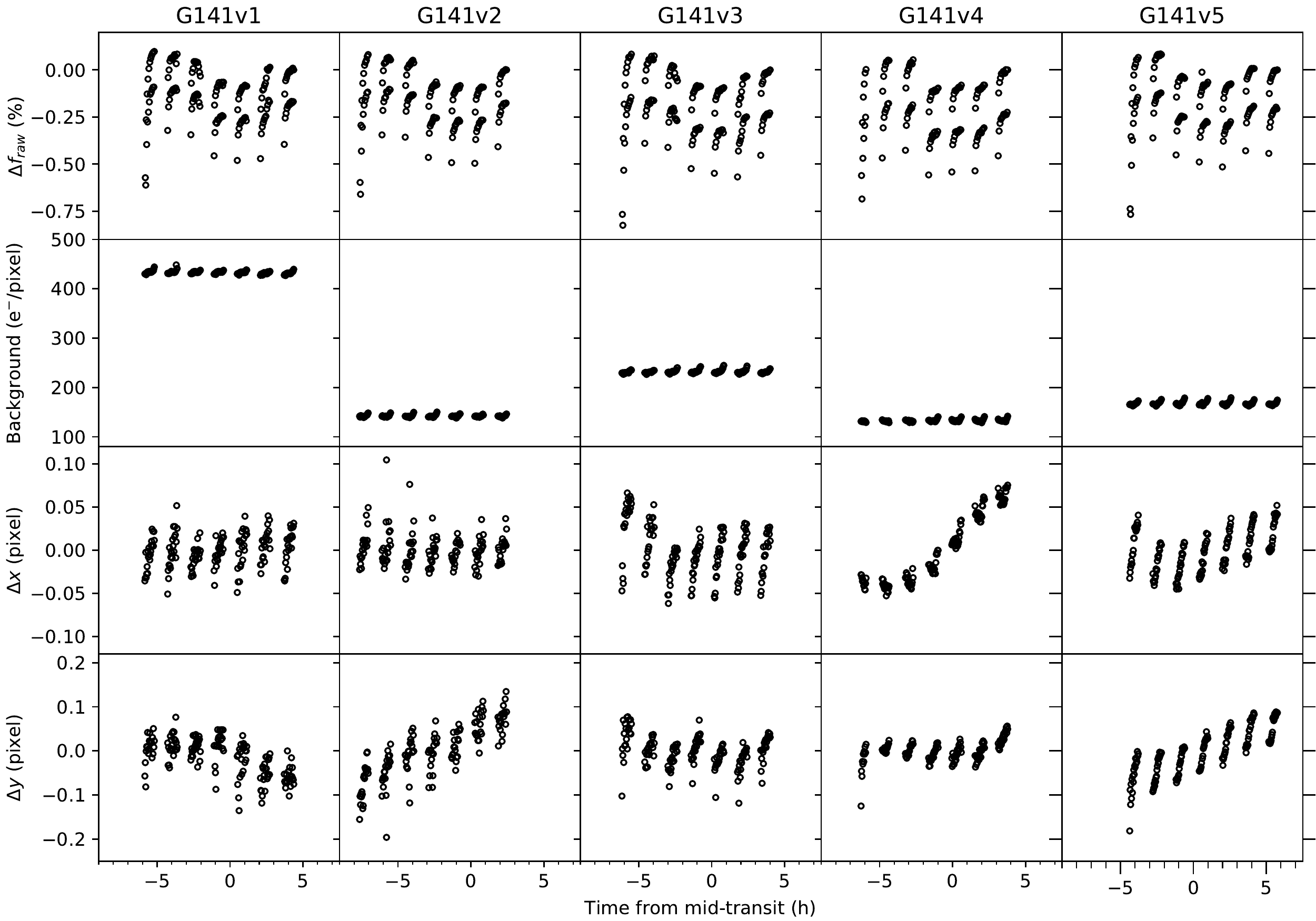}
\caption{Time series extracted from the spatial scan images for each WFC3 visit. Rows from top to bottom are: variation of the broadband relative flux ($\Delta f_{\textnormal{raw}}$); integrated background counts per pixel; median-subtracted drift along the dispersion axis ($\Delta x$); median-subtracted drift along the cross-dispersion axis ($\Delta y$). Note the offset in $f_{raw}$ between successive forward and backward exposures, and the strong ramp systematics in the first HST orbit.}
\label{fig:auxvars}
\end{figure}

\section{Light curve analysis} \label{sec:lcfits}

\subsection{\textit{HST} broadband light curves} \label{sec:hstbroad}

\begin{figure}
\centering  
\includegraphics[width=0.75\columnwidth]{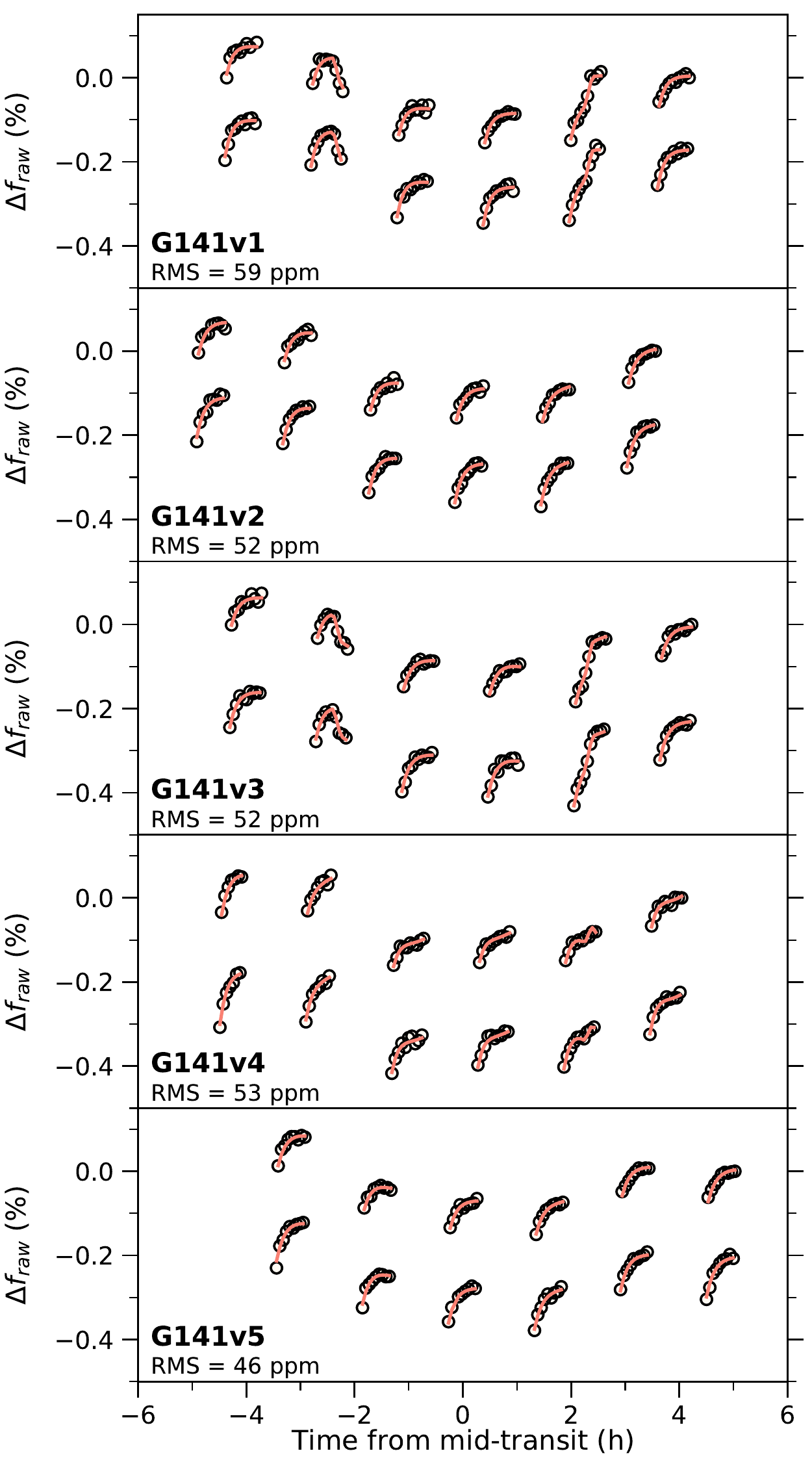}
\caption{Raw WFC3 broadband light curves with best-fit models. Datasets are labeled in the lower left corner of each axis with the root mean square (RMS) of the model residuals. Note that the first HST orbit has been discarded for the light curve fitting, as described in Appendix \ref{app:hst:broad}.}
\label{fig:wfc3raw}
\end{figure}

Broadband light curves were produced for all HST visits by summing each spectrum across the $0.8$-$1.95\,\mu$m wavelength range. The resulting light curves are shown in Figures \ref{fig:auxvars} and \ref{fig:wfc3raw}, exhibiting systematics typical of WFC3 transit datasets \citep[e.g.][]{2014Natur.513..526F,2014Natur.505...66K,2016ApJ...822L...4E,2017Natur.548...58E,2014Natur.505...69K,2020arXiv200607444K}. We fit all five broadband light curves jointly, sharing the planet-to-star radius ratio ($\RpRs$) across all light curves while allowing the transit mid-times ($T_i$) to vary separately for each dataset. Other transit parameters were held fixed to the values listed in Table \ref{table:broadfit}, namely: the normalized semimajor axis ($\aRs$); the orbital impact parameter ($b$), period ($P$), and eccentricity ($e$); and quadratic stellar limb darkening coefficients ($u_1$,\,$u_2$). We also allowed the white noise to vary for each light curve, parameterized as a rescaling of the photon noise level ($\beta_i$ for the $i$th visit). Further details of our light curve fitting methodology, including the sources of the values listed in Table \ref{table:broadfit} and the adopted transit and systematics models, are provided in Appendix \ref{app:hst:broad}.

\begin{figure}
\centering  
\includegraphics[width=\columnwidth]{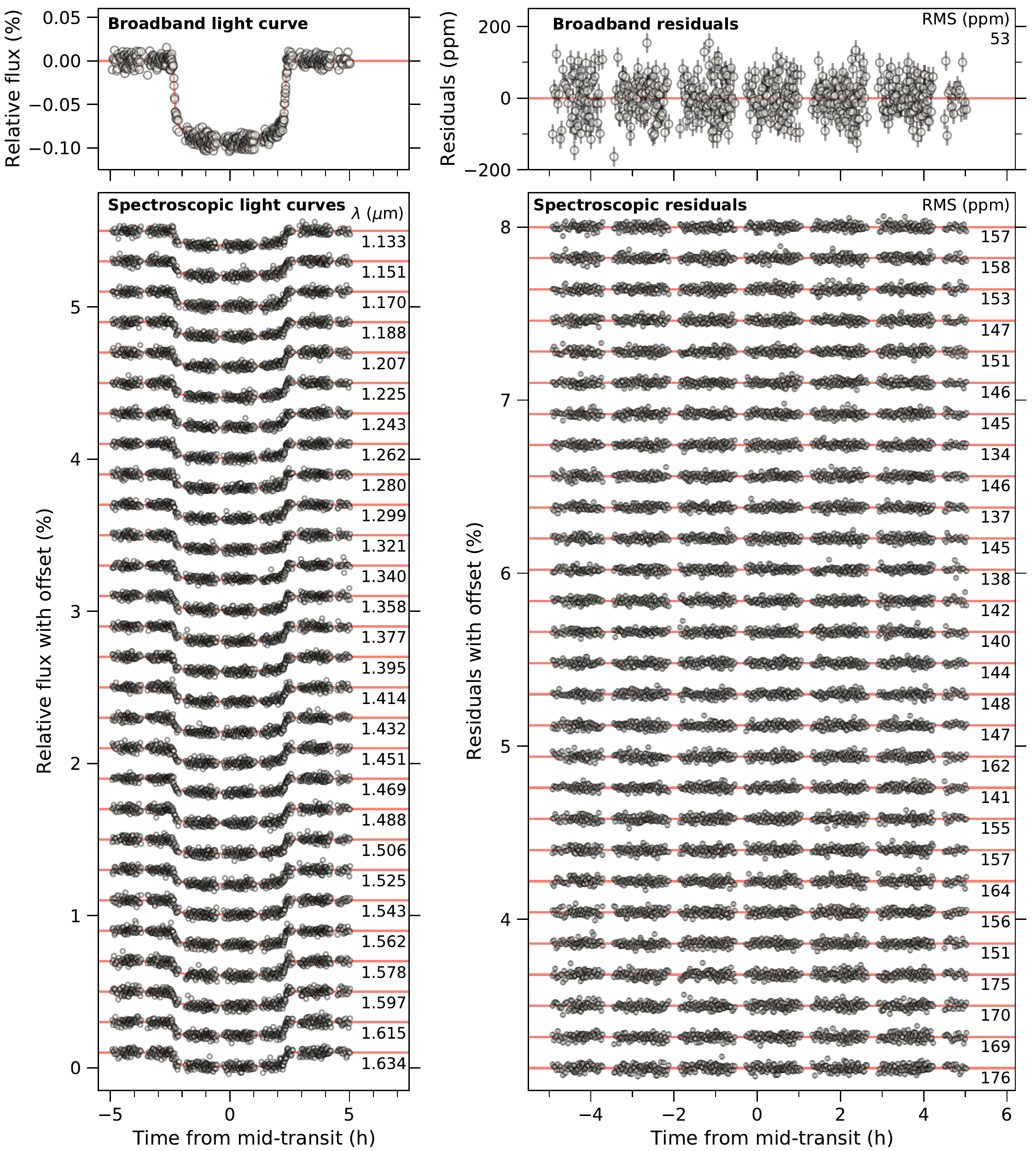}
\caption{Light curve measurements for the combined \textit{HST} WFC3 dataset, after accounting for systematics, with best-fit transit signals. The upper left panel shows the broadband light curve and the upper right panel shows the corresponding model residuals. Lower panels show the same for the spectroscopic light curves.}
\label{fig:wfc3corr}
\end{figure}

We marginalized the posterior distribution of the model parameters using the \texttt{emcee} Python package \citep{2013PASP..125..306F}, which implements affine-invariant Markov chain Monte Carlo (MCMC). The results are summarized in Table \ref{table:broadfit}, with the best-fit models shown in Figures \ref{fig:wfc3raw} and \ref{fig:wfc3corr}. For the transit depth we find $(\RpRs)^2=852_{-55}^{+45}$\,ppm, derived from the posterior sample values for $\RpRs=0.02919_{-0.00095}^{+0.00076}$. Inferred values for the white noise rescaling factors ($\beta_1=1.90_{-0.17}^{+0.17}$, $\beta_2=1.43_{-0.10}^{+0.12}$, $\beta_3=1.68_{-0.14}^{+0.15}$, $\beta_4=1.83_{-0.14}^{+0.13}$, $\beta_5=1.48_{-0.11}^{+0.11}$) indicate a high-frequency source of noise is affecting all datasets at a level $\sim 40$-$90$\% above the photon noise floor. This is evident in the broadband light curve residuals shown in Figure \ref{fig:wfc3corr}.

\begin{table}
\begin{minipage}{\columnwidth}
  \centering
\scriptsize
\caption{HD~3167c properties. Values with uncertainties correspond to MCMC posterior medians and 68\% credible intervals for the WFC3 broadband light curve analysis.  \label{table:broadfit}}
\begin{tabular}{ccc}
  \hline \\
Parameter & Value & Note \medskip \\ \cline{1-3}
&& \\

$\RpRs$ & $0.02919_{-0.00095}^{+0.00076}$ & Free \\
$(\RpRs)^2$ (ppm) & $852_{-55}^{+45}$ & Derived from $\RpRs$ \\
$\Rs$ ($\RS$) & $0.872 \pm 0.057$ & \cite{2017AJ....154..122C} \\ 
$\Rp$ ($\RE$) &$2.77_{+0.20}^{+0.20}$ & Derived from $\RpRs$ and $\Rs$ \\ 

$u_1$ & $0.15$ (fixed) & Fixed \\
$u_2$ & $0.31$ (fixed) & Fixed \\
$P$ (d) & $29.84622$ & Fixed \\ 
$e$ & $0$ & Fixed \\ 
$\aRs$ & $45.9$ & Fixed \\ 
$b$ & $0.35$ & Fixed \\ 
$i$ ($^\circ$) & $89.56$ & Derived from $Rs$ and $b$ \\ 

$T_{1}$ (JD$_{\textnormal{UTC}}$)  & $2458260.52959_{-0.00035}^{+0.00036}$ & Free \\ 
$T_{2}$ (JD$_{\textnormal{UTC}}$)  & $2458320.21401_{-0.00549}^{+0.00522}$ & Free \\ 
$T_{3}$ (JD$_{\textnormal{UTC}}$)  & $2458648.53174_{-0.00037}^{+0.00035}$ & Free \\ 
$T_{4}$ (JD$_{\textnormal{UTC}}$)  & $2458708.19273_{-0.00148}^{+0.00773}$ & Free \\ 
$T_{5}$ (JD$_{\textnormal{UTC}}$)  & $2459036.51367_{-0.00356}^{+0.00361}$ & Free \\ 

$\beta_{1}$ & $1.90_{-0.17}^{+0.17}$ & Free \\ 
$\beta_{2}$ & $1.43_{-0.10}^{+0.12}$ & Free \\ 
$\beta_{3}$ & $1.68_{-0.14}^{+0.15}$ & Free \\ 
$\beta_{4}$ & $1.83_{-0.14}^{+0.13}$ & Free \\ 
$\beta_{5}$ & $1.48_{-0.11}^{+0.11}$ & Free \\ \\ \hline

\end{tabular}
\end{minipage}
\end{table}

\subsection{\textit{HST} spectroscopic light curves} \label{sec:hstspec}

Spectroscopic light curves were produced from the WFC3 data by summing flux within 28 equal-width channels between wavelengths $1.125$-$1.645\,\um$. To do so, we followed a similar methodology to that used previously in \cite{2016ApJ...822L...4E,2017Natur.548...58E} and \cite{2019MNRAS.488.2222M}, which in turn was adapted from an original implementation by \cite{2013ApJ...774...95D}. This involved first cross-correlating each 1D spectrum against a template spectrum, solving for both a lateral shift in wavelength and a wavelength-uniform rescaling of the flux. For our analysis, we adopted the final spectrum of each visit as the template spectrum. The residuals of this cross-correlation were then binned in wavelength to produce a time series for each wavelength channel, before adding in the transit model derived from the broadband fit (Section \ref{sec:hstbroad}) to produce the spectroscopic light curves.

In practice, this process is similar to other methods used in the literature for generating spectroscopic light curves from WFC3 data. Specifically, the flux rescaling step is equivalent to typical common-mode corrections, such as dividing the raw spectroscopic light curves by the broadband light curve (i.e.\ $f_{raw}$ shown in Figures \ref{fig:auxvars} and \ref{fig:wfc3raw}). For example, \cite{2019ApJ...887L..14B} produced spectroscopic lightcurves by dividing the binned fluxes through by the broadband time series. Accounting for the lateral shifts, denoted by $\Delta x$ in Figure \ref{fig:auxvars}, helps to further minimize systematics in the spectroscopic light curves arising due to pointing drift. For comparison, \cite{2019ApJ...887L..14B} instead decorrelated against $\Delta x$ during the light curve fitting stage.

\begin{figure}
\centering  
\includegraphics[width=0.5\columnwidth]{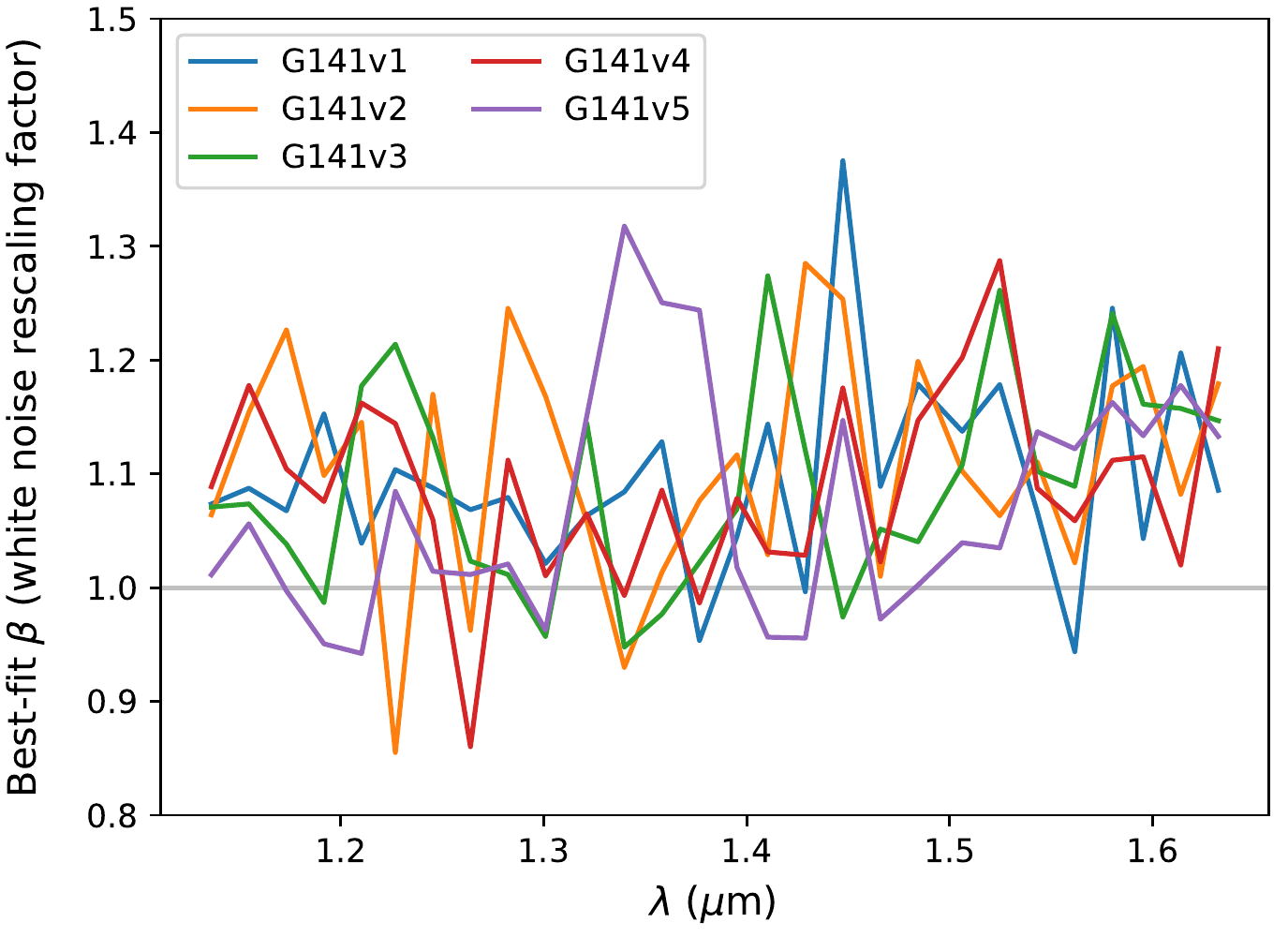}
\caption{Best-fit white noise rescaling factors ($\beta$) inferred for each spectroscopic light curve.}
\label{fig:wnoise}
\end{figure}

We fit the spectroscopic light curves following a similar approach to the broadband light curve fits, with a few differences. In particular, we held the transit mid-times fixed to the the best-fit values determined for the corresponding broadband light curve. Full details are given in Appendix \ref{app:hst:spec}. Inferred values for $\RpRs$ and the transit depth $(\RpRs)^2$ are reported in Table \ref{table:specfit}. The median uncertainty for the inferred transit depths across all channels is $17$\,ppm, ranking among the most precise WFC3 transmission spectra published to date. Systematics-corrected spectroscopic light curves are shown in Figure \ref{fig:wfc3corr}. The median root-mean-square (RMS) of the best-fit residuals is 150\,ppm (per 70\,s exposure), close to the median photon noise level of 138\,ppm. Accordingly, the inferred white noise rescaling factors are typically within $\sim 10$\% of unity across all spectroscopic channels (Figure \ref{fig:wnoise}). We also performed a second independent analyis using the data reduction and light curve fitting code of \cite{2014Natur.505...69K}. As described in Appendix \ref{app:hst:independent}, this gave consistent results, increasing our confidence in the measured transmission spectrum.

\begin{table}
\begin{minipage}{\columnwidth}
  \centering
\scriptsize
\caption{Similar to Table \ref{table:broadfit}, for the WFC3 spectroscopic light curve fits of our primary analysis, and the \textit{K2} and IRAC analyses.   \label{table:specfit}}
\begin{tabular}{cccccc}
  \hline \\
        & $\lambda$ & $\RpRs$ & $(\RpRs)^2$ & $u_1$ & $u_2$ \\ 
Dataset & ($\um$) &  & (ppm) & (fixed) & (fixed) \medskip \\ \cline{1-6}

\textit{K2} & $0.45$-$0.65$ & $0.03005_{-0.00025}^{+0.00025}$ & $903_{-15}^{+15}$ & $0.490$ & $0.194$ \\ \hline

WFC3 & $1.125$-$1.144$ & $0.02960_{-0.00028}^{+0.00029}$ & $879_{-18}^{+19}$ & $0.244$ & $0.247$ \\       
     & $1.144$-$1.162$ & $0.02936_{-0.00024}^{+0.00027}$ & $863_{-16}^{+18}$ & $0.243$ & $0.249$ \\       
     & $1.162$-$1.181$ & $0.02940_{-0.00025}^{+0.00026}$ & $866_{-16}^{+17}$ & $0.233$ & $0.255$ \\       
     & $1.181$-$1.199$ & $0.02919_{-0.00029}^{+0.00028}$ & $852_{-18}^{+18}$ & $0.230$ & $0.255$ \\       
     & $1.199$-$1.218$ & $0.02898_{-0.00027}^{+0.00029}$ & $838_{-17}^{+18}$ & $0.225$ & $0.259$ \\       
     & $1.218$-$1.236$ & $0.02917_{-0.00028}^{+0.00026}$ & $851_{-18}^{+17}$ & $0.220$ & $0.266$ \\       
     & $1.236$-$1.255$ & $0.02874_{-0.00026}^{+0.00024}$ & $823_{-17}^{+15}$ & $0.215$ & $0.269$ \\       
     & $1.255$-$1.273$ & $0.02935_{-0.00026}^{+0.00024}$ & $862_{-16}^{+16}$ & $0.208$ & $0.275$ \\       
     & $1.273$-$1.292$ & $0.02898_{-0.00025}^{+0.00026}$ & $839_{-16}^{+17}$ & $0.187$ & $0.290$ \\       
     & $1.292$-$1.310$ & $0.02888_{-0.00027}^{+0.00024}$ & $832_{-17}^{+15}$ & $0.197$ & $0.285$ \\       
     & $1.310$-$1.329$ & $0.02917_{-0.00025}^{+0.00026}$ & $851_{-16}^{+16}$ & $0.189$ & $0.289$ \\       
     & $1.329$-$1.347$ & $0.02884_{-0.00023}^{+0.00025}$ & $830_{-15}^{+16}$ & $0.182$ & $0.296$ \\       
     & $1.347$-$1.366$ & $0.02928_{-0.00023}^{+0.00021}$ & $858_{-15}^{+14}$ & $0.175$ & $0.302$ \\       
     & $1.366$-$1.384$ & $0.02935_{-0.00025}^{+0.00027}$ & $862_{-16}^{+18}$ & $0.165$ & $0.310$ \\       
     & $1.384$-$1.403$ & $0.02961_{-0.00023}^{+0.00024}$ & $879_{-15}^{+15}$ & $0.158$ & $0.317$ \\       
     & $1.403$-$1.421$ & $0.02945_{-0.00025}^{+0.00027}$ & $869_{-16}^{+17}$ & $0.149$ & $0.323$ \\       
     & $1.421$-$1.440$ & $0.02983_{-0.00023}^{+0.00024}$ & $894_{-15}^{+16}$ & $0.141$ & $0.328$ \\       
     & $1.440$-$1.458$ & $0.02987_{-0.00025}^{+0.00027}$ & $896_{-17}^{+18}$ & $0.130$ & $0.335$ \\       
     & $1.458$-$1.477$ & $0.02921_{-0.00023}^{+0.00022}$ & $853_{-15}^{+14}$ & $0.121$ & $0.344$ \\       
     & $1.477$-$1.495$ & $0.02926_{-0.00026}^{+0.00028}$ & $856_{-16}^{+18}$ & $0.112$ & $0.348$ \\       
     & $1.495$-$1.514$ & $0.02900_{-0.00026}^{+0.00027}$ & $840_{-17}^{+17}$ & $0.105$ & $0.348$ \\       
     & $1.514$-$1.532$ & $0.02961_{-0.00025}^{+0.00027}$ & $879_{-16}^{+18}$ & $0.092$ & $0.361$ \\       
     & $1.532$-$1.551$ & $0.02937_{-0.00029}^{+0.00029}$ & $864_{-19}^{+19}$ & $0.080$ & $0.368$ \\       
     & $1.551$-$1.569$ & $0.02923_{-0.00025}^{+0.00025}$ & $855_{-16}^{+16}$ & $0.075$ & $0.367$ \\       
     & $1.569$-$1.588$ & $0.02972_{-0.00030}^{+0.00033}$ & $886_{-20}^{+22}$ & $0.070$ & $0.363$ \\       
     & $1.588$-$1.606$ & $0.02972_{-0.00026}^{+0.00031}$ & $887_{-17}^{+20}$ & $0.066$ & $0.365$ \\       
     & $1.606$-$1.625$ & $0.02880_{-0.00029}^{+0.00031}$ & $827_{-18}^{+20}$ & $0.055$ & $0.368$ \\       
     & $1.625$-$1.643$ & $0.02934_{-0.00032}^{+0.00028}$ & $861_{-20}^{+18}$ & $0.052$ & $0.368$ \\ \hline

IRAC2 & $4$-$5$ & $0.02983_{-0.00238}^{+0.00233}$ & $890_{-142}^{+139}$ & $0.042$ & $0.129$ \\ \hline
\end{tabular}
\end{minipage}
\end{table}

\subsection{\textit{K2} and \textit{IRAC} light curves} \label{sec:k2irac}

\begin{figure}
\centering
\includegraphics[width=\linewidth]{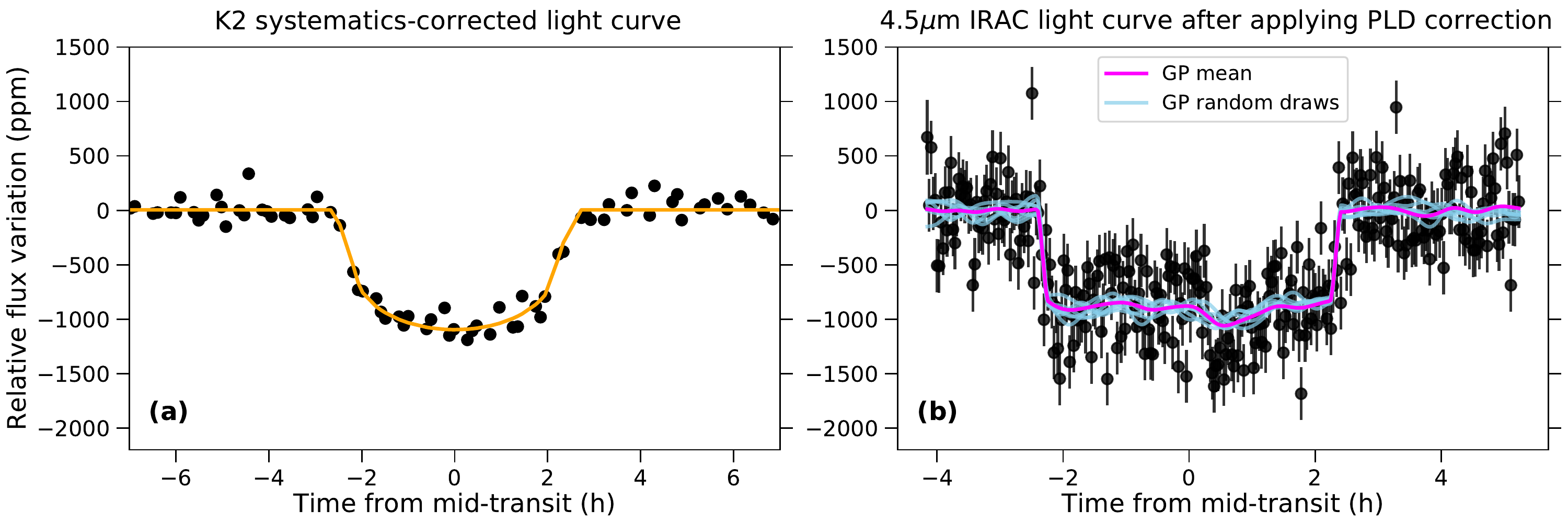}
\caption{(a) Systematics-corrected \textit{K2} light curve with best-fit transit model. (b) IRAC light curve after applying the PLD correction described in the text. Residual correlations with a characteristic amplitude of $\sim 100$\,ppm remain visible in the data, motivating the use of a $t$-dependent GP in the systematics modeling. Purple line shows the best-fit transit signal multiplied by the GP mean function and light blue lines show random draws from the GP.}
\label{fig:k2irac}
\end{figure}

Analysis of the \textit{K2} light curve followed largely the same approach described in \cite{2017AJ....154..122C}, but with two main differences. First, we fixed $\aRs$ and $b$ to the same values used for the broadband WFC3 analysis (Table \ref{table:broadfit}). Second, unlike \cite{2017AJ....154..122C}, we did not allow for secondary light dilution as a free parameter in our fit, as this could bias the measured $\RpRs$ to larger values. This is justified by high-contrast imaging of the HD~3167 system, which confirms the lack of blending with nearby stars in the \textit{K2} photometry (see Appendix \ref{app:hst:datared} for references). To fit the IRAC light curve, we used a GP approach similar to that employed for the HST broadband light curves. Details of this latter fit are provided in Appendix \ref{app:irac}.

\section{Stellar abundance analysis} \label{sec:stellarabund}

In addition to analyzing the transit light curves described in the previous sections, we used three spectra of HD~3167 to conduct an assay of the stellar elemental abundances. These spectra were taken with Keck/HIRES \citep{vogt:1994} on 2016 July 11 to provide the high signal-to-noise stellar template used to calculate the precision radial velocities reported by \cite{2017AJ....154..122C}. The observations each exposed for 166\,s using the B1 decker during good conditions and with an effective instrumental seeing of 0.9\,arcsec; the spectra are publicly available from the Keck Observatory Archive.\footnote{\url{https://koa.ipac.caltech.edu/}}

To measure the stellar abundances, we used the methodology and software tools of \cite{brewer:2016a}, to which readers may refer for a full description of these methods. In Table~\ref{table:stellarabund} we report the measured abundances for fifteen elements along with statistical and systematic uncertainties ($\sigma_\mathrm{stat}$ and $\sigma_\mathrm{sys}$, respectively) for each. The $\sigma_\mathrm{sys}$ uncertainties come from Table 6 of \cite{brewer:2016a}, and essentially describe the reliability of the modelling code used to provide the same result for two nearly identical spectra (not accounting for any additional uncertainty due to the star's distance from solar values). The $\sigma_\mathrm{stat}$ values indicate the standard deviation on the abundances measured from each of the three HIRES spectra; these values are all smaller than $\sigma_\mathrm{sys}$, likely because the three  spectra were all taken on the same night.

The elemental abundance patterns seen in HD~3167 (Table~\ref{table:stellarabund}) are similar to the solar values, with differences of $<0.1$\,dex for all fifteen measured elements. In particular, our measurements indicate a stellar C/O=0.48 and Mg/Si consistent with the solar value to within $1\sigma$, both entirely consistent with the peak of the distribution for local stars \citep{brewer:2016}. This suggests models of planet formation, evolution, and atmospheres assuming solar abundances should be generally applicable to this system as well. We also find $[$Y/Mg$]=-0.08 \pm 0.03$, suggesting the star may be slightly older than the Sun \citep[consistent with earlier studies, e.g.][]{2017AJ....154..122C}. Using the abundance-age relation of \cite{nissen:2020} and propagating all uncertainties using Monte Carlo techniques, we estimate a stellar age of $6.7 \pm 0.8$\,Gyr. This value should be interpreted cautiously, since HD~3167 is about 300\,K cooler than the nearly Solar-like stars used to calibrate that relation. However, it is in agreement with the age of $7.8 \pm 4.3$\,Gyr reported by \cite{2017AJ....154..122C}, which was determined by isochrone fitting.

\begin{table}
    \centering
  \scriptsize
\caption{HD~3167 host star element abundances \label{table:stellarabund}}
\begin{tabular}{l r l l}
  \hline \hline
Abundance &  Value &  $\sigma_\mathrm{stat}$ & $\sigma_\mathrm{sys}$ \\ \hline
 $[$C/H$]$ &  0.01 &  0.003  &  0.026\\
 $[$N/H$]$ & $-$0.03 &  0.013  &  0.042\\
 $[$O/H$]$ &  0.07 &  0.008  &  0.036\\
$[$Na/H$]$ &  0.05 &  0.002  &  0.014\\
$[$Mg/H$]$ &  0.06 &  0.004  &  0.012\\
$[$Al/H$]$ &  0.08 &  0.006  &  0.028\\
$[$Si/H$]$ &  0.07 &  0.003  &  0.008\\
$[$Ca/H$]$ &  0.05 &  0.001  &  0.014\\
$[$Ti/H$]$ &  0.07 &  0.002  &  0.012\\
 $[$V/H$]$ &  0.07 &  0.005  &  0.034\\
$[$Cr/H$]$ &  0.04 &  0.003  &  0.014\\
$[$Mn/H$]$ &  0.03 &  0.001  &  0.020\\
$[$Fe/H$]$ &  0.04 &  0.000  &  0.010\\
$[$Ni/H$]$ &  0.05 &  0.002  &  0.012\\
 $[$Y/H$]$ & $-$0.02 &  0.005  &  0.030\\
 \hline
 \end{tabular}
\end{table}

\section{Atmospheric characterization} \label{sec:atmchar}

\begin{figure}
\centering  
\includegraphics[width=\linewidth]{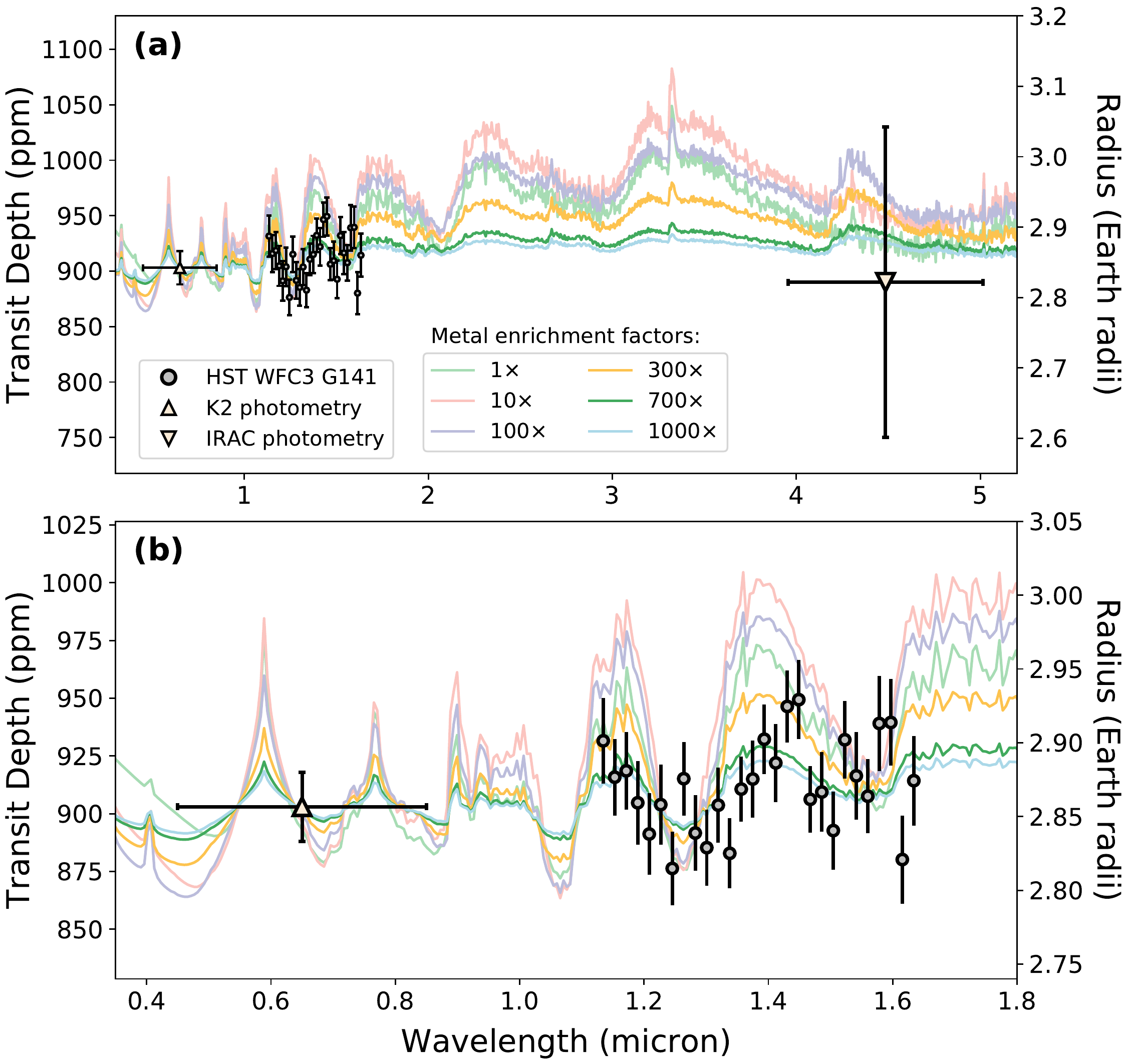}
\caption{(a) Transmission spectrum for HD~3167c measured with \textit{K2}, HST, and \textit{Spitzer}. Colored lines show chemical equilibrium models for different heavy element enrichments relative to solar abundances. Note that a small offset has been applied to the HST data relative to the \textit{K2} and IRAC, within the uncertainty of the broadband level. (b) The same, but covering a narrower wavelength range centered on the \textit{K2} and HST data.}
\label{fig:trspec1}
\end{figure}

\begin{figure}
\centering  
\includegraphics[width=0.9\linewidth]{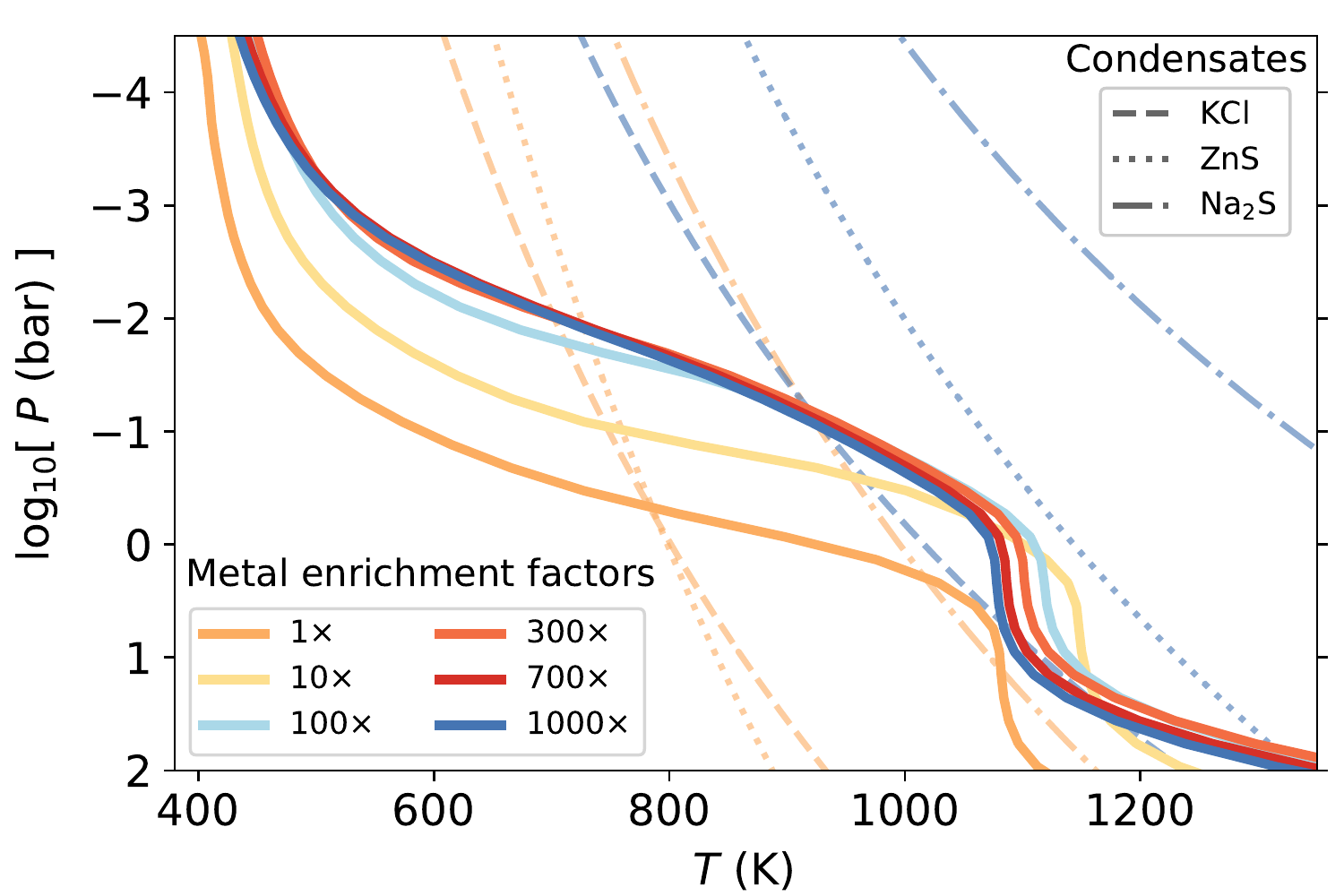}
\caption{Thick solid lines show pressure-temperature profiles derived self-consistently for a selection of atmospheric metallicities, assuming chemical equilibrium. Condensation curves are shown for KCl (dashed lines), ZnS (dotted lines), and Na$_2$S (dot-dashed liens) for $1\times$ and $1000\times$ solar metallicity, using the same color scheme. Condensation curves for intermediate metallicities fall between these two end cases. Note that the condensation curve for H$_2$O is to the left of the axis.}
\label{fig:PT}
\end{figure}

The measured transmission spectrum obtained from the light curve fits presented in Section \ref{sec:hstspec} and \ref{sec:k2irac} is shown in Figure \ref{fig:trspec1}. The \textit{K2} measurement constrains the broadband optical level at a precision of $15$\,ppm, comparable to that achieved for each of the individual WFC3 channels ($\sim\,17$\,ppm; Table \ref{table:specfit}). Note, however, that the overall WFC3 level is less well constrained to a precision of $\sim\,50$\,ppm (Table \ref{table:broadfit}), owing to the flexible GP approach we adopted for modeling the light curve baseline level (Section \ref{sec:hstbroad}). Relative to the \textit{K2} measurement, the IRAC level is also constrained relatively imprecisely ($\sim 140$\,ppm; Table \ref{table:specfit}), due to the presence of time-correlated noise in the light curve that was marginalized with a flexible GP (Figure \ref{fig:k2irac}).

The WFC3 data exhibit an apparent trough at around $1.3\,\um$. The depth of this feature is $\sim 30$-$50\,$ppm, corresponding to a variation of approximately 3-5 atmospheric pressure scale heights, in line with expectations for molecular absorption bands at near-infrared wavelengths. This calculation was made by taking the equilibrium temperature ($\Teq \sim 500$-$600$\,K) and surface gravity ($g \approx 10.7\,$m\,s$^{-2}$) of HD~3167c, and assuming an atmospheric mean molecular weight of $\muatm = 5$ atomic mass units (amu). The latter is consistent with sub-Neptune population synthesis simulations \citep[e.g.][]{2013ApJ...775...80F}, although there is anticipated to be a large spread due to the stochastic nature of planetesimal accretion during formation. In the following sections, we compare the data with detailed models of the planetary atmosphere.

\subsection{Equilibrium chemistry forward models} \label{sec:chemeq}

Model transmission spectra were generated under the assumption that the atmosphere is in both chemical and radiative equilibrium, following the same approach outlined in \cite{2015arXiv150407655B,2019NatAs...3..813B}. To do this, we treated the atmosphere as being one-dimensional (1D) and considered various levels of heavy element enrichment, ranging from $1$-$1000 \times$ solar metallicity. The Bond albedo was set to 30\%, with heat redistribution from the dayside to nightside hemispheres assumed to be uniform. We then iteratively solved for chemical and radiative-convective equilibrium to obtain a self-consistent solution for the 1D globally-averaged chemical composition and pressure-temperature (PT) profile for each metal enrichment.

The resulting model transmission spectra are shown in Figure \ref{fig:trspec1} with corresponding PT profiles in Figure \ref{fig:PT}. We find good agreement is obtained with the data for those models with metallicity $\gtrsim 300 \times$ solar. Specifically, we obtain reduced $\chi^2$ values close to unity for the $700\times$ solar ($\chi^2=1.2$) and $1000\times$ solar ($\chi^2=1.1$) cases. By contrast, the data is marginally inconsistent with the $300\times$ solar model at the $2.3\sigma$ level of significance. The discrepancy increases to $>5\sigma$ for all models with metallicity $\leq 100\times$ solar.

However, we note three important caveats. First, all models shown in Figure \ref{fig:trspec1} assume a carbon-to-oxygen ratio (C/O) equal to the solar value of C/O$=0.54$ \citep{2009ARA&A..47..481A}, broadly consistent with what we measured for HD~3167 (Table \ref{table:stellarabund}).  Nonetheless a C/O ratio differing from that of the host star is plausible, depending on where in the protoplanetary disc HD~3167c formed and the composition of material it subsequently accreted \citep{2011ApJ...743L..16O}. Second, we have ignored the effect of clouds, which can be highly degenerate with atmospheric metallicity at the level of our data precision \citep{2013ApJ...778..153B}. Third, we adopted a globally-averaged self-consistent 1D solution for the PT profile given an assumed albedo. While it has been shown that global-average temperature profiles provide a reasonable approximation to temperature profiles at the planetary limb \citep[e.g.][]{2010ApJ...709.1396F}, the overall temperature will still depend on the unknown albedo. In the next section, we investigate these effects further by performing a retrieval analysis that includes clouds and allows the atmospheric metallicity, C/O ratio, and temperature to vary.

\subsection{Retrieval analysis with chemically-consistent atmosphere models} \label{sec:retrieval:chemeq}

\begin{figure}[h!]
\centering
\includegraphics[width=0.8\linewidth]{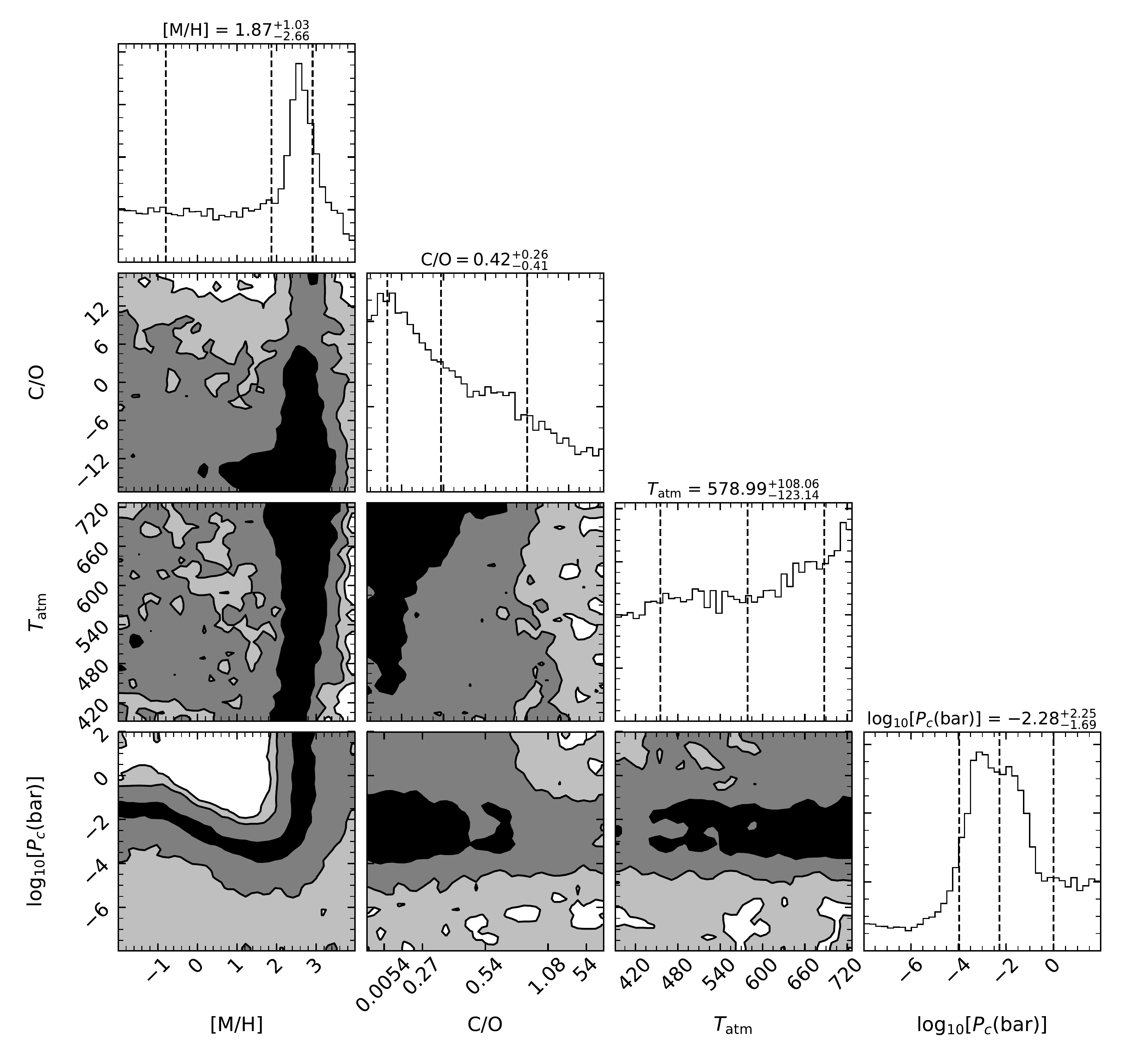}
\caption{Posterior distributions inferred for free parameters of the chemically-consistent \texttt{SCARLET} model: metallicity ([M/H]); carbon-to-oxygen ratio (C/O); and atmospheric temperature ($\Tatm$). Diagonal panels show the fully marginalized distribution for each parameter, with vertical dashed lines showing the median and $1\sigma$ credible ranges. Off-diagonal panels show the marginalized posterior distributions for each pair of model parameters. Contours and shading indicate the $1\sigma$, $2\sigma$, and $3\sigma$ credible regions. }
\label{fig:scarlet:chemeq:posterior}
\end{figure}

\begin{figure}
\centering
\includegraphics[width=0.75\linewidth]{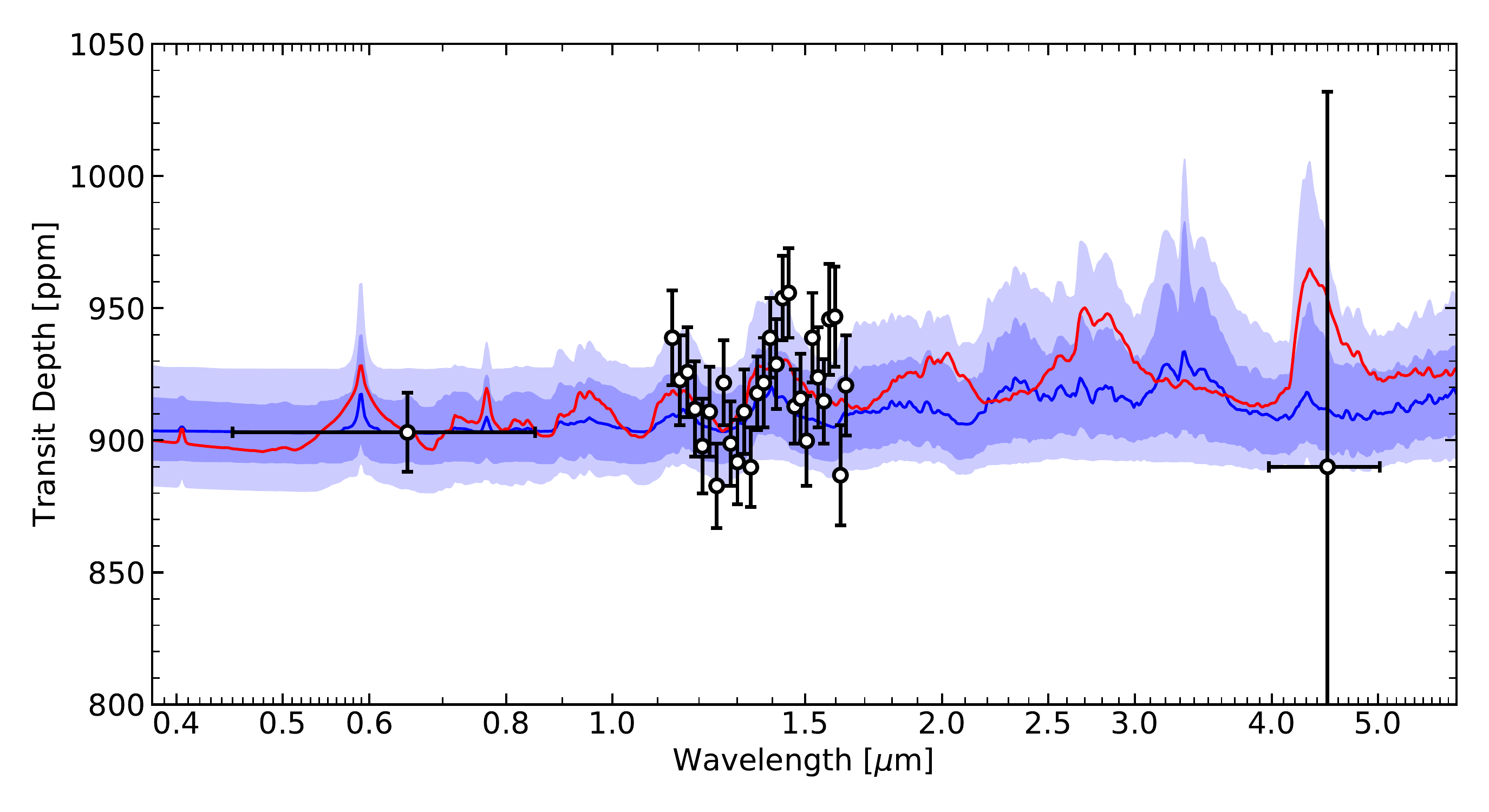}
\caption{Measured transmission spectrum for HD~3167c compared with model spectra obtained from the chemically-consistent \texttt{SCARLET} retrieval analysis. The blue line gives the median of all posterior samples, with blue shading indicating $1\sigma$ and $2\sigma$ credible intervals. The red line shows the model spectrum corresponding to the highest likelihood sample.}
\label{fig:scarlet:chemeq:trspec}
\end{figure}

To investigate a broader range of atmospheric properties for HD~3167c, we performed a retrieval analysis using the \texttt{SCARLET} retrieval framework \citep[e.g.,][]{2012ApJ...753..100B,2013ApJ...778..153B,2014Natur.505...69K,2014Natur.505...66K,2015arXiv150407655B,2019NatAs...3..813B,2019ApJ...887L..14B,2020AJ....159..234W}. For the analysis in this section, we employed the chemically-consistent mode of \texttt{SCARLET} described in \cite{2015arXiv150407655B}, which considers the space of models satisfying thermochemical equilibrium. We assumed an isothermal PT profile and allowed the temperature ($\Tatm$) to vary as a free parameter, along with the metallicity ([M/H]), C/O, and a grey cloud-top pressure ($\Pc$). We adopted a uniform prior for $\Tatm \sim \mathcal{U}(400,800)$\,K, a log-uniform prior for [M/H]$\sim \log\mathcal{U}(-2,4)$ (dex), and a log-uniform prior for $\Pc\sim\log\mathcal{U}(-8,2)$ (dex bar). For C/O, we applied a broad uniform prior on a custom-stretched parameter space, designed to provide the most effective sampling of the parameter space, as described in \cite{2015arXiv150407655B}. We then marginalized over the parameter space using nested sampling \citep{2004AIPC..735..395S} with $10,000$ active samples. The resulting parameter posterior distributions are shown in Figure \ref{fig:scarlet:chemeq:posterior}. The credible distribution of model transmission spectra and highest likelihood model are shown in Figure \ref{fig:scarlet:chemeq:trspec}, providing a good match to the data within the uncertainties.

\begin{figure}
\centering
\includegraphics[width=\linewidth]{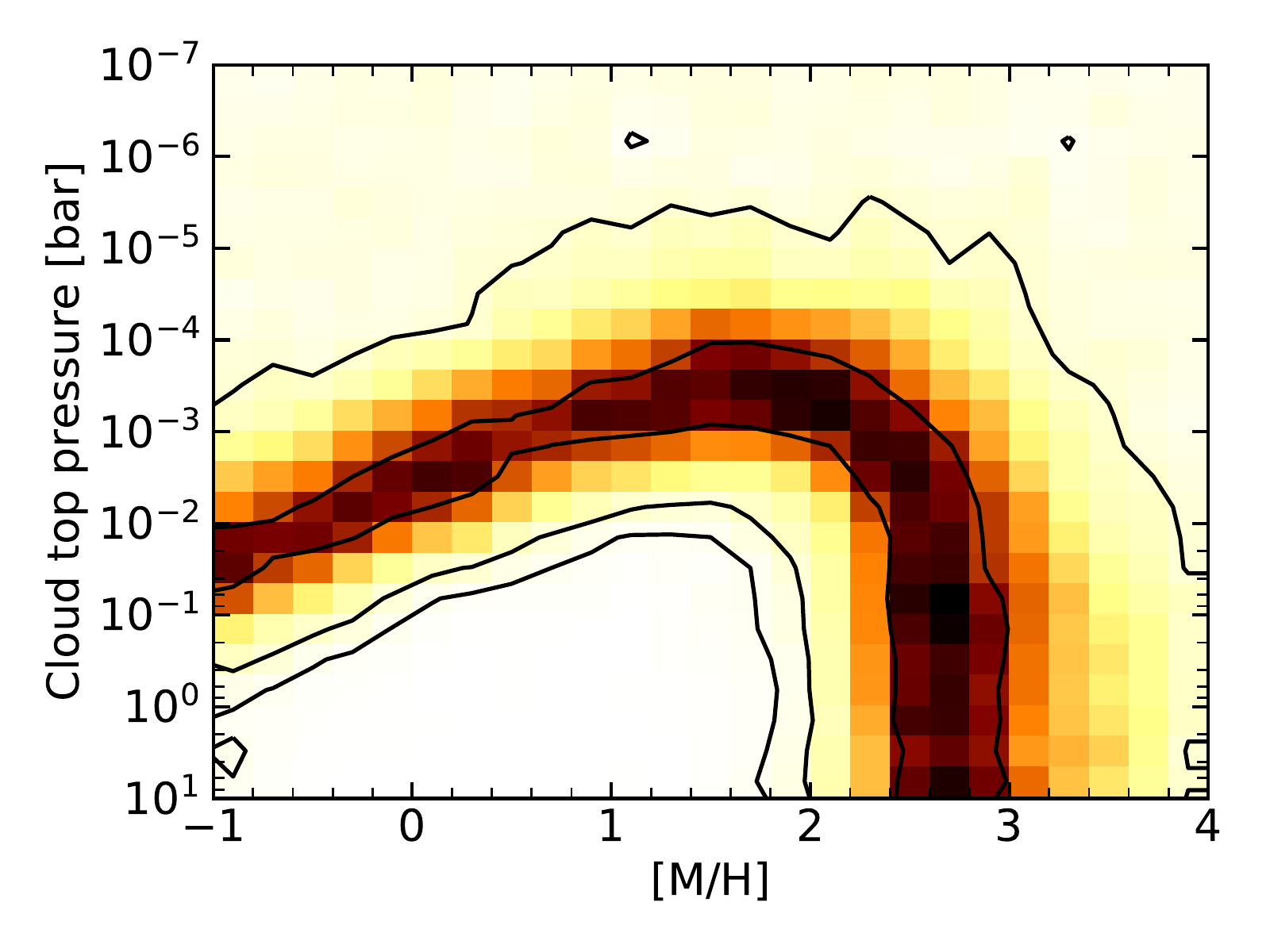}
\caption{Joint constraints on the metallicity versus the cloud top pressure inferred by the chemically-consistent \texttt{SCARLET} retrieval. Colored shading indicates the normalized probability density as a function of water abundance above the clouds and cloud top pressure. Black contours show the $1\sigma$, $2\sigma$, and $3\sigma$ credible regions.}
\label{fig:scarlet:chemeq:MHvsCloud}
\end{figure}

Allowing for the presence of cloud in this manner, we are unable to place a strong constraint on the atmospheric metallicity. As has been discussed extensively elsewhere \citep[e.g.][]{2013ApJ...778..153B}, the effects on the transmission spectrum of varying gas species abundances and the pressure level of a grey cloud deck can be effectively indistinguishable. A close-up view of this degeneracy is shown in Figure \ref{fig:scarlet:chemeq:MHvsCloud}, with metallicities ranging from subsolar to $1000\times$ solar allowed by the data, depending on the unknown role played by cloud. In addition, we are unable to place a meaningful constraint on the C/O ratio. However, this is not surprising, as to do so would require both carbon-bearing and oxygen-bearing species to be resolved by the data. As we explore in the next section, the available data are unable to confidently discriminate between major gases expected at the wavelengths covered by the data, such as H$_2$O, CH$_4$, and CO$_2$.

\subsection{Retrieval analyses with free chemistry} \label{sec:retrieval:free}

In this section, we describe ``free chemistry'' retrieval analyses for which chemical equilibrium was not imposed. Instead, the abundances of individual gas species were allowed to vary without constraint. This was done using two independent codes: \texttt{SCARLET} \citep{2012ApJ...753..100B,2013ApJ...778..153B} and \texttt{petitRADTRANS} \citep{2019A&A...627A..67M}.

\subsubsection{\texttt{SCARLET} free chemistry retrievals} \label{sec:retrieval:free:scarlet}

The \texttt{SCARLET} free chemistry retrievals proceeded in a similar manner as those described for the equilibrium chemistry retrievals described in Section \ref{sec:retrieval:chemeq}, with a number of important differences. In particular, the mole fractions of individual gas species are included as free parameters of the model, rather than [M/H] and C/O. The spectrally active gas species we considered in this manner were H$_2$O, CH$_4$, CO, CO$_2$, NH$_3$, and HCN. We also included the mole fraction of N$_2$ as a free parameter, despite it not having spectral features at the wavelengths covered by our dataset. Nonetheless, N$_2$ is predicted to be abundant in the atmosphere of HD~3167c (Section \ref{sec:consistentdiseq}) and can affect the atmospheric pressure scale height, and thus the amplitude of spectral features. Aside from these gases, we assumed the rest of the atmosphere was composed of H/He in solar proportions. For the mole fractions of the free gases, we employed log-uniform priors between $10^{-10}$ and $1$. As for the chemical equilibrium retrievals, we assumed an isothermal PT profile and allowed $\Tatm$ to vary as a free parameter, along with the pressure level of a grey cloud deck ($P_c$), and adopted the same priors described in Section \ref{sec:retrieval:chemeq}.

\subsubsection{\texttt{petitRADTRANS} free chemistry retrievals} \label{sec:retrieval:free:prt}

A second free chemistry retrieval was conducted using the publicly available\footnote{\url{https://petitradtrans.readthedocs.io}} \texttt{petitRADTRANS} (\texttt{pRT}) Python package \citep{2019A&A...627A..67M}, in a manner similar to the \texttt{SCARLET} analysis described in the previous section. Specifically, we used \texttt{pRT} to take atmospheric properties as input and return model transmission spectra as output. This allowed us to define a log-likelihood function and marginalize over the model parameter space using multimodal nested sampling, as implemented by the publicly available\footnote{\url{https://johannesbuchner.github.io/PyMultiNest}} \texttt{PyMultiNest} package \citep{2014A&A...564A.125B}. As for the \texttt{SCARLET} analysis, we allowed the mole fractions of  H$_2$O, CH$_4$, CO, CO$_2$, NH$_3$, HCN, and N$_2$ to vary with log-uniform priors between $10^{-10}$ and $1$. Again, $\Tatm$ and $P_c$ were included as additional free parameters with the same priors assumed for the \texttt{SCARLET} retrievals.

\subsubsection{Free chemistry retrieval results} \label{sec:retrieval:free:results}

\begin{table}
\begin{minipage}{\columnwidth} 
\centering 
\scriptsize 
\caption{Free chemistry retrieval prior and posterior distributions.  \label{table:retrieval:free:post}} 
\begin{tabular}{rccccc} 
\hline \\ 
&Parameter & Unit & Prior & \texttt{SCARLET} & \texttt{petitRADTRANS}  \medskip \\   \cline{1-6}  
& $\Tatm$ & $K$ & $\mathcal{U}(400,800)$ & ${558}_{-108}^{+109}$ & ${488}_{-66}^{+119}$ \\ 
 & $\log_{10}P_c$ & bar dex & $\mathcal{U}(-8,2)$ & ${-0.9}_{-2.0}^{+1.9}$ & ${-0.4}_{-1.9}^{+1.5}$ \\ \hline 
 Mole fractions & $\log_{10}$H$_2$O & dex & $\mathcal{U}(-10,0)$ & ${-3.6}_{-3.4}^{+2.3}$ & ${-3.8}_{-1.1}^{+1.7}$ \\ 
  & $\log_{10}$HCN & dex & $\mathcal{U}(-10,0)$  & ${-1.6}_{-4.3}^{+1.0}$ & ${-3.7}_{-3.4}^{+1.9}$ \\ 
  & $\log_{10}$CO$_2$ & dex & $\mathcal{U}(-10,0)$  & ${-4.1}_{-3.9}^{+2.5}$ & ${-2.1}_{-4.6}^{+1.0}$ \\ 
  & $\log_{10}$CO & dex & $\mathcal{U}(-10,0)$  & ${-5.0}_{-3.3}^{+3.3}$ & ${-5.3}_{-3.0}^{+3.1}$ \\ 
  & $\log_{10}$N$_2$ & dex & $\mathcal{U}(-10,0)$  & ${-5.0}_{-3.3}^{+3.3}$ & ${-5.5}_{-2.9}^{+3.1}$ \\ 
  & $\log_{10}$CH$_4$ & dex & $\mathcal{U}(-10,0)$  & ${-5.9}_{-2.7}^{+2.7}$ & ${-6.5}_{-2.2}^{+2.2}$ \\ 
  & $\log_{10}$NH$_3$ & dex & $\mathcal{U}(-10,0)$  & ${-7.0}_{-1.9}^{+2.2}$ & ${-7.3}_{-1.7}^{+1.8}$ \\ \\ \hline 
\end{tabular} 
\end{minipage} 
\end{table} 

\begin{figure}
\centering
\includegraphics[width=\linewidth]{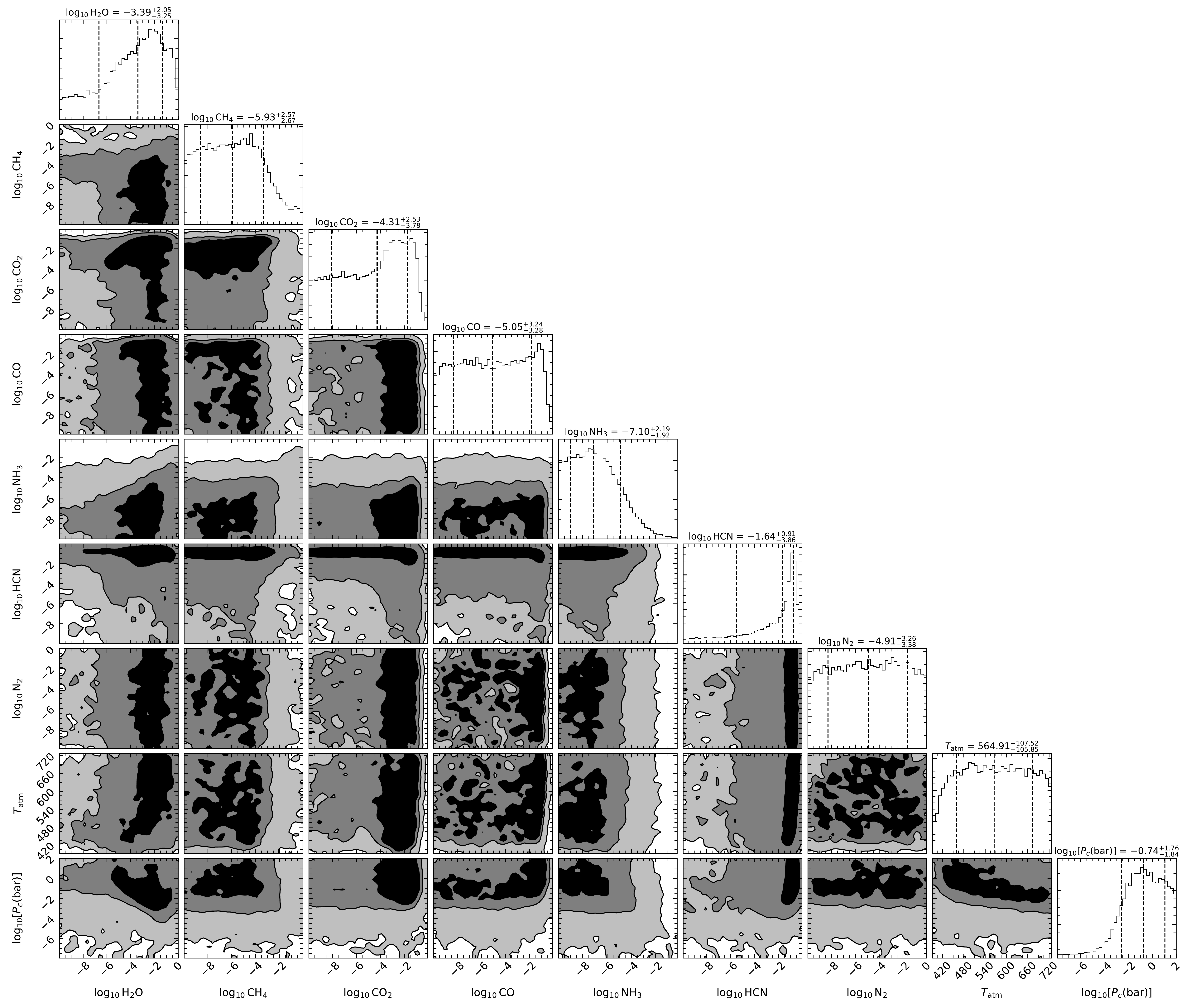}
\caption{Posterior distributions inferred for the free parameters of the \texttt{SCARLET} retrieval with unconstrained chemistry. Diagonal panels show the fully marginalized distribution for each parameter, with vertical dashed lines showing the median and $1\sigma$ credible ranges. Off-diagonal panels show the marginalized posterior distributions for each pair of model parameters. Contours and shading indicate the $1\sigma$, $2\sigma$, and $3\sigma$ credible regions.}
\label{fig:freechem:posterior:scarlet}
\end{figure}

\begin{figure}
\centering
\includegraphics[width=\linewidth]{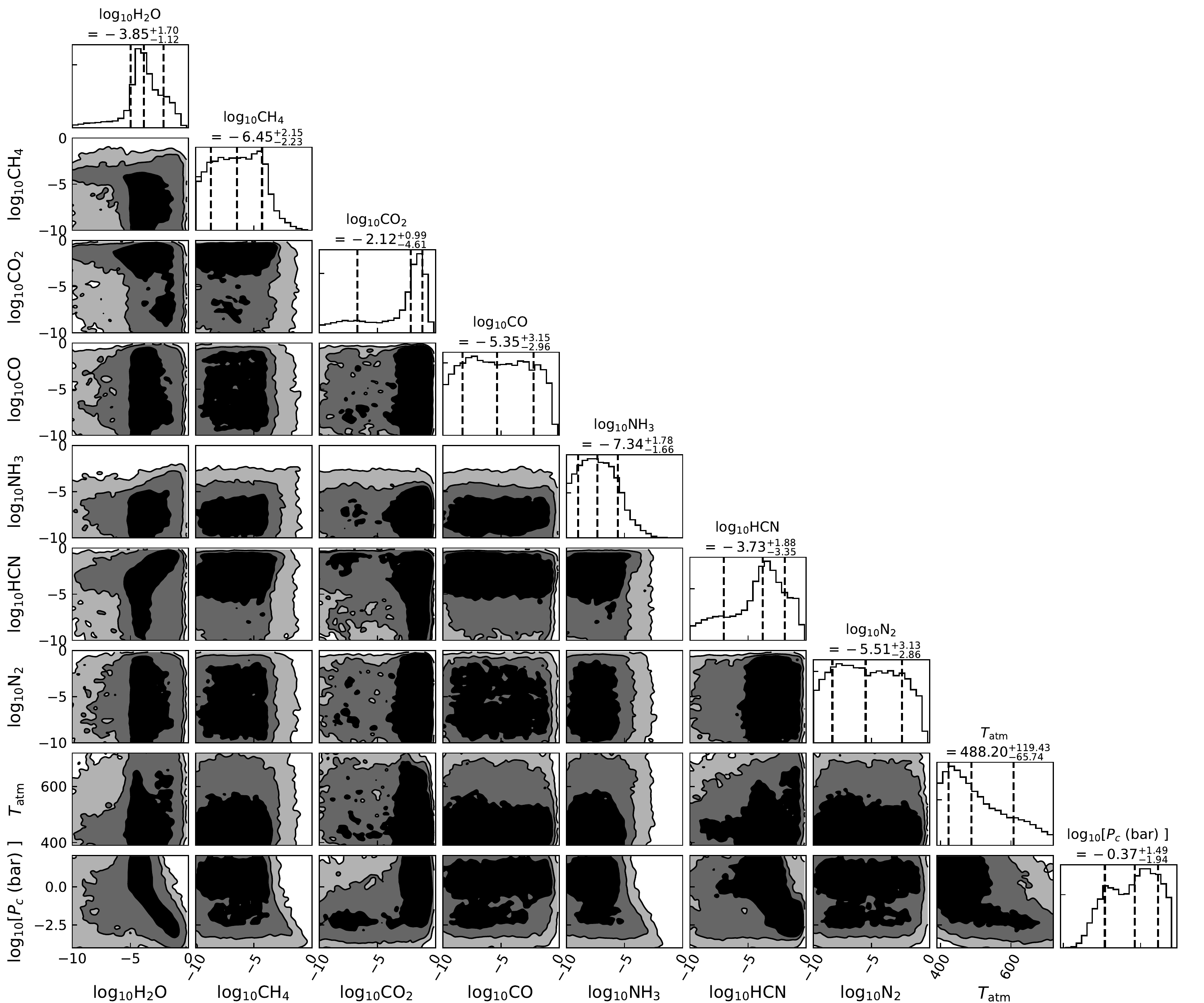}
\caption{Similar to Figure \ref{fig:freechem:posterior:scarlet}, showing the posterior distributions obtained for the \texttt{petitRADTRANS} free chemistry retrieval analysis. Contours and shading indicate the $1\sigma$, $2\sigma$, and $3\sigma$ credible regions.}
\label{fig:freechem:posterior:prt}
\end{figure}

The posterior distributions for the \texttt{SCARLET} and \texttt{petitRADTRANS} free chemistry retrievals are shown in Figures \ref{fig:freechem:posterior:scarlet} and \ref{fig:freechem:posterior:prt}, respectively, and summarized in Table \ref{table:retrieval:free:post}. The results of both retrieval analyses are in broad agreement for most parameters, but there is some tension for the retrieved abundances of HCN and CO$_2$. The \texttt{SCARLET} retrieval favors a higher HCN abundance relative to the \texttt{petitRADTRANS} retrieval, while the opposite is true for the inferred CO$_2$ abundances. However, in all cases the posteriors are broad, with the $1\sigma$ credible ranges of both retrievals overlapping for HCN and CO$_2$.

The posterior distributions reported in Table \ref{table:retrieval:free:post} provide the plausible mole fractions for each gas species that would be consistent with the measured transmission spectrum. Alone, however, they do not allow us to assess the evidence for any particular gas species actually being present in the atmosphere. To address this separate question, we employed the Bayesian model comparison framework outlined in \cite{2013ApJ...778..153B}. This involved systematically excluding one or more gas absorbers from the model and repeating the \texttt{SCARLET} and \texttt{petitRADTRANS} retrieval analyses. Each time, the Bayesian evidence $\mathcal{Z}$ of the resulting model would be recorded, as returned by the nested sampling algorithm. Denoting the evidence of the model with all gas absorbers included as $\mathcal{Z}_0$, the Bayes factor was then computed as $B=\mathcal{Z}_0/\mathcal{Z}$. Under this formulation, $B$ can be considered the likelihood ratio of the removed absorber being present in the atmosphere. This can also be expressed as an equivalent ``$N$ sigma'' ($N_\sigma$) statistical significance under the frequentist paradigm using the approach described in \cite{2008ConPh..49...71T}. The results of this analysis are given in Table \ref{table:retrieval:free:bayes}. Neither retrieval uncovers even tentative ($>2\sigma$) evidence for an individual gas absorber. Therefore, we cannot claim the unambiguous detection of any specific gas species based on the available data.

As reported in Table \ref{table:retrieval:free:bayes}, we also investigated removing two or three gas species at a time and computing the resulting Bayes factors. This approach was motivated by the reasonable expectation that multiple gas absorbers are likely to be present in the atmosphere, rather than a single spectrally active gas. Although the absorption signature of any individual gas may not be statistically significant, the combined signature of multiple gases may be. Indeed, when H$_2$O, CH$_4$, and HCN are removed from the model, the resulting Bayes factors imply positive evidence at $2.7\sigma$ significance for the \texttt{SCARLET} analysis and $2.2\sigma$ significance for the \texttt{petitRADTRANS} analysis. As listed in Table \ref{table:retrieval:free:bayes}, a number of additional molecule combinations were tested with \texttt{petitRADTRANS}, which showed levels of evidence intermediate to these values. Namely, the combinations: H$_2$O+CH$_4$ ($2.4\sigma$); H$_2$O+CO$_2$ ($2.4\sigma$); H$_2$O+HCN ($2.4\sigma$); H$_2$O+CO$_2$+CH$_4$ ($2.5\sigma$); and H$_2$O+CO$_2$+HCN ($2.5\sigma$). From these results, we determine that molecular absorption is detected in the transmission spectrum at $\sim 2.5\sigma$ significance, likely due to one or more of H$_2$O, CH$_4$, CO$_2$, and HCN.

\begin{table}
\begin{minipage}{\columnwidth} 
\centering 
\scriptsize 
\caption{Bayesian model comparison results for the \texttt{SCARLET} and \texttt{petitRADTRANS} free chemistry retrieval analyses. $B$ is the Bayes factor, giving the relative likelihood of the model with all gas absorbers included to the listed model with a subset of absorbers removed. $N_\sigma$ gives the corresponding statistical significance under the frequentist paradigm, following \cite{2008ConPh..49...71T}.  \label{table:retrieval:free:bayes}} 
\begin{tabular}{ccccccc} 
\hline \\ 
\multicolumn{3}{c}{\texttt{SCARLET}} && \multicolumn{3}{c}{\texttt{petitRADTRANS}} \\ \cline{1-3} \cline{5-7}
Model & $B$ & $N_\sigma$    &&     Model & $B$ & $N_\sigma$ \medskip \\ \cline{1-3} \cline{5-7}
H$_2$O, CH$_4$, HCN excluded & 9.6 & 2.7 & &    H$_2$O, CO$_2$, HCN excluded & 7.4 & 2.5  \\
                HCN excluded & 2.5 & 2.0 & & H$_2$O, CO$_2$, CH$_4$ excluded & 6.3 & 2.5  \\
     H$_2$O, CH$_4$ excluded & 2.1 & 1.8 & &            H$_2$O, HCN excluded & 5.9 & 2.4  \\
             H$_2$O excluded & 1.8 & 1.7 & &         H$_2$O, CO$_2$ excluded & 5.9 & 2.4  \\
             CO$_2$ excluded & 1.2 & 1.4 & &         H$_2$O, CH$_4$ excluded & 5.4 & 2.4  \\
                 CO excluded & 1.1 & 1.1 & &    H$_2$O, CH$_4$, HCN excluded & 3.4 & 2.2  \\
              N$_2$ excluded & 1.0 & 1.0 & &         H$_2$O, NH$_3$ excluded & 2.3 & 1.9  \\
      All absorbers included & 1.0 & 0.9 & &            CH$_4$, HCN excluded & 1.7 & 1.7  \\
             CH$_4$ excluded & 0.9 & 0.9 & &                 H$_2$O excluded & 1.5 & 1.6  \\
             NH$_3$ excluded & 0.6 & 0.9 & &                    HCN excluded & 1.5 & 1.6  \\
                             &     &     & &            HCN, NH$_3$ excluded & 1.5 & 1.6  \\
                             &     &     & &         CO$_2$, NH$_3$ excluded & 1.4 & 1.6  \\
                             &     &     & &                     CO excluded & 1.3 & 1.5  \\
                             &     &     & &         CH$_4$, CO$_2$ excluded & 1.2 & 1.3  \\
                             &     &     & &          All absorbers included & 1.0 & 0.9  \\
                             &     &     & &                 CO$_2$ excluded & 0.6 & 0.9  \\
                             &     &     & &                 NH$_3$ excluded & 0.6 & 0.9  \\
                             &     &     & &                 CH$_4$ excluded & 0.6 & 0.9  \\
                             &     &     & &         CH$_4$, NH$_3$ excluded & 0.6 & 0.9  \\ \\ \hline
\end{tabular} 
\end{minipage} 
\end{table}

\subsection{Self-consistent models with disequilibrium chemistry} \label{sec:consistentdiseq}

\begin{figure}
\centering  
\includegraphics[width=0.9\columnwidth]{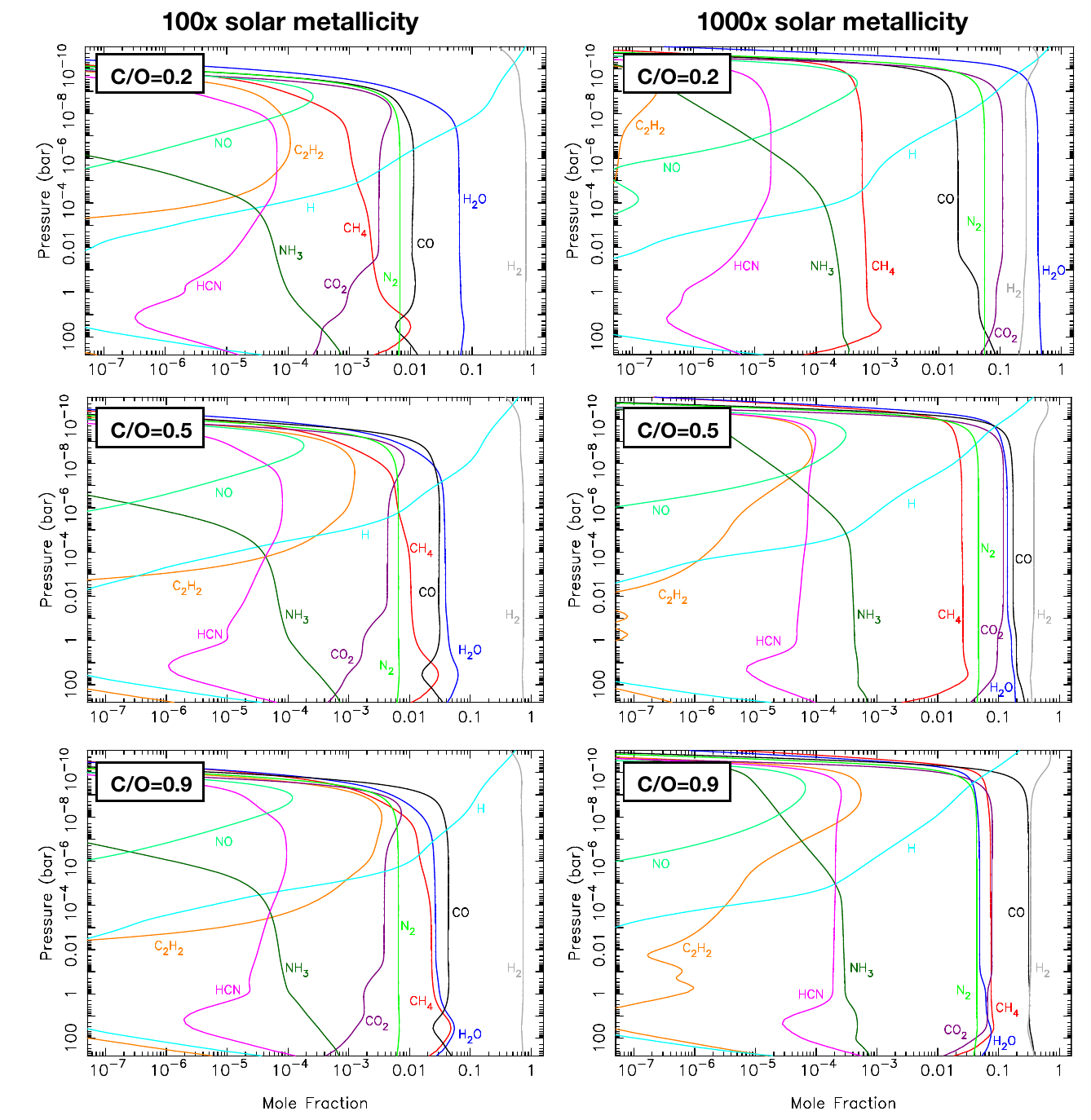}
\caption{Pressure-dependent chemical abundances computed by the thermochemical-photochemical kinetics and transport code described in the text, for a range of metallicities and C/O values. }
\label{fig:diseq}
\end{figure}

To investigate the influence of disequilibrium chemistry on plausible atmospheric compositions for HD~3167c, we ran the thermochemical-photochemical kinetics and transport code described in \citet{2013ApJ...777...34M}. This code accounts for vertical mixing, as well as photochemistry resulting from incident UV photons from the host star. For the latter, we assumed a UV spectrum corresponding to solar-cycle average conditions, taken from \cite{2002GMS...130..221W}. For the vertical mixing, we assumed an eddy diffusion coefficient of $K_{zz}= 2 \times 10^{9}$\,cm\,s$^{-2}$ in the deep convective region of the atmosphere. At lower pressures ($<500$\,bar), we adopted a relation of the form $K_{zz}= \eta \,(1 \textnormal{bar}/P)^{0.65}$\,cm\,s$^{-2}$, based on work by \cite{2013A&A...558A..91P} and \cite{2014A&A...564A..73A} tracking passive tracer particles in 3D general circulation models. We set $\eta = 5 \times 10^6$, which is lower than the values recommended in the latter studies for hot Jupiters, but likely more appropriate for the lower temperature of HD~3167c \citep[e.g.][]{2019ApJ...881..152K}. To restrict the eddy diffusion coefficient to reasonable values at high altitudes, an upper limit of $K_{zz}= 2 \times 10^{10}$\,cm\,s$^{-2}$ was imposed. The models were run for a range of heavy element enrichments ranging from $100$-$1000\times$ solar and C/O values of $0.2$, $0.54$ (i.e. solar), and $0.9$. For each metallicity, we adopt the PT profiles computed in Section \ref{sec:chemeq} under the assumption of chemical equilibrium. 

The resulting pressure-dependent chemical abundances are shown in Figure \ref{fig:diseq}. At the pressures probed by the transmission spectrum (between approximately $10^{-1}$ and $10^{-4}$ bar), the predicted mole fractions are broadly consistent with the abundance constraints derived from the free chemistry retrievals (Table \ref{table:retrieval:free:post}). Admittedly, this is primarily due to the latter being poorly constrained, making them compatible with a large range of atmospheric conditions. Nonetheless, we venture to make a few observations. First, the self-consistent models presented in this section predict the HCN abundance to not exceed $\sim 300$\,ppm. Thus, the peak exhibited by the \texttt{SCARLET} posterior distribution for HCN mole fractions $>1\%$ (Figure \ref{fig:freechem:posterior:scarlet}) seems unlikely from a physical perspective. Instead, we expect the true HCN mole fraction to be contained within the broad lower tail of the \texttt{SCARLET} posterior distribution, which extends down below the 1\,ppm level. Meanwhile, as noted above in Section \ref{sec:retrieval:free:results}, the \texttt{petitRADTRANS} posterior distribution shows an analogous peak around relatively high ($>1000$\,ppm) mole fractions for CO$_2$, rather than HCN. As has been noted previously \citep[e.g.][]{2013ApJ...777...34M,2014RSPTA.37230073M}, the CO$_2$ abundance is highly sensitive to the overall metallicity of the atmosphere. This is clearly evident in Figure \ref{fig:diseq}, with CO$_2$ mole fractions increasing from $<1$\% for $100\times$ solar metallicity to nearly $10$\% for $1000\times$ solar metallicity. As metallicity increases, the H$_2$O abundance is also expected to rise to mole fractions comparable to or exceeding the CO$_2$ mole fraction (Figure \ref{fig:diseq}). As can be seen in Figure \ref{fig:freechem:posterior:scarlet}, the $1\sigma$ credible range for the H$_2$O mole fraction inferred by the \texttt{SCARLET} free chemistry retrieval does indeed extend to nearly 10\%, as is expected for metallicity between $100$-$1000\times$ solar (Figure {fig:diseq}). However, the corresponding range for the H$_2$O mole fraction inferred by the \texttt{petitRADTRANS} retrieval extends up to only $\sim 1\%$ (Table \ref{table:retrieval:free:post}), and would suggest a metallicity $<100\times$ solar. This appears to be in tension with the good match to the data obtained for the chemical equilibrium models with metallicities $>300\times$ solar (Section \ref{sec:chemeq}), and could possibly indicate the \texttt{petitRADTRANS} retrieval has underestimated the H$_2$O abundance.

\begin{figure}[ht!]
\centering  
\includegraphics[width=0.7\linewidth]{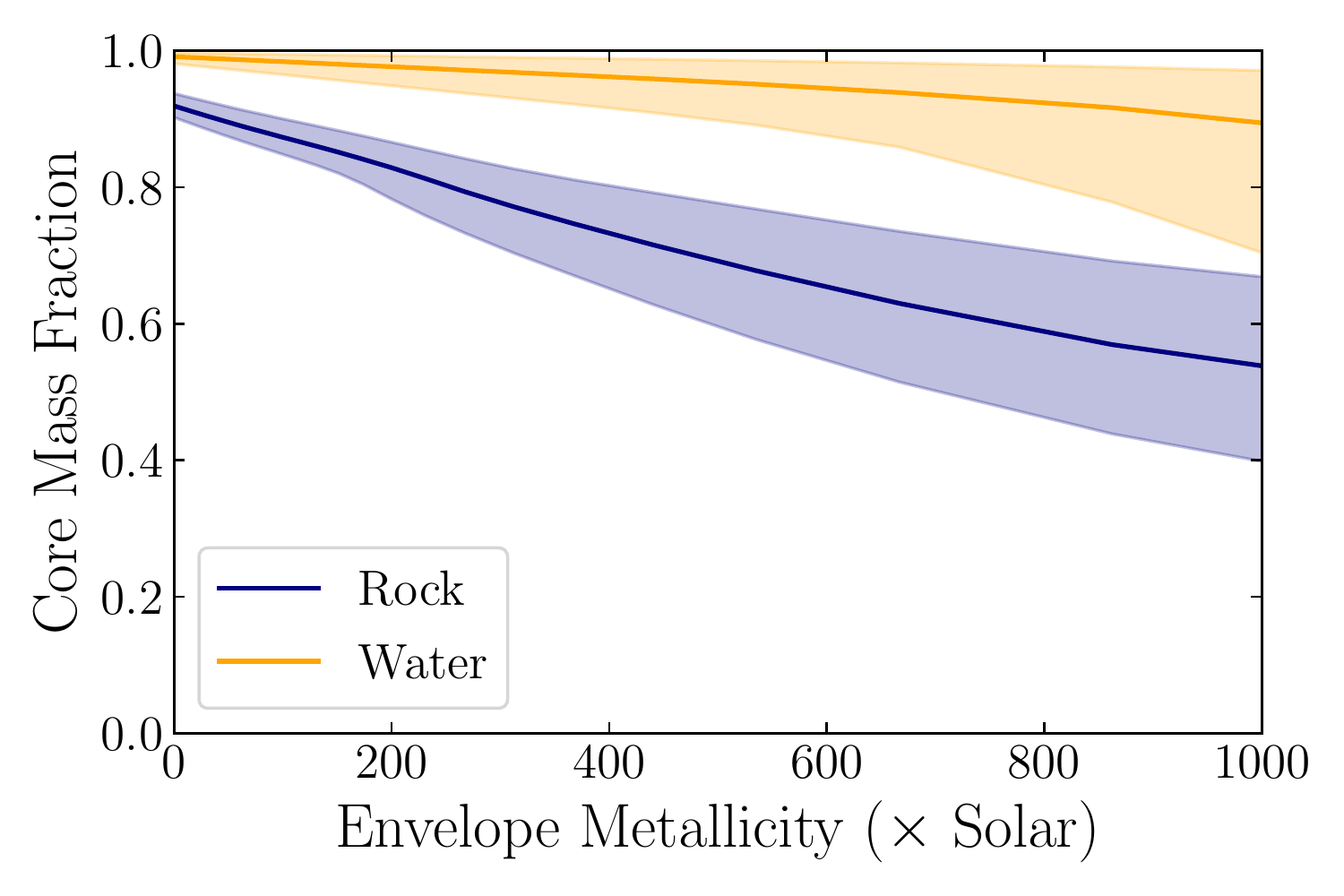}
\caption{Core mass fractions implied by a two-layer interior structure model for a range of atmospheric envelope metallicities. As described in the text, the limiting cases are considered: a pure water core (orange) and a rocky core with a 2-to-1 olivine-to-iron composition (purple). In both cases, the atmospheric envelope is H/He-dominated. Shaded regions indicate $1\sigma$ credible ranges.}
\label{fig:interior}
\end{figure}

\section{Interior structure modeling} \label{sec:interior}

To evaluate the implications of our results for the interior structure of HD~3167c, we take the approach of \cite{2019ApJ...874L..31T} to construct structure evolution models matched to the observed parameters of the planet.  Briefly, these models solve the equations of hydrostatic equilibrium, mass conservation, and an equation of state in one spherically-symmetric dimension to determine the radius from the mass, composition, and envelope specific entropy.  The envelope entropy is evolved over time from a hot initial state using the atmosphere models of \cite{2007ApJ...659.1661F} to regulate cooling.  A detailed description of this model can be found in \cite{2016ApJ...831...64T}.

For simplicity, we assume a two-layer (core and envelope) model and consider two bracketing cases for the core composition: a convective water core and an isothermal olivine-iron core in an Earth-like 2-to-1 ratio \citep[as in][]{2014ApJ...792....1L}.  For our equations of state we use \cite{2019ApJ...872...51C} for H/He, \cite{2019A&A...621A.128M} for the convective water layer, ANEOS \citep{osti_6939284} for the envelope ices and core rock (represented by olivine), and SESAME \citep{ironEOS} for the core iron.  The H/He envelope is approximated as fully adiabatic.  Although sub-Neptunes may have composition gradients which cause semi-convection in some regions, the fully-convecting envelope approximation has produced good results for Neptune \citep{2010SSRv..152..423F}, and has been usefully applied to other sub-Neptunes \citep{2014ApJ...792....1L,2011ApJ...733....2N}.

For HD 3167c, we adopt the following properties:\ radius $2.77 \pm 0.20 \,\RE$ (Table \ref{table:broadfit}); mass $9.8 \pm 1.3 \,\ME$ \citep{2017AJ....154..122C}; and age $6.7 \pm 0.8$\,Gyr (Section \ref{sec:stellarabund}).  We then use MCMC to estimate the core mass required to match these properties as a function of atmospheric envelope metallicity. The results are shown in Figure \ref{fig:interior}. Note that the atmospheric metallicity results from the transmission spectrum are not used directly in the model, but help us to interpret the result, as upward-convecting envelope material becomes atmoshpere material. In particular, the transmission spectrum favors atmospheric metallicities up to $1000\times$ solar (Figure \ref{fig:scarlet:chemeq:MHvsCloud}). We find that for atmospheric envelope metallicities across this range, HD~3167c must have a core mass fraction of at least $\sim 40$\% for a rock-dominated core and at least $\sim 60$\% for a water-dominated core at the $1\sigma$ level (Figure \ref{fig:interior}). The real core is probably a mixture of these and so its minimum mass likely lies between these two limiting cases.

\section{Discussion} \label{sec:discussion}

\begin{figure}
\centering  
\includegraphics[width=0.95\linewidth]{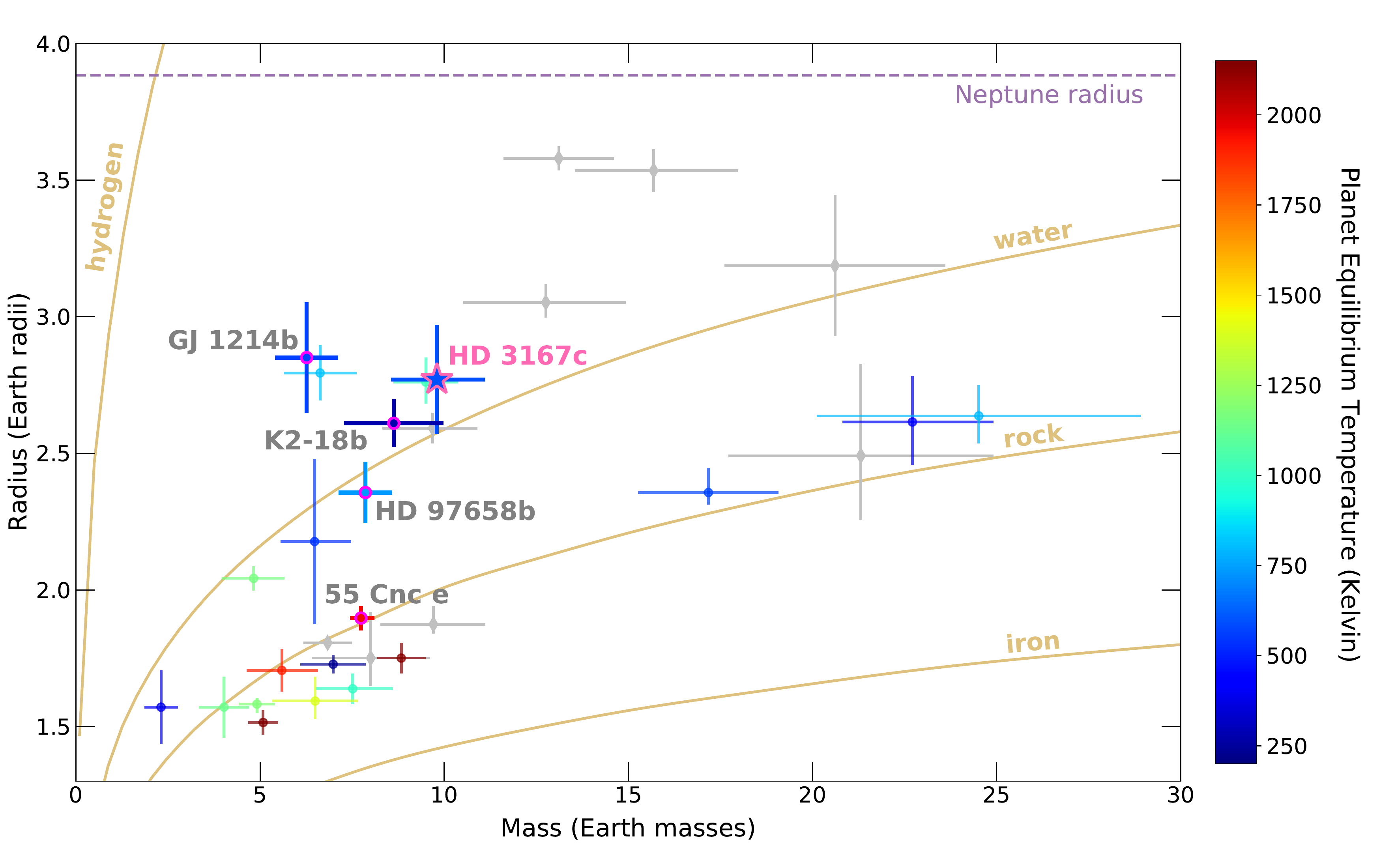}
\caption{Masses and radii for known sub-Neptunes and super-Earths. Error bars show $1\sigma$ measurement uncertainties. Only planets with published masses measured to better than 20\% precision are shown. Colored markers indicate planets orbiting bright stars ($J<10$mag), typically a prerequisite for atmosphere characterization observations, while grey diamonds indicate fainter systems. The color scale for bright systems corresponds to planet equilibrium temperature, assuming zero Bond albedo and uniform heat redistribution from dayside to nightside hemispheres. Light brown contours are for 100\% iron, rock, water, and hydrogen compositions. The five sub-Neptunes with published HST transmission spectra are labeled and highlighted by pink halos. 
\label{fig:massradius}}
\end{figure}

The HST transmission spectrum presented here for HD~3167c adds to a modest sample acquired to date for sub-Neptunes above the radius valley, with radii between $1.8$-$4\,\RE$. As shown in Figure \ref{fig:massradius}, the other sub-Neptunes are GJ~1214b \citep{2014Natur.505...69K}, 55~Cnc~e \citep{2016ApJ...820...99T}, HD~97658b \citep{2020AJ....159..239G}, and K2-18b \citep{2019ApJ...887L..14B}. In addition to those shown in Figure \ref{fig:massradius}, there have been a few spectra published for planets somewhat larger than Neptune ($4$-$5\,\RE$), namely: HAT-P-11b \citep{2014Natur.513..526F}, GJ~436b \citep{2014Natur.505...66K}, GJ~3470b \citep{2019NatAs...3..813B}, and HD~106315c \citep{2020arXiv200607444K}. Already, the diversity predicted for the atmospheric properties of this population \citep[e.g.][]{2008ApJ...685.1237E,2013ApJ...775...80F} is being borne out by observations, reminiscent of that which has been well documented for the hot Jupiters \citep[e.g.][]{2016Natur.529...59S}.

As described in Section \ref{sec:atmchar}, the currently available data do not allow us to discriminate between high metallicity and cloudy atmosphere scenarios for HD~3167c. However, due to the sensitivity of transmission spectra to even trace amounts of cloud \citep{2005MNRAS.364..649F}, this degeneracy is a widespread issue that not only affects other Neptune-sized planets \citep[e.g.][]{2019ApJ...887L..14B,2020arXiv200607444K}, but many hot Jupiters \citep[e.g.][]{2015arXiv150407655B}.

For the composition of a putative cloud layer, broadly speaking there are two possibilities:\ equilibrium condensates and photochemical hazes. Given the temperature of HD~3167c, the most likely equilibrium condensate is KCl, which was proposed by \cite{2012ApJ...756..172M} as an important cloud species in the atmospheres of T dwarfs. Indeed, for all but the highest metallicities ($<1000\times$ solar), the PT profile of HD~3167c is expected to first cross the KCl condensation curve at pressures $\lesssim 1$\,bar (Figure \ref{fig:PT}), coinciding with pressures probed by the transmission spectrum \citep[e.g.][]{2019A&A...627A..67M}. The PT profile is also expected to cross the condensation curve of ZnS (Figure \ref{fig:PT}), which was suggested by \cite{2012ApJ...756..172M} as another potentially important cloud species. However, recent modeling work by \cite{2020NatAs.tmp..142G} suggests the formation of significant ZnS cloud mass is challenging, owing to high nucleation energy barriers. In the same study, Gao et al.\ note that although the formation of KCl cloud should be efficient, at temperatures below 950\,K photochemical haze is likely to to be the dominant aerosol opacity source. This is due to the increasing abundance of CH$_4$ under chemical equilibrium, which can be photodissociated by incoming UV photons to generate hydrocarbon haze \citep[e.g.][]{2015ApJ...815..110M,2018ApJ...853....7K,2019ApJ...877..109K}.

It is interesting to compare HD~3167c with GJ~1214b, as these two planets have similar radii and temperatures (Figure \ref{fig:massradius}). Although the GJ~1214b transmission spectrum measured with WFC3 \citep{2014Natur.505...69K} is not as precise as that presented here for HD~3167c (Table \ref{table:specfit}), the smaller radius of GJ~1214 \citep[$\GJRS$;][]{2012ApJ...747...35B} relative to HD~3167 \citep[$\HDRS$;][]{2017AJ....154..122C} results in tighter constraints for the atmospheric properties of the former. Specifically, the only plausible explanation for the lack of spectral features detected in the GJ~1214b spectrum is a high-altitude aerosol layer \citep{2014Natur.505...69K,2015ApJ...815..110M}. However, we find $2.5\sigma$ evidence for molecular absorption in the HD~3167c transmission spectrum (Section \ref{sec:retrieval:free:results}), suggesting if cloud is present it does not extend as high in the atmosphere as for GJ~1214b. The fact that HD~3167c is about 50\% more massive \citep[$\HDMP$;][]{2017AJ....154..122C} than GJ~1214b \citep[$6.26 \pm 0.86\,\ME$;][]{2013A&A...549A..10H}, may provide a clue to understanding this difference. The higher gravitational acceleration could result in an increased sedimentation efficiency for photochemical hazes formed in upper layers of the atmosphere and/or a reduced eddy diffusion coefficient ($K_{zz}$) for mixing equilibrium condensates from deeper layers of the atmosphere. Also worth mentioning in this regard is Kepler-51b, which has a similar equilibrium temperature \citep[$\KTEQ $\,K;][]{2014ApJ...783...53M} to HD~3167c and GJ~1214b, and with a mass of \citep[$\KMP$;][]{2014ApJ...783...53M} could reasonably be considered a sub-Neptune. However, with a radius of \citep[$\KRP$;][]{2014ApJ...783...53M}, Kepler-51b is significantly larger than Neptune and thus has a much lower density than both HD~3167c and GJ1214b. As for GJ~1214b, the measured transmission spectrum is featureless, implying a high-altitude aerosol layer \citep{2020AJ....159...57L}, which would be consistent with the picture that lower density planets are more likely to have photochemical hazes persist at high altitudes, or equilibrium condensates lofted from deeper layers.

The reality is undoubtedly more complicated. The hot Jupiters, for which more transmission spectra have been measured, exhibit no discernible correlation between bulk properties such as mass, radius, and temperature, and the presence of equilibrium condensates \citep{2016Natur.529...59S}. We might expect the atmospheric properties of sub-Neptunes to exhibit a similar level of stochasticity. Along these lines, \cite{2018ApJ...853....7K,2019ApJ...877..109K} have highlighted how photochemical haze production sensitively depends on numerous factors, such as the host star UV spectrum, atmospheric metallicity, and C/O ratio, and variations in vertical mixing efficiences. Existing sub-Neptune observations provide only limited ability to constrain models in this multi-dimensional parameter space. Only by continuing to acquire additional observations for a larger sample of sub-Neptunes might this situation be rectified, while at present it seems impossible to predict the degree to which a given transmission spectrum will be affected by aersols prior to observations being made.

Aside from allowing for an aerosol layer, the transmission spectrum is consistent with high metallicities of $\sim 100$-$1000\times$ solar (Figure \ref{fig:scarlet:chemeq:MHvsCloud}). On this point, the hint of CO$_2$ inferred by the \texttt{petitRADTRANS} retrieval is perhaps worth flagging (Section \ref{sec:retrieval:free:results}), as CO$_2$ is expected to become an important absorber at high metallicities \citep{2013ApJ...777...34M}. However, it must be reiterated that the indication of CO$_2$ in the existing data is extremely tentative, particularly since the \texttt{SCARLET} retrieval did not favor its presence (Table \ref{table:retrieval:free:bayes}). Unfortunately, the uncertainty on the IRAC transit depth is too large to make out the strong CO$_2$ absorption band in the $4$-$5\,\um$ wavelength range, if it is present \citep[see for example,][]{2019arXiv191108859S}. Observations made with either the NIRSpec or NIRCAM instruments on the \textit{James Webb Space Telescope} should shed conclusive light on this \citep[e.g.][]{2016ApJ...817...17G}.

\section{Conclusion} \label{sec:conclusion}

We have presented a transmission spectrum for the warm Neptune HD~3167c measured using HST WFC3, combined with broadband \textit{K2} and IRAC photometry. Our results rule out cloud-free models with metallicities $<100 \times$ solar at high confidence. Instead, the data can be well explained by cloud-free equilibrium chemistry models with metallicities $>700\times$ solar. However, when clouds are considered, the data are also consistent with much lower metallicities, including subsolar. We find evidence for at least one of H$_2$O, CO$_2$, HCN, and/or CH$_4$ at the equivalent of $2.5\sigma$ significance when using free-chemistry retrievals. However, the available data do not allow the unambiguous identification of any single gas species in the data. We found the broad allowed ranges for all considered gas species are consistent with predictions made by self-consistent models that account for photochemistry and vertical mixing. Two-layered interior structure modeling suggests the core mass fraction is at least 40\%, independent of the assumed core composition and atmospheric envelope metallicities up to $1000\times$ solar.

\clearpage

\bibliographystyle{apj}
\bibliography{hd3167}

\begin{thebibliography}{}
\expandafter\ifx\csname natexlab\endcsname\relax\def\natexlab#1{#1}\fi

\bibitem[{{Adams} {et~al.}(2008){Adams}, {Seager}, \&
  {Elkins-Tanton}}]{2008ApJ...673.1160A}
{Adams}, E.~R., {Seager}, S., \& {Elkins-Tanton}, L. 2008, \apj, 673, 1160

\bibitem[{{Ag{\'u}ndez} {et~al.}(2014){Ag{\'u}ndez}, {Parmentier}, {Venot},
  {Hersant}, \& {Selsis}}]{2014A&A...564A..73A}
{Ag{\'u}ndez}, M., {Parmentier}, V., {Venot}, O., {Hersant}, F., \& {Selsis},
  F. 2014, \aap, 564, A73

\bibitem[{{Alam} {et~al.}(2020){Alam}, {Lopez-Morales}, {Nikolov}, {Sing},
  {Henry}, {Baxter}, {Desert}, {Barstow}, {Mikal-Evans}, {Bourrier}, {Lavvas},
  {Wakeford}, {Williamson}, {Sanz-Forcada}, {Buchhave}, {Cohen}, \& {Garcia
  Munoz}}]{2020arXiv200511293A}
{Alam}, M.~K., {Lopez-Morales}, M., {Nikolov}, N., {et~al.} 2020, arXiv
  e-prints, arXiv:2005.11293

\bibitem[{{Arcangeli} {et~al.}(2019){Arcangeli}, {D{\'e}sert}, {Parmentier},
  {Stevenson}, {Bean}, {Line}, {Kreidberg}, {Fortney}, \&
  {Showman}}]{2019A&A...625A.136A}
{Arcangeli}, J., {D{\'e}sert}, J.-M., {Parmentier}, V., {et~al.} 2019, \aap,
  625, A136

\bibitem[{{Asplund} {et~al.}(2009){Asplund}, {Grevesse}, {Sauval}, \&
  {Scott}}]{2009ARA&A..47..481A}
{Asplund}, M., {Grevesse}, N., {Sauval}, A.~J., \& {Scott}, P. 2009, \araa, 47,
  481

\bibitem[{{Astropy Collaboration} {et~al.}(2013){Astropy Collaboration},
  {Robitaille}, {Tollerud}, {Greenfield}, {Droettboom}, {Bray}, {Aldcroft},
  {Davis}, {Ginsburg}, {Price-Whelan}, {Kerzendorf}, {Conley}, {Crighton},
  {Barbary}, {Muna}, {Ferguson}, {Grollier}, {Parikh}, {Nair}, {Unther},
  {Deil}, {Woillez}, {Conseil}, {Kramer}, {Turner}, {Singer}, {Fox}, {Weaver},
  {Zabalza}, {Edwards}, {Azalee Bostroem}, {Burke}, {Casey}, {Crawford},
  {Dencheva}, {Ely}, {Jenness}, {Labrie}, {Lim}, {Pierfederici}, {Pontzen},
  {Ptak}, {Refsdal}, {Servillat}, \& {Streicher}}]{2013A&A...558A..33A}
{Astropy Collaboration}, {Robitaille}, T.~P., {Tollerud}, E.~J., {et~al.} 2013,
  \aap, 558, A33

\bibitem[{{Astropy Collaboration} {et~al.}(2018){Astropy Collaboration},
  {Price-Whelan}, {Sip{\H{o}}cz}, {G{\"u}nther}, {Lim}, {Crawford}, {Conseil},
  {Shupe}, {Craig}, {Dencheva}, {Ginsburg}, {Vand erPlas}, {Bradley},
  {P{\'e}rez-Su{\'a}rez}, {de Val-Borro}, {Aldcroft}, {Cruz}, {Robitaille},
  {Tollerud}, {Ardelean}, {Babej}, {Bach}, {Bachetti}, {Bakanov}, {Bamford},
  {Barentsen}, {Barmby}, {Baumbach}, {Berry}, {Biscani}, {Boquien}, {Bostroem},
  {Bouma}, {Brammer}, {Bray}, {Breytenbach}, {Buddelmeijer}, {Burke},
  {Calderone}, {Cano Rodr{\'\i}guez}, {Cara}, {Cardoso}, {Cheedella}, {Copin},
  {Corrales}, {Crichton}, {D'Avella}, {Deil}, {Depagne}, {Dietrich}, {Donath},
  {Droettboom}, {Earl}, {Erben}, {Fabbro}, {Ferreira}, {Finethy}, {Fox},
  {Garrison}, {Gibbons}, {Goldstein}, {Gommers}, {Greco}, {Greenfield},
  {Groener}, {Grollier}, {Hagen}, {Hirst}, {Homeier}, {Horton}, {Hosseinzadeh},
  {Hu}, {Hunkeler}, {Ivezi{\'c}}, {Jain}, {Jenness}, {Kanarek}, {Kendrew},
  {Kern}, {Kerzendorf}, {Khvalko}, {King}, {Kirkby}, {Kulkarni}, {Kumar},
  {Lee}, {Lenz}, {Littlefair}, {Ma}, {Macleod}, {Mastropietro}, {McCully},
  {Montagnac}, {Morris}, {Mueller}, {Mumford}, {Muna}, {Murphy}, {Nelson},
  {Nguyen}, {Ninan}, {N{\"o}the}, {Ogaz}, {Oh}, {Parejko}, {Parley}, {Pascual},
  {Patil}, {Patil}, {Plunkett}, {Prochaska}, {Rastogi}, {Reddy Janga},
  {Sabater}, {Sakurikar}, {Seifert}, {Sherbert}, {Sherwood-Taylor}, {Shih},
  {Sick}, {Silbiger}, {Singanamalla}, {Singer}, {Sladen}, {Sooley},
  {Sornarajah}, {Streicher}, {Teuben}, {Thomas}, {Tremblay}, {Turner},
  {Terr{\'o}n}, {van Kerkwijk}, {de la Vega}, {Watkins}, {Weaver}, {Whitmore},
  {Woillez}, {Zabalza}, \& {Astropy Contributors}}]{2018AJ....156..123A}
{Astropy Collaboration}, {Price-Whelan}, A.~M., {Sip{\H{o}}cz}, B.~M., {et~al.}
  2018, \aj, 156, 123

\bibitem[{{Bate} {et~al.}(2010){Bate}, {Lodato}, \&
  {Pringle}}]{2010MNRAS.401.1505B}
{Bate}, M.~R., {Lodato}, G., \& {Pringle}, J.~E. 2010, \mnras, 401, 1505

\bibitem[{{Batygin}(2012)}]{2012Natur.491..418B}
{Batygin}, K. 2012, \nat, 491, 418

\bibitem[{{Baumeister} {et~al.}(2020){Baumeister}, {Padovan}, {Tosi},
  {Montavon}, {Nettelmann}, {MacKenzie}, \& {Godolt}}]{2020ApJ...889...42B}
{Baumeister}, P., {Padovan}, S., {Tosi}, N., {et~al.} 2020, \apj, 889, 42

\bibitem[{{Benneke}(2015)}]{2015arXiv150407655B}
{Benneke}, B. 2015, arXiv e-prints, arXiv:1504.07655

\bibitem[{{Benneke} \& {Seager}(2012)}]{2012ApJ...753..100B}
{Benneke}, B., \& {Seager}, S. 2012, \apj, 753, 100

\bibitem[{{Benneke} \& {Seager}(2013)}]{2013ApJ...778..153B}
---. 2013, \apj, 778, 153

\bibitem[{{Benneke} {et~al.}(2019{\natexlab{a}}){Benneke}, {Knutson},
  {Lothringer}, {Crossfield}, {Moses}, {Morley}, {Kreidberg}, {Fulton},
  {Dragomir}, {Howard}, {Wong}, {D{\'e}sert}, {McCullough}, {Kempton},
  {Fortney}, {Gilliland }, {Deming}, \& {Kammer}}]{2019NatAs...3..813B}
{Benneke}, B., {Knutson}, H.~A., {Lothringer}, J., {et~al.} 2019{\natexlab{a}},
  Nature Astronomy, 3, 813

\bibitem[{{Benneke} {et~al.}(2019{\natexlab{b}}){Benneke}, {Wong}, {Piaulet},
  {Knutson}, {Lothringer}, {Morley}, {Crossfield}, {Gao}, {Greene}, {Dressing},
  {Dragomir}, {Howard}, {McCullough}, {Kempton}, {Fortney}, \&
  {Fraine}}]{2019ApJ...887L..14B}
{Benneke}, B., {Wong}, I., {Piaulet}, C., {et~al.} 2019{\natexlab{b}}, \apjl,
  887, L14

\bibitem[{{Berta} {et~al.}(2012){Berta}, {Charbonneau}, {D{\'e}sert},
  {Miller-Ricci Kempton}, {McCullough}, {Burke}, {Fortney}, {Irwin}, {Nutzman},
  \& {Homeier}}]{2012ApJ...747...35B}
{Berta}, Z.~K., {Charbonneau}, D., {D{\'e}sert}, J.-M., {et~al.} 2012, \apj,
  747, 35

\bibitem[{{Brewer} \& {Fischer}(2016)}]{brewer:2016}
{Brewer}, J.~M., \& {Fischer}, D.~A. 2016, \apj, 831, 20

\bibitem[{{Brewer} {et~al.}(2016){Brewer}, {Fischer}, {Valenti}, \&
  {Piskunov}}]{brewer:2016a}
{Brewer}, J.~M., {Fischer}, D.~A., {Valenti}, J.~A., \& {Piskunov}, N. 2016,
  \apjs, 225, 32

\bibitem[{{Bruno} {et~al.}(2020){Bruno}, {Lewis}, {Alam}, {L{\'o}pez-Morales},
  {Barstow}, {Wakeford}, {Sing}, {Henry}, {Ballester}, {Bourrier}, {Buchhave},
  {Cohen}, {Mikal-Evans}, {Garc{\'\i}a Mu{\~n}oz}, {Lavvas}, \&
  {Sanz-Forcada}}]{2020MNRAS.491.5361B}
{Bruno}, G., {Lewis}, N.~K., {Alam}, M.~K., {et~al.} 2020, \mnras, 491, 5361

\bibitem[{{Buchner} {et~al.}(2014){Buchner}, {Georgakakis}, {Nandra}, {Hsu},
  {Rangel}, {Brightman}, {Merloni}, {Salvato}, {Donley}, \&
  {Kocevski}}]{2014A&A...564A.125B}
{Buchner}, J., {Georgakakis}, A., {Nandra}, K., {et~al.} 2014, \aap, 564, A125

\bibitem[{{Carone} {et~al.}(2020){Carone}, {Molli{\`e}re}, {Zhou}, {Bouwman},
  {Yan}, {Baeyens}, {Apai}, {Espinoza}, {Rackham}, {Jord{\'a}n}, {Angerhausen},
  {Decin}, {Lendl}, {Venot}, \& {Henning}}]{2020arXiv200605382C}
{Carone}, L., {Molli{\`e}re}, P., {Zhou}, Y., {et~al.} 2020, arXiv e-prints,
  arXiv:2006.05382

\bibitem[{{Carter} {et~al.}(2020){Carter}, {Nikolov}, {Sing}, {Alam}, {Goyal},
  {Mikal-Evans}, {Wakeford}, {Henry}, {Morrell}, {L{\'o}pez-Morales},
  {Smalley}, {Lavvas}, {Barstow}, {Garc{\'\i}a Mu{\~n}oz}, {Gibson}, \&
  {Wilson}}]{2020MNRAS.494.5449C}
{Carter}, A.~L., {Nikolov}, N., {Sing}, D.~K., {et~al.} 2020, \mnras, 494, 5449

\bibitem[{{Chabrier} {et~al.}(2019){Chabrier}, {Mazevet}, \&
  {Soubiran}}]{2019ApJ...872...51C}
{Chabrier}, G., {Mazevet}, S., \& {Soubiran}, F. 2019, \apj, 872, 51

\bibitem[{{Chachan} {et~al.}(2019){Chachan}, {Knutson}, {Gao}, {Kataria},
  {Wong}, {Henry}, {Benneke}, {Zhang}, {Barstow}, {Bean}, {Mikal-Evans},
  {Lewis}, {Mansfield}, {L{\'o}pez-Morales}, {Nikolov}, {Sing}, \&
  {Wakeford}}]{2019AJ....158..244C}
{Chachan}, Y., {Knutson}, H.~A., {Gao}, P., {et~al.} 2019, \aj, 158, 244

\bibitem[{{Christiansen} {et~al.}(2017){Christiansen}, {Vanderburg}, {Burt},
  {Fulton}, {Batygin}, {Benneke}, {Brewer}, {Charbonneau}, {Ciardi}, {Collier
  Cameron}, {Coughlin}, {Crossfield}, {Dressing}, {Greene}, {Howard}, {Latham},
  {Molinari}, {Mortier}, {Mullally}, {Pepe}, {Rice}, {Sinukoff}, {Sozzetti},
  {Thompson}, {Udry}, {Vogt}, {Barman}, {Batalha}, {Bouchy}, {Buchhave},
  {Butler}, {Cosentino}, {Dupuy}, {Ehrenreich}, {Fiorenzano}, {Hansen},
  {Henning}, {Hirsch}, {Holden}, {Isaacson}, {Johnson}, {Knutson}, {Kosiarek},
  {L{\'o}pez-Morales}, {Lovis}, {Malavolta}, {Mayor}, {Micela}, {Motalebi},
  {Petigura}, {Phillips}, {Piotto}, {Rogers}, {Sasselov}, {Schlieder},
  {S{\'e}gransan}, {Watson}, \& {Weiss}}]{2017AJ....154..122C}
{Christiansen}, J.~L., {Vanderburg}, A., {Burt}, J., {et~al.} 2017, \aj, 154,
  122

\bibitem[{{Cloutier} \& {Menou}(2020)}]{2020AJ....159..211C}
{Cloutier}, R., \& {Menou}, K. 2020, \aj, 159, 211

\bibitem[{{Col{\'o}n} {et~al.}(2020){Col{\'o}n}, {Kreidberg}, {Line},
  {Welbanks}, {Madhusudhan}, {Beatty}, {Tamburo}, {Stevenson}, {Mandell},
  {Rodriguez}, {Barclay}, {Lopez}, {Stassun}, {Angerhausen}, {Fortney},
  {James}, {Pepper}, {Ahlers}, {Plavchan}, {Awiphan}, {Kotnik}, {McLeod},
  {Murawski}, {Chotani}, {LeBrun}, {Matzko}, {Rea}, {Vidaurri}, {Webster},
  {Williams}, {Sheraden Cox}, {Tan}, \& {Gilbert}}]{2020arXiv200505153C}
{Col{\'o}n}, K.~D., {Kreidberg}, L., {Line}, M., {et~al.} 2020, arXiv e-prints,
  arXiv:2005.05153

\bibitem[{{Crossfield} \&
  {Kreidberg}(2017{\natexlab{a}})}]{2017hst..prop15333C}
{Crossfield}, I. J.~M., \& {Kreidberg}, L. 2017{\natexlab{a}}, {The Atmospheric
  Diversity of Mini-Neptunes in Multi-planet Systems}, HST Proposal

\bibitem[{{Crossfield} \&
  {Kreidberg}(2017{\natexlab{b}})}]{2017AJ....154..261C}
---. 2017{\natexlab{b}}, \aj, 154, 261

\bibitem[{{Dai} {et~al.}(2019){Dai}, {Masuda}, {Winn}, \&
  {Zeng}}]{2019ApJ...883...79D}
{Dai}, F., {Masuda}, K., {Winn}, J.~N., \& {Zeng}, L. 2019, \apj, 883, 79

\bibitem[{{Dalal} {et~al.}(2019){Dalal}, {H{\'e}brard}, {Lecavelier des
  {\'E}tangs}, {Petit}, {Bourrier}, {Laskar}, {K{\"o}nig}, \&
  {Correia}}]{2019A&A...631A..28D}
{Dalal}, S., {H{\'e}brard}, G., {Lecavelier des {\'E}tangs}, A., {et~al.} 2019,
  \aap, 631, A28

\bibitem[{{de Wit} {et~al.}(2018){de Wit}, {Wakeford}, {Lewis}, {Delrez},
  {Gillon}, {Selsis}, {Leconte}, {Demory}, {Bolmont}, {Bourrier}, {Burgasser},
  {Grimm}, {Jehin}, {Lederer}, {Owen}, {Stamenkovi{\'c}}, \&
  {Triaud}}]{2018NatAs...2..214D}
{de Wit}, J., {Wakeford}, H.~R., {Lewis}, N.~K., {et~al.} 2018, Nature
  Astronomy, 2, 214

\bibitem[{{Deming} {et~al.}(2019){Deming}, {Louie}, \&
  {Sheets}}]{2019PASP..131a3001D}
{Deming}, D., {Louie}, D., \& {Sheets}, H. 2019, \pasp, 131, 013001

\bibitem[{{Deming} {et~al.}(2013){Deming}, {Wilkins}, {McCullough}, {Burrows},
  {Fortney}, {Agol}, {Dobbs-Dixon}, {Madhusudhan}, {Crouzet}, {Desert},
  {Gilliland}, {Haynes}, {Knutson}, {Line}, {Magic}, {Mandell}, {Ranjan},
  {Charbonneau}, {Clampin}, {Seager}, \& {Showman}}]{2013ApJ...774...95D}
{Deming}, D., {Wilkins}, A., {McCullough}, P., {et~al.} 2013, \apj, 774, 95

\bibitem[{{Deming} {et~al.}(2015){Deming}, {Knutson}, {Kammer}, {Fulton},
  {Ingalls}, {Carey}, {Burrows}, {Fortney}, {Todorov}, {Agol}, {Cowan},
  {Desert}, {Fraine}, {Langton}, {Morley}, \& {Showman}}]{2015ApJ...805..132D}
{Deming}, D., {Knutson}, H., {Kammer}, J., {et~al.} 2015, \apj, 805, 132

\bibitem[{{Deming} \& {Seager}(2017)}]{2017JGRE..122...53D}
{Deming}, L.~D., \& {Seager}, S. 2017, Journal of Geophysical Research
  (Planets), 122, 53

\bibitem[{{Dorn} {et~al.}(2017){Dorn}, {Venturini}, {Khan}, {Heng}, {Alibert},
  {Helled}, {Rivoldini}, \& {Benz}}]{2017A&A...597A..37D}
{Dorn}, C., {Venturini}, J., {Khan}, A., {et~al.} 2017, \aap, 597, A37

\bibitem[{{dos Santos} {et~al.}(2019){dos Santos}, {Ehrenreich}, {Bourrier},
  {Lecavelier des Etangs}, {L{\'o}pez-Morales}, {Sing}, {Ballester},
  {Ben-Jaffel}, {Buchhave}, {Garc{\'\i}a Mu{\~n}oz}, {Henry}, {Kataria},
  {Lavie}, {Lavvas}, {Lewis}, {Mikal-Evans}, {Sanz-Forcada}, \&
  {Wakeford}}]{2019A&A...629A..47D}
{dos Santos}, L.~A., {Ehrenreich}, D., {Bourrier}, V., {et~al.} 2019, \aap,
  629, A47

\bibitem[{{Dressing} \& {Charbonneau}(2013)}]{2013ApJ...767...95D}
{Dressing}, C.~D., \& {Charbonneau}, D. 2013, \apj, 767, 95

\bibitem[{{Ehrenreich} {et~al.}(2014){Ehrenreich}, {Bonfils}, {Lovis},
  {Delfosse}, {Forveille}, {Mayor}, {Neves}, {Santos}, {Udry}, \&
  {S{\'e}gransan}}]{2014A&A...570A..89E}
{Ehrenreich}, D., {Bonfils}, X., {Lovis}, C., {et~al.} 2014, \aap, 570, A89

\bibitem[{{Elkins-Tanton} \& {Seager}(2008)}]{2008ApJ...685.1237E}
{Elkins-Tanton}, L.~T., \& {Seager}, S. 2008, \apj, 685, 1237

\bibitem[{{Evans} {et~al.}(2015){Evans}, {Aigrain}, {Gibson}, {Barstow},
  {Amundsen}, {Tremblin}, \& {Mourier}}]{2015MNRAS.451..680E}
{Evans}, T.~M., {Aigrain}, S., {Gibson}, N., {et~al.} 2015, \mnras, 451, 680

\bibitem[{{Evans} {et~al.}(2016){Evans}, {Sing}, {Wakeford}, {Nikolov},
  {Ballester}, {Drummond}, {Kataria}, {Gibson}, {Amundsen}, \&
  {Spake}}]{2016ApJ...822L...4E}
{Evans}, T.~M., {Sing}, D.~K., {Wakeford}, H.~R., {et~al.} 2016, \apjl, 822, L4

\bibitem[{{Evans} {et~al.}(2017){Evans}, {Sing}, {Kataria}, {Goyal}, {Nikolov},
  {Wakeford}, {Deming}, {Marley}, {Amundsen}, {Ballester}, {Barstow},
  {Ben-Jaffel}, {Bourrier}, {Buchhave}, {Cohen}, {Ehrenreich}, {Garc{\'{\i}}a
  Mu{\~n}oz}, {Henry}, {Knutson}, {Lavvas}, {Lecavelier Des Etangs}, {Lewis},
  {L{\'o}pez-Morales}, {Mandell}, {Sanz-Forcada}, {Tremblin}, \&
  {Lupu}}]{2017Natur.548...58E}
{Evans}, T.~M., {Sing}, D.~K., {Kataria}, T., {et~al.} 2017, \nat, 548, 58

\bibitem[{{Fazio}(2004)}]{2004ApJS..154...10F}
{Fazio}, G.~G. 2004, \apjs, 154, 10

\bibitem[{{Foreman-Mackey} {et~al.}(2013){Foreman-Mackey}, {Hogg}, {Lang}, \&
  {Goodman}}]{2013PASP..125..306F}
{Foreman-Mackey}, D., {Hogg}, D.~W., {Lang}, D., \& {Goodman}, J. 2013, \pasp,
  125, 306

\bibitem[{{Foreman-Mackey} {et~al.}(2014){Foreman-Mackey}, {Hogg}, \&
  {Morton}}]{2014ApJ...795...64F}
{Foreman-Mackey}, D., {Hogg}, D.~W., \& {Morton}, T.~D. 2014, \apj, 795, 64

\bibitem[{{Fortney}(2005)}]{2005MNRAS.364..649F}
{Fortney}, J.~J. 2005, \mnras, 364, 649

\bibitem[{{Fortney} {et~al.}(2007){Fortney}, {Marley}, \&
  {Barnes}}]{2007ApJ...659.1661F}
{Fortney}, J.~J., {Marley}, M.~S., \& {Barnes}, J.~W. 2007, \apj, 659, 1661

\bibitem[{{Fortney} {et~al.}(2013){Fortney}, {Mordasini}, {Nettelmann},
  {Kempton}, {Greene}, \& {Zahnle}}]{2013ApJ...775...80F}
{Fortney}, J.~J., {Mordasini}, C., {Nettelmann}, N., {et~al.} 2013, \apj, 775,
  80

\bibitem[{{Fortney} \& {Nettelmann}(2010)}]{2010SSRv..152..423F}
{Fortney}, J.~J., \& {Nettelmann}, N. 2010, \ssr, 152, 423

\bibitem[{{Fortney} {et~al.}(2010){Fortney}, {Shabram}, {Showman}, {Lian},
  {Freedman}, {Marley}, \& {Lewis}}]{2010ApJ...709.1396F}
{Fortney}, J.~J., {Shabram}, M., {Showman}, A.~P., {et~al.} 2010, \apj, 709,
  1396

\bibitem[{{Fraine} {et~al.}(2014){Fraine}, {Deming}, {Benneke}, {Knutson},
  {Jord{\'a}n}, {Espinoza}, {Madhusudhan}, {Wilkins}, \&
  {Todorov}}]{2014Natur.513..526F}
{Fraine}, J., {Deming}, D., {Benneke}, B., {et~al.} 2014, \nat, 513, 526

\bibitem[{{Fressin} {et~al.}(2013){Fressin}, {Torres}, {Charbonneau}, {Bryson},
  {Christiansen}, {Dressing}, {Jenkins}, {Walkowicz}, \&
  {Batalha}}]{2013ApJ...766...81F}
{Fressin}, F., {Torres}, G., {Charbonneau}, D., {et~al.} 2013, \apj, 766, 81

\bibitem[{{Fu} {et~al.}(2020){Fu}, {Deming}, {Lothringer}, {Nikolov}, {Sing},
  {Kempton}, {Ih}, {Evans}, {Stevenson}, {Wakeford}, {Rodriguez}, {Eastman},
  {Stassun}, {Henry}, {L{\'o}pez-Morales}, {Lendl}, {Conti}, {Stockdale},
  {Collins}, {Kielkopf}, {Barstow}, {Sanz-Forcada}, {Ehrenreich}, \&
  {Bourrier}}]{2020arXiv200502568F}
{Fu}, G., {Deming}, D., {Lothringer}, J., {et~al.} 2020, arXiv e-prints,
  arXiv:2005.02568

\bibitem[{{Fulton} {et~al.}(2017){Fulton}, {Petigura}, {Howard}, {Isaacson},
  {Marcy}, {Cargile}, {Hebb}, {Weiss}, {Johnson}, {Morton}, {Sinukoff},
  {Crossfield}, \& {Hirsch}}]{2017AJ....154..109F}
{Fulton}, B.~J., {Petigura}, E.~A., {Howard}, A.~W., {et~al.} 2017, \aj, 154,
  109

\bibitem[{{Gaia Collaboration} {et~al.}(2018){Gaia Collaboration}, {Brown},
  {Vallenari}, {Prusti}, {de Bruijne}, {Babusiaux}, {Bailer-Jones}, {Biermann},
  {Evans}, {Eyer}, \& et~al.}]{2018A&A...616A...1G}
{Gaia Collaboration}, {Brown}, A.~G.~A., {Vallenari}, A., {et~al.} 2018, \aap,
  616, A1

\bibitem[{{Gandolfi} {et~al.}(2017){Gandolfi}, {Barrag{\'a}n}, {Hatzes},
  {Fridlund}, {Fossati}, {Donati}, {Johnson}, {Nowak}, {Prieto-Arranz},
  {Albrecht}, {Dai}, {Deeg}, {Endl}, {Grziwa}, {Hjorth}, {Korth}, {Nespral},
  {Saario}, {Smith}, {Antoniciello}, {Alarcon}, {Bedell}, {Blay}, {Brems},
  {Cabrera}, {Csizmadia}, {Cusano}, {Cochran}, {Eigm{\"u}ller}, {Erikson},
  {Gonz{\'a}lez Hern{\'a}ndez}, {Guenther}, {Hirano}, {Su{\'a}rez
  Mascare{\~n}o}, {Narita}, {Palle}, {Parviainen}, {P{\"a}tzold}, {Persson},
  {Rauer}, {Saviane}, {Schmidtobreick}, {Van Eylen}, {Winn}, \&
  {Zakhozhay}}]{2017AJ....154..123G}
{Gandolfi}, D., {Barrag{\'a}n}, O., {Hatzes}, A.~P., {et~al.} 2017, \aj, 154,
  123

\bibitem[{{Gao} {et~al.}(2020){Gao}, {Thorngren}, {Lee}, {Fortney}, {Morley},
  {Wakeford}, {Powell}, {Stevenson}, \& {Zhang}}]{2020NatAs.tmp..142G}
{Gao}, P., {Thorngren}, D.~P., {Lee}, G. K.~H., {et~al.} 2020, Nature
  Astronomy, doi:10.1038/s41550-020-1114-3

\bibitem[{{Greene} {et~al.}(2016){Greene}, {Line}, {Montero}, {Fortney},
  {Lustig-Yaeger}, \& {Luther}}]{2016ApJ...817...17G}
{Greene}, T.~P., {Line}, M.~R., {Montero}, C., {et~al.} 2016, \apj, 817, 17

\bibitem[{{Guo} {et~al.}(2020){Guo}, {Crossfield}, {Dragomir}, {Kosiarek},
  {Lothringer}, {Mikal-Evans}, {Rosenthal}, {Benneke}, {Knutson}, {Dalba},
  {Kempton}, {Henry}, {McCullough}, {Barman}, {Blunt}, {Chontos}, {Fortney},
  {Fulton}, {Hirsch}, {Howard}, {Isaacson}, {Matthews}, {Mocnik}, {Morley},
  {Petigura}, \& {Weiss}}]{2020AJ....159..239G}
{Guo}, X., {Crossfield}, I. J.~M., {Dragomir}, D., {et~al.} 2020, \aj, 159, 239

\bibitem[{{Gupta} \& {Schlichting}(2019)}]{2019MNRAS.487...24G}
{Gupta}, A., \& {Schlichting}, H.~E. 2019, \mnras, 487, 24

\bibitem[{{Harps{\o}e} {et~al.}(2013){Harps{\o}e}, {Hardis}, {Hinse},
  {J{\o}rgensen}, {Mancini}, {Southworth}, {Alsubai}, {Bozza}, {Browne},
  {Burgdorf}, {Calchi Novati}, {Dodds}, {Dominik}, {Fang}, {Finet}, {Gerner},
  {Gu}, {Hundertmark}, {Jessen-Hansen}, {Kains}, {Kerins}, {Kjeldsen},
  {Liebig}, {Lund}, {Lundkvist}, {Mathiasen}, {Nesvorn{\'y}}, {Nikolov},
  {Penny}, {Proft}, {Rahvar}, {Ricci}, {Sahu}, {Scarpetta}, {Sch{\"a}fer},
  {Sch{\"o}nebeck}, {Snodgrass}, {Skottfelt}, {Surdej}, {Tregloan-Reed}, \&
  {Wertz}}]{2013A&A...549A..10H}
{Harps{\o}e}, K.~B.~W., {Hardis}, S., {Hinse}, T.~C., {et~al.} 2013, \aap, 549,
  A10

\bibitem[{{Horne}(1986)}]{1986PASP...98..609H}
{Horne}, K. 1986, \pasp, 98, 609

\bibitem[{{Howard} {et~al.}(2010){Howard}, {Marcy}, {Johnson}, {Fischer},
  {Wright}, {Isaacson}, {Valenti}, {Anderson}, {Lin}, \&
  {Ida}}]{2010Sci...330..653H}
{Howard}, A.~W., {Marcy}, G.~W., {Johnson}, J.~A., {et~al.} 2010, Science, 330,
  653

\bibitem[{{Howe} {et~al.}(2014){Howe}, {Burrows}, \&
  {Verne}}]{2014ApJ...787..173H}
{Howe}, A.~R., {Burrows}, A., \& {Verne}, W. 2014, \apj, 787, 173

\bibitem[{{Howell} {et~al.}(2014){Howell}, {Sobeck}, {Haas}, {Still},
  {Barclay}, {Mullally}, {Troeltzsch}, {Aigrain}, {Bryson}, {Caldwell},
  {Chaplin}, {Cochran}, {Huber}, {Marcy}, {Miglio}, {Najita}, {Smith},
  {Twicken}, \& {Fortney}}]{2014PASP..126..398H}
{Howell}, S.~B., {Sobeck}, C., {Haas}, M., {et~al.} 2014, \pasp, 126, 398

\bibitem[{{Hunter}(2007)}]{2007CSE.....9...90H}
{Hunter}, J.~D. 2007, Computing in Science and Engineering, 9, 90

\bibitem[{{Jin} \& {Mordasini}(2018)}]{2018ApJ...853..163J}
{Jin}, S., \& {Mordasini}, C. 2018, \apj, 853, 163

\bibitem[{{Kawashima} \& {Ikoma}(2018)}]{2018ApJ...853....7K}
{Kawashima}, Y., \& {Ikoma}, M. 2018, \apj, 853, 7

\bibitem[{{Kawashima} \& {Ikoma}(2019)}]{2019ApJ...877..109K}
---. 2019, \apj, 877, 109

\bibitem[{{Knutson} {et~al.}(2014{\natexlab{a}}){Knutson}, {Benneke}, {Deming},
  \& {Homeier}}]{2014Natur.505...66K}
{Knutson}, H.~A., {Benneke}, B., {Deming}, D., \& {Homeier}, D.
  2014{\natexlab{a}}, \nat, 505, 66

\bibitem[{{Knutson} {et~al.}(2014{\natexlab{b}}){Knutson}, {Dragomir},
  {Kreidberg}, {Kempton}, {McCullough}, {Fortney}, {Bean}, {Gillon}, {Homeier},
  \& {Howard}}]{2014ApJ...794..155K}
{Knutson}, H.~A., {Dragomir}, D., {Kreidberg}, L., {et~al.} 2014{\natexlab{b}},
  \apj, 794, 155

\bibitem[{{Komacek} {et~al.}(2019){Komacek}, {Showman}, \&
  {Parmentier}}]{2019ApJ...881..152K}
{Komacek}, T.~D., {Showman}, A.~P., \& {Parmentier}, V. 2019, \apj, 881, 152

\bibitem[{{Kreidberg}(2015)}]{2015PASP..127.1161K}
{Kreidberg}, L. 2015, \pasp, 127, 1161

\bibitem[{{Kreidberg} {et~al.}(2014){Kreidberg}, {Bean}, {D{\'e}sert},
  {Benneke}, {Deming}, {Stevenson}, {Seager}, {Berta-Thompson}, {Seifahrt}, \&
  {Homeier}}]{2014Natur.505...69K}
{Kreidberg}, L., {Bean}, J.~L., {D{\'e}sert}, J.-M., {et~al.} 2014, \nat, 505,
  69

\bibitem[{{Kreidberg} {et~al.}(2020){Kreidberg}, {Molli{\`e}re}, {Crossfield},
  {Thorngren}, {Kawashima}, {Morley}, {Benneke}, {Mikal-Evans}, {Berardo},
  {Kosiarek}, {Gorjian}, {Ciardi}, {Christiansen}, {Dragomir}, {Dressing},
  {Fortney}, {Fulton}, {Greene}, {Hardegree-Ullman}, {Howard}, {Howell},
  {Isaacson}, {Krick}, {Livingston}, {Lothringer}, {Morales}, {Petigura},
  {Rodriguez}, {Schlieder}, \& {Weiss}}]{2020arXiv200607444K}
{Kreidberg}, L., {Molli{\`e}re}, P., {Crossfield}, I. J.~M., {et~al.} 2020,
  arXiv e-prints, arXiv:2006.07444

\bibitem[{{Kubyshkina} {et~al.}(2019){Kubyshkina}, {Cubillos}, {Fossati},
  {Erkaev}, {Johnstone}, {Kislyakova}, {Lammer}, {Lendl}, {Odert}, \&
  {G{\"u}del}}]{2019ApJ...879...26K}
{Kubyshkina}, D., {Cubillos}, P.~E., {Fossati}, L., {et~al.} 2019, \apj, 879,
  26

\bibitem[{{Kurucz}(1993)}]{1993KurCD..13.....K}
{Kurucz}, R. 1993, ATLAS9 Stellar Atmosphere Programs and 2 km/s grid.~Kurucz
  CD-ROM No.~13.~ Cambridge, Mass.: Smithsonian Astrophysical Observatory

\bibitem[{{Libby-Roberts} {et~al.}(2020){Libby-Roberts}, {Berta-Thompson},
  {D{\'e}sert}, {Masuda}, {Morley}, {Lopez}, {Deck}, {Fabrycky}, {Fortney},
  {Line}, {Sanchis-Ojeda}, \& {Winn}}]{2020AJ....159...57L}
{Libby-Roberts}, J.~E., {Berta-Thompson}, Z.~K., {D{\'e}sert}, J.-M., {et~al.}
  2020, \aj, 159, 57

\bibitem[{{Ligi} {et~al.}(2018){Ligi}, {Demangeon}, {Barros}, {Mesa},
  {Bonavita}, {Vigan}, {Bonnefoy}, {Gratton}, \&
  {Deleuil}}]{2018AJ....156..182L}
{Ligi}, R., {Demangeon}, O., {Barros}, S., {et~al.} 2018, \aj, 156, 182

\bibitem[{{Lopez} \& {Fortney}(2013)}]{2013ApJ...776....2L}
{Lopez}, E.~D., \& {Fortney}, J.~J. 2013, \apj, 776, 2

\bibitem[{{Lopez} \& {Fortney}(2014)}]{2014ApJ...792....1L}
---. 2014, \apj, 792, 1

\bibitem[{{Lothringer} {et~al.}(2018){Lothringer}, {Benneke}, {Crossfield},
  {Henry}, {Morley}, {Dragomir}, {Barman}, {Knutson}, {Kempton}, {Fortney},
  {McCullough}, \& {Howard}}]{2018AJ....155...66L}
{Lothringer}, J.~D., {Benneke}, B., {Crossfield}, I. J.~M., {et~al.} 2018, \aj,
  155, 66

\bibitem[{{Lyon} \& {Johnson}(1992)}]{ironEOS}
{Lyon}, S.~P., \& {Johnson}, J.~D. 1992, {SESAME: The Los Alamos National
  Laboratory Equation of State Database}, Tech. rep.

\bibitem[{{Masuda}(2014)}]{2014ApJ...783...53M}
{Masuda}, K. 2014, \apj, 783, 53

\bibitem[{{Mazevet} {et~al.}(2019){Mazevet}, {Licari}, {Chabrier}, \&
  {Potekhin}}]{2019A&A...621A.128M}
{Mazevet}, S., {Licari}, A., {Chabrier}, G., \& {Potekhin}, A.~Y. 2019, \aap,
  621, A128

\bibitem[{{Mikal-Evans} {et~al.}(2020){Mikal-Evans}, {Sing}, {Kataria},
  {Wakeford}, {Mayne}, {Lewis}, {Barstow}, \& {Spake}}]{2020MNRAS.496.1638M}
{Mikal-Evans}, T., {Sing}, D.~K., {Kataria}, T., {et~al.} 2020, \mnras, 496,
  1638

\bibitem[{{Mikal-Evans} {et~al.}(2019){Mikal-Evans}, {Sing}, {Goyal},
  {Drummond}, {Carter}, {Henry}, {Wakeford}, {Lewis}, {Marley}, {Tremblin},
  {Nikolov}, {Kataria}, {Deming}, \& {Ballester}}]{2019MNRAS.488.2222M}
{Mikal-Evans}, T., {Sing}, D.~K., {Goyal}, J.~M., {et~al.} 2019, \mnras, 488,
  2222

\bibitem[{{Molli{\`e}re} {et~al.}(2019){Molli{\`e}re}, {Wardenier}, {van
  Boekel}, {Henning}, {Molaverdikhani}, \& {Snellen}}]{2019A&A...627A..67M}
{Molli{\`e}re}, P., {Wardenier}, J.~P., {van Boekel}, R., {et~al.} 2019, \aap,
  627, A67

\bibitem[{{Morley} {et~al.}(2012){Morley}, {Fortney}, {Marley}, {Visscher},
  {Saumon}, \& {Leggett}}]{2012ApJ...756..172M}
{Morley}, C.~V., {Fortney}, J.~J., {Marley}, M.~S., {et~al.} 2012, \apj, 756,
  172

\bibitem[{{Morley} {et~al.}(2015){Morley}, {Fortney}, {Marley}, {Zahnle},
  {Line}, {Kempton}, {Lewis}, \& {Cahoy}}]{2015ApJ...815..110M}
---. 2015, \apj, 815, 110

\bibitem[{{Morley} {et~al.}(2017){Morley}, {Knutson}, {Line}, {Fortney},
  {Thorngren}, {Marley}, {Teal}, \& {Lupu}}]{2017AJ....153...86M}
{Morley}, C.~V., {Knutson}, H., {Line}, M., {et~al.} 2017, \aj, 153, 86

\bibitem[{{Moses}(2014)}]{2014RSPTA.37230073M}
{Moses}, J.~I. 2014, Philosophical Transactions of the Royal Society of London
  Series A, 372, 20130073

\bibitem[{{Moses} {et~al.}(2013){Moses}, {Line}, {Visscher}, {Richardson},
  {Nettelmann}, {Fortney}, {Barman}, {Stevenson}, \&
  {Madhusudhan}}]{2013ApJ...777...34M}
{Moses}, J.~I., {Line}, M.~R., {Visscher}, C., {et~al.} 2013, \apj, 777, 34

\bibitem[{{Mousis} {et~al.}(2020){Mousis}, {Deleuil}, {Aguichine}, {Marcq},
  {Naar}, {Aguirre}, {Brugger}, \& {Gon{\c{c}}alves}}]{2020ApJ...896L..22M}
{Mousis}, O., {Deleuil}, M., {Aguichine}, A., {et~al.} 2020, \apjl, 896, L22

\bibitem[{{Mulders} {et~al.}(2019){Mulders}, {Mordasini}, {Pascucci}, {Ciesla},
  {Emsenhuber}, \& {Apai}}]{2019ApJ...887..157M}
{Mulders}, G.~D., {Mordasini}, C., {Pascucci}, I., {et~al.} 2019, \apj, 887,
  157

\bibitem[{{Nettelmann} {et~al.}(2011){Nettelmann}, {Fortney}, {Kramm}, \&
  {Redmer}}]{2011ApJ...733....2N}
{Nettelmann}, N., {Fortney}, J.~J., {Kramm}, U., \& {Redmer}, R. 2011, \apj,
  733, 2

\bibitem[{{Nissen} {et~al.}(2020){Nissen}, {Christensen-Dalsgaard},
  {Mosumgaard}, {Silva Aguirre}, {Spitoni}, \& {Verma.}}]{nissen:2020}
{Nissen}, P.~E., {Christensen-Dalsgaard}, J., {Mosumgaard}, J.~R., {et~al.}
  2020, arXiv e-prints, arXiv:2006.06013

\bibitem[{{{\"O}berg} {et~al.}(2011){{\"O}berg}, {Murray-Clay}, \&
  {Bergin}}]{2011ApJ...743L..16O}
{{\"O}berg}, K.~I., {Murray-Clay}, R., \& {Bergin}, E.~A. 2011, \apjl, 743, L16

\bibitem[{{Owen} \& {Wu}(2013)}]{2013ApJ...775..105O}
{Owen}, J.~E., \& {Wu}, Y. 2013, \apj, 775, 105

\bibitem[{{Owen} \& {Wu}(2017)}]{2017ApJ...847...29O}
---. 2017, \apj, 847, 29

\bibitem[{{Parmentier} {et~al.}(2013){Parmentier}, {Showman}, \&
  {Lian}}]{2013A&A...558A..91P}
{Parmentier}, V., {Showman}, A.~P., \& {Lian}, Y. 2013, \aap, 558, A91

\bibitem[{{Petigura} {et~al.}(2013){Petigura}, {Marcy}, \&
  {Howard}}]{2013ApJ...770...69P}
{Petigura}, E.~A., {Marcy}, G.~W., \& {Howard}, A.~W. 2013, \apj, 770, 69

\bibitem[{{Rogers}(2015)}]{2015ApJ...801...41R}
{Rogers}, L.~A. 2015, \apj, 801, 41

\bibitem[{{Rogers} \& {Seager}(2010{\natexlab{a}})}]{2010ApJ...712..974R}
{Rogers}, L.~A., \& {Seager}, S. 2010{\natexlab{a}}, \apj, 712, 974

\bibitem[{{Rogers} \& {Seager}(2010{\natexlab{b}})}]{2010ApJ...716.1208R}
---. 2010{\natexlab{b}}, \apj, 716, 1208

\bibitem[{{Sing} {et~al.}(2016){Sing}, {Fortney}, {Nikolov}, {Wakeford},
  {Kataria}, {Evans}, {Aigrain}, {Ballester}, {Burrows}, {Deming},
  {D{\'e}sert}, {Gibson}, {Henry}, {Huitson}, {Knutson}, {Etangs}, {Pont},
  {Showman}, {Vidal-Madjar}, {Williamson}, \& {Wilson}}]{2016Natur.529...59S}
{Sing}, D.~K., {Fortney}, J.~J., {Nikolov}, N., {et~al.} 2016, \nat, 529, 59

\bibitem[{{Sing} {et~al.}(2019){Sing}, {Lavvas}, {Ballester}, {Lecavelier des
  Etangs}, {Marley}, {Nikolov}, {Ben-Jaffel}, {Bourrier}, {Buchhave}, {Deming},
  {Ehrenreich}, {Mikal-Evans}, {Kataria}, {Lewis}, {L{\'o}pez-Morales},
  {Garc{\'\i}a Mu{\~n}oz}, {Henry}, {Sanz-Forcada}, {Spake}, {Wakeford}, \&
  {PanCET Collaboration}}]{2019AJ....158...91S}
{Sing}, D.~K., {Lavvas}, P., {Ballester}, G.~E., {et~al.} 2019, \aj, 158, 91

\bibitem[{{Skilling}(2004)}]{2004AIPC..735..395S}
{Skilling}, J. 2004, in American Institute of Physics Conference Series, Vol.
  735, American Institute of Physics Conference Series, ed. R.~{Fischer},
  R.~{Preuss}, \& U.~V. {Toussaint}, 395--405

\bibitem[{{Sotzen} {et~al.}(2020){Sotzen}, {Stevenson}, {Sing}, {Kilpatrick},
  {Wakeford}, {Filippazzo}, {Lewis}, {H{\"o}rst}, {L{\'o}pez-Morales}, {Henry},
  {Buchhave}, {Ehrenreich}, {Fraine}, {Garc{\'\i}a Mu{\~n}oz}, {Jayaraman},
  {Lavvas}, {Lecavelier des Etangs}, {Marley}, {Nikolov}, {Rathcke}, \&
  {Sanz-Forcada}}]{2020AJ....159....5S}
{Sotzen}, K.~S., {Stevenson}, K.~B., {Sing}, D.~K., {et~al.} 2020, \aj, 159, 5

\bibitem[{{Spake} {et~al.}(2019){Spake}, {Sing}, {Wakeford}, {Nikolov},
  {Mikal-Evans}, {Deming}, {Barstow}, {Anderson}, {Carter}, {Gillon}, {Goyal},
  {Hebrard}, {Hellier}, {Kataria}, {Lam}, \& {Triaud}}]{2019arXiv191108859S}
{Spake}, J.~J., {Sing}, D.~K., {Wakeford}, H.~R., {et~al.} 2019, arXiv
  e-prints, arXiv:1911.08859

\bibitem[{{Stevenson} \& {Fowler}(2019)}]{2019wfc..rept...12S}
{Stevenson}, K.~B., \& {Fowler}, J. 2019, {Analyzing Eight Years of Transiting
  Exoplanet Observations Using WFC3's Spatial Scan Monitor}, Tech. rep.

\bibitem[{{STScI Development Team}(2013)}]{2013ascl.soft03023S}
{STScI Development Team}. 2013, {pysynphot: Synthetic photometry software
  package}, ascl:1303.023

\bibitem[{Thompson(1990)}]{osti_6939284}
Thompson, S.~L. 1990, doi:10.2172/6939284

\bibitem[{{Thorngren} \& {Fortney}(2019)}]{2019ApJ...874L..31T}
{Thorngren}, D., \& {Fortney}, J.~J. 2019, \apjl, 874, L31

\bibitem[{{Thorngren} {et~al.}(2016){Thorngren}, {Fortney}, {Murray-Clay}, \&
  {Lopez}}]{2016ApJ...831...64T}
{Thorngren}, D.~P., {Fortney}, J.~J., {Murray-Clay}, R.~A., \& {Lopez}, E.~D.
  2016, \apj, 831, 64

\bibitem[{{Trotta}(2008)}]{2008ConPh..49...71T}
{Trotta}, R. 2008, Contemporary Physics, 49, 71

\bibitem[{{Tsiaras} {et~al.}(2016){Tsiaras}, {Rocchetto}, {Waldmann}, {Venot},
  {Varley}, {Morello}, {Damiano}, {Tinetti}, {Barton}, {Yurchenko}, \&
  {Tennyson}}]{2016ApJ...820...99T}
{Tsiaras}, A., {Rocchetto}, M., {Waldmann}, I.~P., {et~al.} 2016, \apj, 820, 99

\bibitem[{{Valencia} {et~al.}(2013){Valencia}, {Guillot}, {Parmentier}, \&
  {Freedman}}]{2013ApJ...775...10V}
{Valencia}, D., {Guillot}, T., {Parmentier}, V., \& {Freedman}, R.~S. 2013,
  \apj, 775, 10

\bibitem[{{Valencia} {et~al.}(2007){Valencia}, {Sasselov}, \&
  {O'Connell}}]{2007ApJ...665.1413V}
{Valencia}, D., {Sasselov}, D.~D., \& {O'Connell}, R.~J. 2007, \apj, 665, 1413

\bibitem[{{van der Walt} {et~al.}(2011){van der Walt}, {Colbert}, \&
  {Varoquaux}}]{numpy2011}
{van der Walt}, S., {Colbert}, S.~C., \& {Varoquaux}, G. 2011, Computing in
  Science Engineering, 13, 22

\bibitem[{{Van Eylen} {et~al.}(2018){Van Eylen}, {Agentoft}, {Lundkvist},
  {Kjeldsen}, {Owen}, {Fulton}, {Petigura}, \& {Snellen}}]{2018MNRAS.479.4786V}
{Van Eylen}, V., {Agentoft}, C., {Lundkvist}, M.~S., {et~al.} 2018, \mnras,
  479, 4786

\bibitem[{{Vanderburg} \& {Johnson}(2014)}]{2014PASP..126..948V}
{Vanderburg}, A., \& {Johnson}, J.~A. 2014, \pasp, 126, 948

\bibitem[{{Vanderburg} {et~al.}(2016){Vanderburg}, {Bieryla}, {Duev},
  {Jensen-Clem}, {Latham}, {Mayo}, {Baranec}, {Berlind}, {Kulkarni}, {Law},
  {Nieberding}, {Riddle}, \& {Salama}}]{2016ApJ...829L...9V}
{Vanderburg}, A., {Bieryla}, A., {Duev}, D.~A., {et~al.} 2016, \apjl, 829, L9

\bibitem[{{Virtanen} {et~al.}(2020){Virtanen}, {Gommers}, {Oliphant},
  {Haberland}, {Reddy}, {Cournapeau}, {Burovski}, {Peterson}, {Weckesser},
  {Bright}, {van der Walt}, {Brett}, {Wilson}, {Jarrod Millman}, {Mayorov},
  {Nelson}, {Jones}, {Kern}, {Larson}, {Carey}, {Polat}, {Feng}, {Moore}, {Vand
  erPlas}, {Laxalde}, {Perktold}, {Cimrman}, {Henriksen}, {Quintero}, {Harris},
  {Archibald}, {Ribeiro}, {Pedregosa}, {van Mulbregt}, \&
  {Contributors}}]{2020SciPy-NMeth}
{Virtanen}, P., {Gommers}, R., {Oliphant}, T.~E., {et~al.} 2020, Nature
  Methods, 17, 261

\bibitem[{{Vogt} {et~al.}(1994){Vogt}, {Allen}, {Bigelow}, {Bresee}, {Brown},
  {Cantrall}, {Conrad}, {Couture}, {Delaney}, {Epps}, {Hilyard}, {Hilyard},
  {Horn}, {Jern}, {Kanto}, {Keane}, {Kibrick}, {Lewis}, {Osborne},
  {Pardeilhan}, {Pfister}, {Ricketts}, {Robinson}, {Stover}, {Tucker}, {Ward},
  \& {Wei}}]{vogt:1994}
{Vogt}, S.~S., {Allen}, S.~L., {Bigelow}, B.~C., {et~al.} 1994, in \procspie,
  Vol. 2198, Instrumentation in Astronomy VIII, ed. D.~L. {Crawford} \& E.~R.
  {Craine}, 362

\bibitem[{{Wakeford} {et~al.}(2020){Wakeford}, {Sing}, {Stevenson}, {Lewis},
  {Pirzkal}, {Wilson}, {Goyal}, {Kataria}, {Mikal-Evans}, {Nikolov}, \&
  {Spake}}]{2020AJ....159..204W}
{Wakeford}, H.~R., {Sing}, D.~K., {Stevenson}, K.~B., {et~al.} 2020, \aj, 159,
  204

\bibitem[{{Weiss} \& {Marcy}(2014)}]{2014ApJ...783L...6W}
{Weiss}, L.~M., \& {Marcy}, G.~W. 2014, \apjl, 783, L6

\bibitem[{{Werner} {et~al.}(2016){Werner}, {Crossfield}, {Akeson}, {Beichman},
  {Benneke}, {Christiansen}, {Ciardi}, {Deck}, {Dressing}, {Howard}, {Howell},
  {Knutson}, {Krick}, {Livingston}, {Morales}, {Petigura}, {Schlieder}, \&
  {Gorjian}}]{2016sptz.prop13052W}
{Werner}, M., {Crossfield}, I., {Akeson}, R., {et~al.} 2016, {Spitzer v. K2:
  Part II}, Spitzer Proposal

\bibitem[{{Wong} {et~al.}(2020){Wong}, {Benneke}, {Gao}, {Knutson}, {Chachan},
  {Henry}, {Deming}, {Kataria}, {Lee}, {Nikolov}, {Sing}, {Ballester},
  {Baskin}, {Wakeford}, \& {Williamson}}]{2020AJ....159..234W}
{Wong}, I., {Benneke}, B., {Gao}, P., {et~al.} 2020, \aj, 159, 234

\bibitem[{{Woods} \& {Rottman}(2002)}]{2002GMS...130..221W}
{Woods}, T.~N., \& {Rottman}, G.~J. 2002, Washington DC American Geophysical
  Union Geophysical Monograph Series, 130, 221

\bibitem[{{Zahnle} \& {Catling}(2017)}]{2017ApJ...843..122Z}
{Zahnle}, K.~J., \& {Catling}, D.~C. 2017, \apj, 843, 122

\bibitem[{{Zeng} {et~al.}(2019){Zeng}, {Jacobsen}, {Sasselov}, {Petaev},
  {Vanderburg}, {Lopez-Morales}, {Perez-Mercader}, {Mattsson}, {Li}, {Heising},
  {Bonomo}, {Damasso}, {Berger}, {Cao}, {Levi}, \&
  {Wordsworth}}]{2019PNAS..116.9723Z}
{Zeng}, L., {Jacobsen}, S.~B., {Sasselov}, D.~D., {et~al.} 2019, Proceedings of
  the National Academy of Science, 116, 9723

\end{thebibliography}

\acknowledgements 

The authors are grateful to the referee for constructive feedback that improved the quality of the manuscript. Support for HST program GO-15333 was provided by NASA through a grant from the Space Telescope Science Institute, which is operated by the Association of Universities for Research in Astronomy, Inc., under NASA contract NAS 5-26555. This work is based [in part] on observations made with the Spitzer Space Telescope, which was operated by the Jet Propulsion Laboratory, California Institute of Technology under a contract with NASA. This research has made use of the SIMBAD database, operated at CDS, Strasbourg, France. Some of the data presented in this paper were obtained from the Mikulski Archive for Space Telescopes (MAST). STScI is operated by the Association of Universities for Research in Astronomy, Inc., under NASA contract NAS5-26555. Support for MAST for non-HST data is provided by the NASA Office of Space Science via grant NNX13AC07G and by other grants and contracts. P.M. acknowledges support from the European Research Council under the European Union's Horizon 2020 research and innovation program under grant agreement No. 832428. D. D. acknowledges support from the TESS Guest Investigator Program grant 80NSSC19K1727 and NASA Exoplanet Research Program grant 18-2XRP18 2-0136. M.R.K is supported by the NSF Graduate Research Fellowship, grant No. DGE 1339067. 

\software{NumPy \citep{numpy2011}, SciPy \citep{2020SciPy-NMeth}, Matplotlib \citep{2007CSE.....9...90H}, emcee \citep{2013PASP..125..306F}, batman \citep{2015PASP..127.1161K}, Astropy \citep{2013A&A...558A..33A,2018AJ....156..123A}, pysynphot \citep{2013ascl.soft03023S}, petitRadtrans \citep{2019A&A...627A..67M}, PyMultinest \citep{2014A&A...564A.125B}}

\facilities{ HST(WFC3), Spitzer(IRAC), Kepler(K2), Keck(HIRES) }

\appendix

\section{Data reduction: HST WFC3} \label{app:hst:datared}

\begin{figure}
\centering  
\includegraphics[width=0.4\columnwidth]{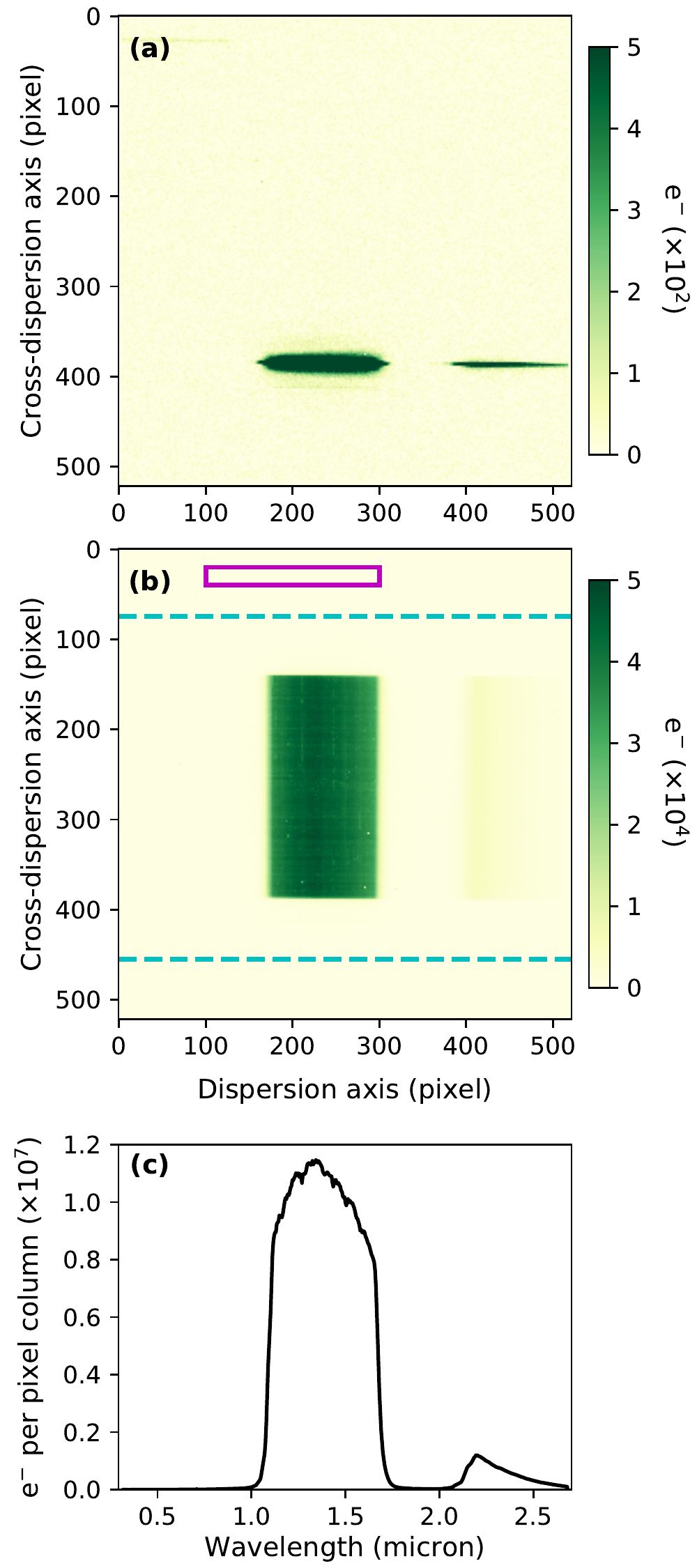}
\caption{(a) First non-destructive read for an example IMA frame. Colorscale has been stretched to highlight the lack of neighboring spectra overlapping the target spectrum. The first-order spectrum of HD~3167 spans dispersion columns 170-300 and the fainter second-order spectrum spans columns 390-520. (b) Final non-destructive read of the same IMA frame. Dashed cyan lines indicate the photometric aperture, which was summed along the cross-dispersion axis. Background counts were obtained by taking the median within the purple box. (c) Corresponding 1D spectrum as integrated electron counts versus wavelength.}
\label{fig:dframe}
\end{figure}

We used the IMA files produced by the \textit{calwf3} pipeline version 3.5.0, which already have basic calibrations such as flat fielding and bias subtraction applied. The target flux was extracted from each exposure by taking the difference between successive non-destructive reads. To do this, we first estimated and subtracted the background flux for each read, by taking the median pixel count within the box defined by cross-dispersion rows 20-40 and dispersion columns 100-300 (Figure \ref{fig:dframe}). Typical background levels per pixel integrated over the full $70\,$s exposures were: $\sim 430\,\textnormal{e}^{-}$ for G141v1; $\sim 140\,\textnormal{e}^{-}$ for G141v2; $\sim 230\,\textnormal{e}^{-}$ for G141v3; $\sim 130\,\textnormal{e}^{-}$ for G141v4; and $170\,\textnormal{e}^{-}$ for G141v5 (Figure \ref{fig:auxvars}). We also experimented with background estimates obtained using different boxes located away from the target spectrum, but found the count levels to be similar with no appreciable effect on the final results.

For each read-difference frame, we then determined the flux-weighted center of the scanned spectrum along the cross-dispersion axis. All pixel values located more than 130 pixels above and below this row were set to zero, effectively removing flux contributions from nearby contaminant stars and cosmic ray strikes outside a rectangular aperture. We note, however, that visual inspection of the data frames did not suggest any apparent contamination of the target scans by nearby stars (Figure \ref{fig:dframe}). This is supported by ground-based direct imaging observations, which place deep limits on the minimum contrast of any nearby stars within $3$\,arcsec of HD~3167 \citep{2016ApJ...829L...9V,2017AJ....154..122C,2018AJ....156..182L}\footnote{e.g.\ see Figure 1 of \cite{2018AJ....156..182L}.}, and visual inspection of the HST acquisition images for wider separations. Final reconstructed frames were produced by adding together the read-differences produced in this manner. During this process, we also estimated how the spectrum drifted across the detector over the course of the observations. For all visits, we found the scanned spectrum drifted by no more than $\sim 0.1$\,pixel along the dispersion axis and $\sim 0.2$\,pixel along the cross-dispersion axis (Figure \ref{fig:auxvars}).

The target spectrum was then extracted from each frame by summing the flux within a rectangular aperture spanning the full dispersion axis and 380 pixels along the cross-dispersion axis, centered on the central cross-dispersion row of the scan (Figure \ref{fig:dframe}). Before settling on this aperture size, we experimented with other apertures ranging from 300-400 pixels and determined this choice had a negligible effect on our final results. The wavelength solution was determined by cross-correlating each of the spectra extracted with the 380 pixel aperture against a model stellar spectrum modulated by the throughput of the G141 grism, as in \cite{2016ApJ...822L...4E,2017Natur.548...58E}. For the stellar spectrum, we used the \texttt{pysynphot} Python package \citep{2013ascl.soft03023S} to interpolate the \cite{1993KurCD..13.....K} model grid for properties appropriate to the HD~3167 host star. 

\section{Fitting the HST WFC3 broadband light curve} \label{app:hst:broad}

Following standard practice, we discarded the entire first HST orbit of each visit and the first round-trip scan of each subsequent orbit from our analysis, as these are affected by especially strong systematics. We then fit all light curves simultaneously, defining a log-likelihood function of the form:
\begin{align}
\ln p & = \sum_{i}^{N}{\ln p_i}
\end{align}
where $p_i$ is the log-likelihood of the $i$th dataset. For the latter, we assumed a Gaussian process (GP) log-likelihood of the form:
\begin{align}
\ln p_i & \sim  \ln \mathcal{N}( \boldsymbol{\mu_i}, \boldsymbol{ \Sigma_i} )
\end{align}
where $\mathcal{N}$ is a Gaussian distribution with deterministic mean function $\boldsymbol{\mu_i}$ and covariance matrix $\boldsymbol{ \Sigma_i}$. For $\boldsymbol{\mu_i}$, we used:
\begin{align}
  \boldsymbol{\mu_i} = \begin{cases}
    f_i \, M(\,t\,;\,\boldsymbol{\alpha_i}\,) \, R(\,t\, , \, \phi \,;\, \boldsymbol{ \gamma_i }\, ) & \text{for forward scans} \\
    b_i \, M(\,t\,;\,\boldsymbol{\alpha_i}\,) \, R( \,t\, , \, \phi \,;\, \boldsymbol{ \gamma_i }\, ) & \text{for backward scans}    
    \end{cases}
\end{align}
where $t$ is time, $\phi$ is HST orbital phase, $f_i$ and $b_i$ are normalization constants for the two spatial scan directions, $M$ is a transit function, which we implemented using the \texttt{batman} Python package \citep{2015PASP..127.1161K}, and $R$ is an analytic model for the WFC3 detector systematics, described below.

The transit function was parameterized by $\boldsymbol{\alpha_i}=[ \, \RpRs, \, T_i \, ]$, with the planet-to-star radius ratio ($\RpRs$) shared across all five datasets while the transit mid-times ($T_i$) were allowed to vary separately for each dataset. We experimented with allowing the stellar limb darkening coefficients to vary as free parameters and holding them fixed to values determined from stellar models. We found this choice did not affect the final results, and so adopted a quadratic limb darkening law with coefficients $(u_1,u_2)$ fixed to values computed using the online ExoCTK tool.\footnote{https://exoctk.stsci.edu} As noted in Section \ref{sec:hstbroad}, we also fixed the normalized semimajor axis ($\aRs$) and orbital impact parameter ($b$) to the values listed in Table \ref{table:broadfit}, obtained from a global fit to all available \textit{K2} and \textit{Spitzer} data for HD~3167b and HD~3167c (Hardegree-Ullman et al., in prep). We note that these values are within the $1\sigma$ credible ranges reported by both \cite{2017AJ....154..122C} and \cite{2017AJ....154..123G}. The planetary orbital period was fixed to $P=29.84622$ day, as reported by \cite{2017AJ....154..123G}. We assumed a circular orbit, given the lack of compelling evidence for a nonzero eccentricity \citep{2017AJ....154..122C,2017AJ....154..123G} and the minimal impact it has on the shape of the transit at the level of data precision.

For the deterministic component of the detector systematics, we used an analytic ramp model almost identical to that introduced by \cite{2018NatAs...2..214D}, given by:
\begin{align}
  R(\,t\, , \, \phi \, ;\, \boldsymbol{ \gamma }\,) &= r_v(\,t\,;\,\gamma_1,\,\gamma_2\,)\,r_o(\, \phi \, ;\,\gamma_3,\,\gamma_4,\,\gamma_5\,) \label{eq:ramp}
\end{align}
where:
\begin{align}
  r_v(\,t\,;\,\gamma_1,\,\gamma_2\,) &= 1 + \gamma_{1}\exp\left[ -\gamma_{2}t \right]
\end{align}
and
\begin{align}
  r_o(\, \phi \, ;\,\gamma_3,\,\gamma_4,\,\gamma_5\,) &= 1 + \gamma_{3}\exp\left[ -\left( \frac{\phi-\gamma_{5}}{\gamma_{4}r_v} \right) \right] \ .
\end{align}

We also used the covariance matrix $\boldsymbol{ \Sigma_i}$ to encode additional systematics and noise properties of the data. The $jk$th entry was defined by:
\begin{align}
\boldsymbol{ \Sigma_i}_{jk} & = K_{\text{SE}}( \, t_j, \, t_k \, ; \, A_{t,i}, \, \eta_{t,i} \, ) + K_{\text{SE}}( \, \phi_j, \, \phi_k \, ; \, A_{\phi,i}, \, \eta_{\phi,i} \, ) + \delta_{jk} \beta_i^2\sigma_i^2 \ , \label{eq:broadKijk}
\end{align}
where $\delta_{jk}$ is the Kronecker delta, $\sigma_i$ is the photon noise value, and $K_{\text{SE}}$ denotes a squared-exponential covariance kernel of the form:
\begin{align}
K_{\text{SE}}( \, \epsilon_j, \, \epsilon_k \, ; \, A, \, \eta \, ) & = A^2\,\exp\left[ -\frac{1}{2} \eta^2 ( \epsilon_j -  \epsilon_k )^2 \right] \ .
\end{align}
where $A$ is the covariance amplitude and $\eta$ is the inverse correlation length scale. In Equation \ref{eq:broadKijk}, the first $K_{\text{SE}}$ kernel was employed to capture the $t$-dependent baseline trend for each visit. The second $K_{\text{SE}}$ kernel was included to account for any systematics repeating from orbit-to-orbit not captured by the analytic ramp model given by Equation \ref{eq:ramp}. Finally, the $\beta_i$ term allows for the uncorrelated noise value to be higher than the formal photon noise $\sigma_i$, which is useful if additional high-frequency noise sources are present in the data that can be approximated as white.

We note in particular that the $t$-dependent $K_{\text{SE}}$ kernel used for the baseline trend is significantly more flexible than the low-order polynomial trends that are typically adopted for WFC3 light curve analyses \citep[e.g.][]{2014Natur.513..526F,2014Natur.505...66K,2016ApJ...822L...4E,2017Natur.548...58E,2019NatAs...3..813B,2019ApJ...887L..14B,2014Natur.505...69K,2020arXiv200607444K}. This was motivated by a visual inspection of the visit-long trends in the broadband light curves, which do not appear to be well approximated by a low-order $t$-dependent polynomial. This is unsurprising given the long duration of the transit, resulting in individual visit durations of $\sim 8.5$ hours after discarding the first HST orbit. A similar effect was noted by \cite{2020AJ....159..239G} and \cite{2020arXiv200505153C} for transit observations of HD~97658b and KELT-11b, respectively. Those authors demonstrated how the broadband $\RpRs$ level varied significantly when different analytic expressions were assumed for the baseline trend, including linear, quadratic, logarithmic, and exponential functions of time. Our approach of using a flexible $t$-dependent GP for the baseline trend attempts to address this by marginalizing across a broad function space.

For the $A_{i,\phi}$ parameters, we adopted Gamma priors of the form $p(A_{i,\phi}) \propto \exp[-10A_{i,\phi}]$, while for the $A_{i,t}$ parameters, stronger priors of the form $p(A_{i,t}) \propto \exp[-100A_{i,t}]$. This was done to prevent the $t$-dependent GP baseline being too flexible and completely degenerate with the transit signal. As in previous work, we fit for the natural logarithm of the inverse correlation length scales ($\ln \eta_{t,i},\ln \eta_{\phi,i}$). For these parameters, and all remaining free parameters, we assumed uniform priors.

\section{Fitting the HST WFC3 spectroscopic light curves} \label{app:hst:spec}

For each spectroscopic channel, we fit the corresponding light curves for all five visits simultaneously, adopting a similar GP methodology to that described in Section \ref{app:hst:broad} for the broadband light curve. However, since the common-mode correction removed most of the systematics prior to light curve fitting, we adopted a somewhat simplified model, with mean function and covariance matrix for the $i$th dataset defined by:
\begin{align}
  \boldsymbol{\mu_i} = \begin{cases}
    ( f_i + m_i t ) \, M(\,t\,;\,\boldsymbol{\alpha_i}\,)  & \text{for forward scans} \\
    ( b_i + m_i t ) \, M(\,t\,;\,\boldsymbol{\alpha_i}\,) & \text{for backward scans}    
    \end{cases}
\end{align}
and:
\begin{align}
\boldsymbol{ \Sigma_i}_{jk} & = K_{\text{SE}}( \, \phi_j, \, \phi_k \, ; \, A_{\phi,i}, \, \eta_{\phi,i} \, ) + \delta_{jk} \beta_i^2\sigma_i^2 \ . \label{eq:specKijk}
\end{align}
In particular, we replace the $t$-dependent $K_{\text{SE}}$ kernel with a linear-$t$ trend for the baseline, and remove the analytic ramp model $R$. For the transit signal $M$, we also fix the mid-times ($T_i$) to the best-fit values determined for the broadband light curve, leaving $\RpRs$ as the only free parameter for each spectroscopic channel, i.e. $\boldsymbol{\alpha_i}=[ \, \RpRs \, ]$. As for the broadband light curve fit, we adopted a quadratic limb darkening law and fixed the coefficients to values determined using the ExoCTK tool, which are listed in Table \ref{table:specfit}. Marginalization was again performed using affine-invariant MCMC.

\section{Independent analysis of the HST WFC3 dataset} \label{app:hst:independent}

To verify the WFC3 analysis presented in Sections \ref{sec:obsdatred} and \ref{sec:lcfits}, a second independent analysis was performed using the method of \cite{2014Natur.505...69K}. In brief, we extracted the spectra from each up-the-ramp sample in an exposure using optimal extraction \citep{1986PASP...98..609H}. The extraction window had a height of 250 pixels. The background was estimated from the median of pixel counts between cross-dispersion rows 6-70 and dispersion columns 450-500. To obtain a final spectrum for each exposure, the up-the-ramp samples were co-added. The data were binned into spectroscopic light curves with similar wavelength spacing as the primary data reduction.

\begin{figure}
\centering  
\includegraphics[width=0.8\columnwidth]{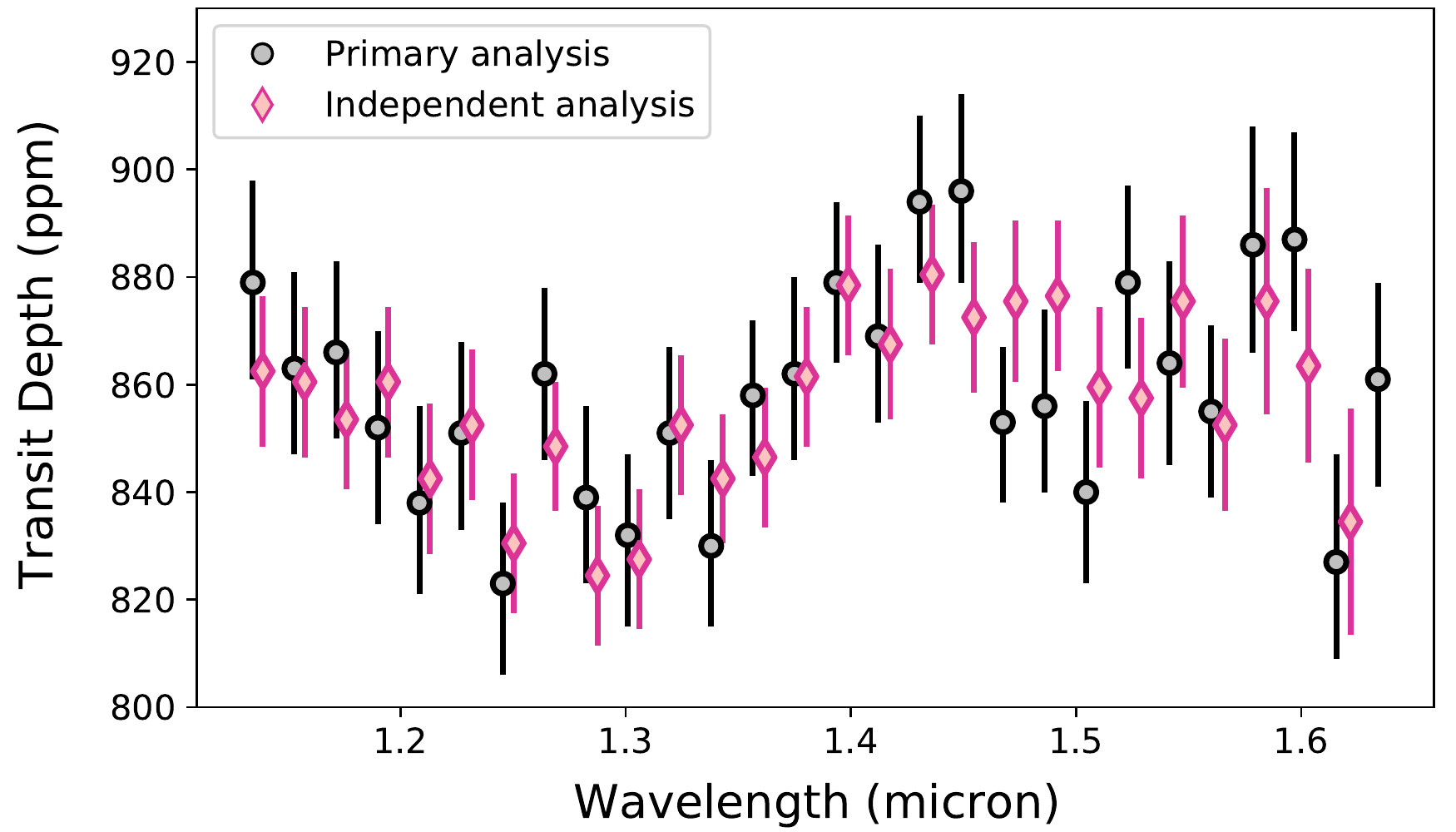}
\caption{Transmission spectra obtained for HD~3167c from the primary analysis described in Sections \ref{sec:obsdatred} and \ref{sec:lcfits}, and a second independent analysis employing the methodology of \cite{2014Natur.505...69K}.}
\label{fig:independent}
\end{figure}

To fit the broadband light curve, we used a transit model together with the \texttt{model-ramp} analytic parameterization for the orbit-long systematics described in \cite{2014Natur.505...69K}. We also simultaneously fit a quadratic function of time for the baseline trend of each visit. We then fit the spectroscopic light curves with the \texttt{divide-white} common-mode systematics model of \cite{2014Natur.505...69K}. In total, the spectroscopic transit light curve fits had free parameters for the planet-to-star radius ratio ($\RpRs$), a linear limb darkening parameter ($u_1$), and a visit-long linear slope in time. We obtained uncertainties on the fit parameters with affine-invariante MCMC using the \texttt{emcee} Python package \citep{2013PASP..125..306F}. The uncertainties on the individual data points were rescaled prior to the final MCMC analysis such that the final reduced $\chi^2$ for the light curve fit was unity; this increased the size of the error bars by a median of $2\%$. We confirmed that the spectroscopic light curve rms bins down with the square root of the number of points per bin, as expected for white, Gaussian-distributed noise. The resulting transmission spectrum is shown in Figure \ref{fig:independent} and is in good agreement with the primary analysis.

\section{Fitting the Spitzer IRAC broadband light curve} \label{app:irac}

For the mean function, we used:
\begin{align}
  \boldsymbol{\mu} &= c_0\,M(\,t\,;\,\boldsymbol{\alpha}\,) \, S(\,t\, , \, f_i \,;\, m \, , \, c_j \, )
\end{align}
where $M$ is a transit model, $c_0$ is a normalization factor, and $S$ is a linear decorrelation of the form:
\begin{align}
  S &= 1 + m t + \sum_{i=1}^{9}{c_if_i(t)} \ , \label{eq:pld}
\end{align}
where $m$ is the slope of a linear time trend, $f_i$ is the time series of the $j$th pixel in a $3 \times 3$ grid centered on the target, and $c_i$ are associated linear coefficients. Equation \ref{eq:pld} corresponds to the pixel level decorrelation (PLD) framework introduced by \cite{2015ApJ...805..132D} to account for intrapixel sensitivity variations that generate the dominant systematics in IRAC time series. For the covariance matrix, we used:
\begin{align}
\boldsymbol{ \Sigma}_{jk} & = K_{\text{SE}}( \, t_j, \, t_k \, ; \, A_{t}, \, \eta_{t} \, )  + \delta_{jk} \beta^2\sigma^2 \ . \label{eq:iracKijk}
\end{align}
We found the $t$-dependent $K_{\text{SE}}$ kernel was required to capture residual correlations in the light curve that were not fully corrected by the PLD.

In total, the free parameters of our IRAC light curve model were:\ $\boldsymbol{\alpha}=[ \, \RpRs, \, T \, ]$ for the transit signal; $c_0$, $m$, and the nine PLD coefficients $c_i$; $A_{t}$ and $\eta_{t}$ for the covariance; and the white noise rescaling factor $\beta$. As for the \textit{K2} fit, we applied the WFC3 posterior constraints as Gaussian priors for $\aRs$ and $b$. For the covariance amplitude, we used a Gamma prior of the form $p(A_{t}) \propto \exp[-100A_{t}]$. Uniform priors were adopted for the remaining free parameters.

To make the GP likelihood computations tractable, we binned the light curve into 2 minute bins, reducing the size of the dataset from $\mathcal{O}(10^5)$ points to $\mathcal{O}(10^2)$ points. Marginalization was performed using \texttt{emcee}. Preliminary fits indicated that the first $\sim 2$ hours of the light curve were poorly accounted for by the PLD, so we chose to discard this segment of the light curve in our final analysis. This reduced the pre-transit baseline to approximately 1.8 hours, which was still sufficient to constrain the baseline flux level.

As for the WFC3 analysis, a quadratic limb darkening law was adopted for both the \textit{K2} and IRAC fits, with limb darkening coefficients fixed to values calculated using the online ExoCTK tool. The latter are listed in Table \ref{table:specfit}, along with the inferred values for $\RpRs$ and the corresponding transits depths $(\RpRs)^2$. The systematics-corrected light curves and best-fit transit models for both datasets are shown in Figure \ref{fig:k2irac}.

\end{document}